\documentclass[amsmath,11pt]{report}
\usepackage{amsfonts,amsmath,amssymb,graphics,graphicx,epsfig,setspace,chapterbib,appendix,multirow,epigraph}
\usepackage{enumerate}\usepackage{theorem}
\usepackage{multicol}
\usepackage[nottoc]{tocbibind}
\usepackage{xcolor}

\begin{document}

\begin{figure}
\centering
\includegraphics[scale=0.8]{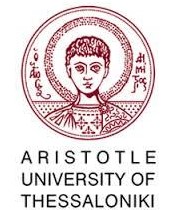}
\end{figure}
\vspace{-15cm}

\date{\empty}

\title{\bf\Huge{Aspects of the Einstein-Maxwell-Weyl coupling}}

\author{\bf\Large{Panagiotis Mavrogiannis}\\\\
{\normalsize Section of Astrophysics, Astronomy and Mechanics,}\\
{\normalsize Department of Physics}\\
{\normalsize Aristotle University of Thessaloniki}\\\\
A dissertation submitted for the degree of \textit{Doctor of Philosophy}}

\maketitle

\pagebreak{}
\tableofcontents{}
\pagebreak{}

\addcontentsline{toc}{chapter}{\protect\numberline{}Preface}

\chapter*{Preface}\label{chap:preface}

\epigraph{
    To the clear sky of Cyprus,\\
    and the sun}

Interesting phenomena and problems arising from the coupling of large-scale electromagnetic fields and spacetime curvature, are introduced and studied within this thesis. From electromagnetic wave propagation in curved spacetime to envisaging a gravito-electromagnetic equivalence on large scales; from magnetic fields' cosmic evolution, and magnetised gravitational collapse in astrophysical environments, to the interaction between electromagnetic and gravitational radiation; the present research work explores some unique characteristics and properties of the gravito-electromagnetic coexistence.\\
Its theoretical framework is generally provided by classical (Maxwellian) electrodynamics in the (4-dimensional) Riemannian spacetime of General Relativity. Chapter 1, dealing with the perspective of a gravito-electromagnetic equivalence in a metric affine geometry (involving torsion and non-metricity), forms an exception to the aforementioned general framework.\\
Regarding its structure, the manuscript consists of four chapters divided in two parts. The first two chapters (forming Part I) are concerned with the Maxwell field in curved, Riemannian (Chapter 1) and (generalised) metric affine (Chapter 2), spacetime framework. Aspects of the Weyl-Maxwell coupling (where the Weyl field refers to long-range curvature) are also included in Chapter 1. The last two chapters (forming Part II) treat magnetohydrodynamics in the general relativistic framework.
In detail, the individual subjects treated per chapter, are the following:\\
As an introduction to Part I, the status of classical electromagnetism in Riemannian spacetime is briefly revisited under a refreshing perspective.\\
\textbf{Chapter 1}: The Maxwell field in Riemannian spacetime is studied basically in terms of potentials, and through the covariant approach to relativity. The general electromagnetic wave equations (involving both kinematic and curvature effects) are derived and subsequently applied to a cosmological and an astrophysical problem. The former regards electromagnetic fields on a linearly perturbed Friedmann-Robertson-Walker background; the latter concerns the interaction between an electromagnetic and a gravitational wave on a Minkowski background (an example of the Weyl-Maxwell coupling).\\
\textbf{Chapter 2}: The interest passes then to a generalised, metric affine geometric framework. Within that, a component of spacetime curvature, associated with length changes, is shown to obey both sets of Maxwell equations. The sourceless one is satisfied by geometric construction while the sourceful one via the consideration of a simple geometric action, same in form with that of classical electrodynamics in Riemannian spacetime. A concept of gravito-electromagnetic equivalence arises thus, putting to question the status of electromagnetism, as a source of gravity, on large scales.\\
A general solution to Faraday's equation at the magnetohydrodynamic limit is derived at the introduction to Part II, and then applied to magnetised gravitational collapse and magnetic elasticity (Chapter 3), as well as to the nucleosynthesis constraint for cosmic magnetic fields (Chapter 4).\\
\textbf{Chapter 3}: Kinematically and gravitationally induced magnetic tension stresses are presented and described, approaching the latter as key factors in curved space magnetohydrodynamics. The aforementioned approach along with our magnetic evolution formula (see the introduction to Part II) are deployed in suggesting a non-collapse criterion (for a magnetised fluid), and in calculating the fracture limit of magnetic forcelines under their gravitational contraction.\\
\textbf{Chapter 4}: Based on the restriction imposed by cosmic nucleosynthesis, our general (parametric dependent) magnetic evolution formula is used to constrain the rate of change of cosmic magnetic fields. Subsequently, we focus our attention to the magnetised Bianchi I cosmological model. In particular, we derive the model's precise evolution formulae, using a novel, convenient and clarifying parametrisation. Moreover, we provide a novel (as far as we know) and detailed qualitative description of the model's small and large scale limits.

\newpage

\part{}

\section{Electromagnetism and spacetime geometry}\label{EMSG}
The electromagnetic field is the only known energy source of vector nature. This feature facilitates a dual coupling between the Maxwell field and the geometry of the host spacetime, mediated by Einstein's equations, on the one hand, and by the Ricci identities on the other.\\

\subsection{Einstein-Maxwell coupling}\label{ssec:Einstein-Maxwell}

Firstly, according to Einstein's equations of gravitation, electromagnetic fields, as a form of energy, along with (ordinary) matter contribute to the formation of spacetime geometry. Hence, in a sense, Maxwell's electromagnetism is incorporated into (or makes part of) General Relativity by providing a kind of source for the gravitational field. Let us write here for reference the relativistic equations for gravity\footnote{We adopt the geometrised units system (refer to the appendix F in~\cite{W1}-Chapter~$2$ bibliography-for details), i.e. $8\pi G=1=c$ ($G$ is the gravitational constant and $c$ is the speed of light), in which all quantities (ordinarily measured in terms of the fundamental units of length $L$, time $T$ and mass $M$) have dimensions expressed as integer powers of length. In particular, we note that mass, time, electric charge and energy have dimensions of length; velocity, force, action and Maxwell potential are dimensionless; Faraday electromagnetic field has inverse length dimensions whilst energy density and electric current density are measured in inverse square length units.},
\begin{equation}
    R_{ab}-\frac{1}{2}Rg_{ab}=T_{ab}\,,
    \label{EFE}
\end{equation}
where the (symmetric) Ricci tensor $R_{ab}$ encodes the local-gravitational field (the Ricci scalar $R=R^{a}{}_{a}$ provides a measure of the average local gravity), and the energy-momentum tensor $T_{ab}$ (Noether conserved currents under translations and rotations) represents the energy sources of gravitation. For ordinary matter and electromagnetic fields, the aforementioned tensor reads: $T_{ab}=T^{\text{(m)}}{}_{ab}+T^{\text{(EM)}}{}_{ab}=T^{\text{(m)}}{}_{ab}+F_{ac}F^{c}{}_{b}+(1/4)F_{cd}F^{cd}g_{ab}$, where $F_{ab}=2\partial_{[a}A_{b]}=\partial_{a}A_{b}-\partial_{b}A_{a}$ is the (antisymmetric) Faraday tensor, written in terms of the Maxwell 4-potential $A_{a}$. Finally, the metric (also symmetric) tensor $g_{ab}$ encodes the geometric properties of spacetime and is used to calculate lengths and angles in Riemannian manifolds. Through Einstein's equations $R_{ab}$ and $g_{ab}$ represent different aspects of the same beingness. In particular, $g_{ab}$ is an agent for forming the Ricci scalar, basically the Lagrangian of relativistic gravitation, and it does not refer to inherent properties of spacetime but it is determined by the coordinate description of the physical system in question.\\

\subsection{Ricci-Maxwell coupling}\label{ssec:Ricci-Maxwell}

Secondly, due to their geometric (vector/tensor) nature, electromagnetic fields directly couple with spacetime curvature via the so-called Ricci identities. The aforementioned coupling is a purely geometrical interaction that goes beyond the usual interplay between matter and geometry monitored by Einstein's equations. In particular, the Ricci identities for the Maxwell 4-potential and the Faraday tensor fields, read (in Riemannian geometry):
\begin{equation}
    2\nabla_{[a}\nabla_{b]}A_{c}=R_{abcd}A^{d} \hspace{15mm} \text{and} \hspace{15mm} 2\nabla_{[a}\nabla_{b]}F_{cd}=-2R_{ab[c}{}^{e}F_{d]e}\,,
    \label{eqn:Ricci-id-4pot}
\end{equation}
respectively, where $R_{abcd}$ is the Riemann curvature tensor encoding the total spacetime curvature. The above relations differ from the gravitational field equations in that firstly, they permit the coupling of only geometric quantities with spacetime curvature. Secondly, they associate geometric fields with the total spacetime curvature (involving the Weyl, long-range curvature, field as well) apart from the local one (as encoded by the Ricci tensor). Finally, note that in a metric affine geometry, the Riemann tensor in the above equations includes contributions from torsion and non-metricity (see the introduction to Chapter~$2$). Also, the metric affine version of eqs~\eqref{eqn:Ricci-id-4pot} includes an extra additive term due to torsion, (e.g. see eqs.~(1.152) and~(1.154) in~\cite{I}-bibliography of Chapter~$2$). \\
The individual timelike ($\Phi$) and spacelike ($\mathcal{A}_{a}$) components (see Chapter~$1$) of the $4$-potential (i.e. $A_{a}=\Phi u_{a}+\mathcal{A}_{a}$, where $u_{a}$ is a timelike $4$-vector, defining a temporal direction) satisfy the following $3$-D versions of the Ricci identities:
\begin{equation}
{\rm D}_{[a}{\rm D}_{b]}\Phi= -\dot{\Phi}\omega_{ab} \label{Ricci-scalar-pot}
\end{equation}
and
\begin{equation}
2{\rm D}_{[a}{\rm D}_{b]}\mathcal{A}_{c}= -2\omega_{ab}\dot{\mathcal{A}}_{\langle c\rangle}+ \mathcal{R}_{dcba}\mathcal{A}^{d}\,,  \label{Ricci-vec-pot}
\end{equation}
respectively, where $\mathcal{R}_{abcd}$ denotes the $3$-D Riemann tensor, determining the geometry of spatial sections. Note that, according to eq.~(\ref{Ricci-scalar-pot}), the spatial gradients of scalars do not commute in rotating spacetimes. This result, as well as the vorticity term on the right-hand side of (\ref{Ricci-vec-pot}), are direct consequences of the Frobenius theorem, which ensures that rotating spacetimes do not contain integrable spacelike hypersurfaces (e.g.~see~\cite{EBH} for a discussion-bibliography of Chapter~$1$).\\
Finally, we mention here for reference the $4$ and $3$-Ricci identities for the electric component of the Maxwell field,
\begin{equation}
2\nabla_{[a}\nabla_{b]}E_{c}= R_{abcd}E^{d}\hspace{6mm}\text{and}\hspace{6mm} 2{\rm D}_{[a}{\rm D}_{b]}E_{c}= -2\omega_{ab}\dot{E}_{\langle c\rangle}+ \mathcal{R}_{dcba}E^{d}\,,
\label{eqn:Ricc-Electr}
\end{equation}
as well as for the shear field,
\begin{equation}
    2{\rm D}_{[a}{\rm D}_{b]}\sigma_{cd}= -2\omega_{ab}\dot{\sigma}_{\langle cd\rangle}+ \mathcal{R}_{ecba}\sigma^{e}{}_{d}+\mathcal{R}_{edba}\sigma^{e}{}_{c}\,.
    \label{eqn:3-Ricci-2nd-rank-tensor}
\end{equation}
 It goes without saying that relations exactly analogous to~\eqref{eqn:Ricc-Electr} hold for the magnetic vector as well.
The aforementioned equations (i.e.~\eqref{eqn:Ricc-Electr} and~\eqref{eqn:3-Ricci-2nd-rank-tensor}) will be used extensively in Chapter~$1$, when deriving the Maxwell field wave equations in Riemannian spacetime, and discussing the interaction between a gravitational and an electromagnetic signal. Also, the gravito-electromagnetic coupling, monitored via the Ricci identities, sets the starting point for our discussion of electromagnetic fields in a metric affine framework (Chapter~2).\\

\newpage

\include{chap:1}

\chapter{\bf The Maxwell field in Riemannian spacetime}\label{chap:1}

Electromagnetic potentials allow for an alternative description of the Maxwell field, the electric and magnetic components of which emerge as gradients of the vector and the scalar potential. We provide a general relativistic analysis of these potentials, by deriving their wave equations in an arbitrary Riemannian spacetime containing a generalised imperfect fluid. Some of the driving agents in the resulting wave formulae are explicitly due to the curvature of the host spacetime. Focusing on the implications of non-Euclidean geometry, we look into the linear evolution of the vector potential in Friedmann universes with nonzero spatial curvature. Our results reveal a qualitative difference in the evolution of the potential between the closed and the open Friedmann models, solely triggered by the different spatial geometry of these spacetimes. We then consider the interaction between gravitational and electromagnetic radiation. We study the effects of the former upon the latter in terms of potentials, and the inverse phenomenon in terms of $E$ and $B$ components. In so doing, we apply the wave formulae of both potentials to a Minkowski background and study the Weyl-Maxwell coupling at the second and third perturbative level respectively. Our solutions, which apply to low-density interstellar environments away from massive compact stars, allow for the resonant amplification, on the one hand, of an electromagnetic signal by gravitational-wave distortions, and on the other hand, of a gravitational signal by electromagnetic radiation.\\

\section{Introduction}\label{sI}
Although potentials are not measurable quantities themselves, they have traditionally been used in physics as an alternative to fields.\footnote{A debate regarding the physicality of the  electromagnetic potentials in quantum mechanics has been raised and driven by the interpretation efforts of the so-called Aharonov-Bohm effect.} The introduction and use of potentials is based on the principle that their differentiation leads to the realisation of the fields themselves.\footnote{There are generally more than one functional forms, the derivatives of which lead to the same result. This fact reflects the well known `gauge freedom' related to the choice of potentials.} Moreover, unless differentiation simplifies the mathematical form of a given function, potentials are generally expected to make the calculations easier to handle. In addition to  their role as auxiliary quantities that can simplify operations, it has been pointed out that electromagnetic potentials could also be introduced directly by fundamental principles (e.g.~charge conservation and action principle), as primary quantities before the fields, with the latter derived subsequently as auxiliary entities (see~\cite{LMEN2,HH} for recent discussions and references therein).

One of the best known potentials in theoretical physics is that of the Maxwell field. In the context of relativity, the electromagnetic potential comes in the form of a 4-vector, with the latter comprising a timelike scalar accompanied by a 3-vector spatial component. The temporal and spatial gradients of these two entities give rise to the electric and the magnetic parts of the Maxwell field. In principle, one could use either the potential or the field description when studying electromagnetic phenomena. Most of the available work, however, employs the latter rather than the former. As a result, certain aspects regarding the behaviour of the electromagnetic potentials are still missing from the available literature. One of the relatively less explored aspects is the coupling between the Maxwell potentials and the geometry of their host spacetime. This area also appears to be one of the most challenging, since the study of electromagnetism in curved spaces has led to rather unconventional phenomenology in a number of occasions (e.g.~see~\cite{DWB}-\cite{PM} for a representative though incomplete list)

One would like to know, in particular, whether and how the Ricci and the Weyl curvature, namely the local and the far components of the gravitational field, couple to the electromagnetic potentials. The interaction with the intrinsic and the extrinsic curvature of the 3-dimensional space hosting the Maxwell field and its implications are additional questions as well. This study attempts to shed light upon these matters, by providing the first (to the best of our knowledge) general relativistic wave-equations of both the vector and the scalar electromagnetic potentials in an arbitrary Riemannian spacetime containing a general imperfect fluid. In so doing, we pay special attention to the geometrical features of the emerging wave formulae and more specifically to that of the vector potential. These features result from the geometrical interpretation of gravity that Einstein's theory introduces and from the vector nature of the Maxwell field. The most compact expression, reflecting the aforementioned coupling between electromagnetism and spacetime curvature, is perhaps the wave equation of the 4-potential. In the absence of sources, the latter reads:
\begin{equation}
\left(\square\delta^a{}_b- R^a{}_b\right)A_a= 0\,,  \label{D'Alen}
\end{equation}
where $\square$ is the d' Alembertian operator, $\delta^a{}_b$ the Kronecker delta, $R^a{}_b$ is the Ricci curvature tensor and $A_a$ is the electromagnetic 4-potential~\cite{MTW}-\cite{DFC}. Note that $\square\delta^a{}_b- R^a{}_b$ defines the so-called de Rham wave-operator, which acts as the generalised d' Alembertian in curved spacetimes.\footnote{Following~\cite{W}, the Ricci curvature term on the left-hand side of (\ref{D'Alen}) underlines the particular attention one needs to pay when obtaining the general relativistic electromagnetic equations from their (flat space) special relativistic counterparts by means of the so-called ``minimal substitution rule''.}

The evolution of the electric and the magnetic components of the Maxwell field in a general spacetime was studied in~\cite{T1}, by means of the 1+3 covariant approach to general relativity~\cite{TCM,EMM}, with the propagation equations given in the form of wave-like formulae. Typically, these describe forced oscillations traveling at the speed of light, with driving terms that include, among others, the curvature of the host spacetime~\cite{T1}. Here, we provide an alternative (though still 1+3 covariant) treatment involving the electromagnetic potential, instead of the $E$ and $B$ components. Starting from Maxwell's equations and adopting the Lorenz gauge, we arrive at a pair of wave-like formulae for the vector and the scalar potentials of the Maxwell field. These hold in any Riemannian spacetime, just like the ones for the electric and the magnetic fields obtained in~\cite{T1}. The qualitative difference is that, there, the matter component was represented by a perfect fluid, whereas here it has the form of a generalised imperfect medium.

The wave equations derived in~\cite{T1} were linearised around a Friedmann-Robertson-Walker (FRW) cosmology with nonzero spatial curvature. In the absence of charges, the solutions revealed that spatial-curvature effects could enhance the amplitude of electromagnetic waves in Friedmann models with hyperbolic spatial geometry. Here, we provide an alternative treatment involving the electromagnetic potential, instead of the actual fields. By linearising the wave-like formula of the vector potential around a Friedmann universe with non-Euclidean spatial hypersurfaces, we arrive at analytic solutions representing generalised forced oscillations. The frequency and the amplitude of the latter depend on the spatial geometry of the background FRW model. The amplitude of the vector potential, in particular, is enhanced when the background Friedmann universe has negative 3-curvature (in full agreement with~\cite{T1} -- see also~\cite{BT}).

We then consider the coupling between the Weyl and the Maxwell fields, namely the interaction between gravitational and electromagnetic radiation. More specifically, we analyse the effects of the former on the propagation of the latter using potentials; while the effects of the latter on the propagation of the former using the $E$ and $B$ components of the Maxwell field. Adopting the Minkowski space as our background, we consider the aforementioned Weyl-Maxwell coupling at the second and third perturbative order respectively. Employing the wave-like formulae of the Maxwell field, we find that the interaction between gravitational and electromagnetic waves can lead to resonant amplification of the initial waveform. In most realistic physical situations, these resonances should result into the linear growth of the electromagnetic signal, in agreement with the earlier studies of~\cite{T3}-\cite{CCG}. In reference to the inverse effect though (i.e. electromagnetic amplification of a gravitational signal), the resonances lead to a quadratic (parabolic) growth of the original gravitational wave. As far as we know, a covariant description of the latter effect has not appeared in the literature. Note that, given the Minkowski nature of the adopted background spacetime, our analysis and our results apply away from massive compact stars to the low-density interstellar/intergalactic environments, where the gravitational field is expected to be weak.

The manuscript starts with a brief presentation of electromagnetic fields and potentials in the framework of the 1+3 covariant formalism, with emphasis on the coupling between electromagnetism and spacetime geometry. The wave-formulae for the vector and the scalar potentials are extracted from Maxwell's equations in~\S~\ref{sMEWF}. These are subsequently linearised around a FRW background with nonzero spatial curvature in~\S~\ref{ssEMPFRWSs} and then employed to study the Weyl-Maxwell coupling at second and third order in~\S~\ref{ssG-WEEMSs} and~\S~\ref{ssec:EM-rad-effects-on-GWs}. We also note that the interested reader can find the necessary background information on the 1+3 covariant formalism in Appendix~\ref{AppA}. Additional technical details and guidance for reproducing the wave formulae for the vector and the scalar electromagnetic potentials are given in Appendix~\ref{AppB}. Finally, in Appendix~\ref{AppC} can be found some details regarding the derivation of the electromagnetically enhanced gravitational waveform.\\

\section{Electromagnetic fields and potentials}\label{sEMFPs}
Maxwell's equations allow for the existence of the electromagnetic 4-potential, the gradients of which lead to the electric and magnetic fields measured by the observers. Therefore, depending on the problem in hand, one is free to choose either representation of the Maxwell field.\\

\subsection{Electric and magnetic vectors}\label{ssEMVs}
The electromagnetic field is covariantly described by the antisymmetric Faraday tensor ($F_{ab}=F_{[ab]}$). Relative to a family of observers, moving along timelike worldlines tangent to the 4-velocity $u_a$ (with $u_au^a=-1$ -- see also Appendix~\ref{AppA1}), the Faraday tensor splits according to:
\begin{equation}
F_{ab}= 2u_{[a}E_{b]}+ \epsilon_{abc}B^{c}\,,  \label{Fab1}
\end{equation}
with $E_a=F_{ab}u^b$ and $B_a=\epsilon_{abc}F^{bc}/2$ representing its electric and magnetic components. Note that $\epsilon_{abc}= \epsilon_{[abc]}$ is the Levi Civita tensor of the 3-D space orthogonal to the observers' worldlines (i.e.~$\epsilon_{abc}u^c=0$ -- see Appendix~\ref{AppA2} as well). The latter guarantees that both the electric and the magnetic fields are 3-D spacelike vectors with $E_au^a=0=B_au^a$.\\
The Faraday tensor also determines the electromagnetic energy-momentum tensor, which satisfies the covariant expression:
\begin{equation}
T^{(\text{em})}_{ab}= -F_{ac}F^c{}_b- \frac{1}{4}\,F_{cd}F^{cd}g_{ab}\,,  \label{Tem}
\end{equation}
where $g_{ab}$ is the spacetime metric with signature $(-,+,+,+)$. Equivalently, we may use decomposition  (\ref{Fab1}) to write
\begin{equation}
T^{(\text{em})}_{ab}= \frac{1}{2}\left(E^{2}+B^{2}\right)u_{a}u_{b}+ \frac{1}{6}\left(E^{2}+B^{2}\right)h_{ab}+ 2\mathcal{Q}_{(a}u_{b)}+ \Pi_{ab}\,, \label{eqn:T-EM1}
\end{equation}
thus explicitly involving the electric and magnetic components. In the above, $E^2=E_aE^a$, $B^2=B_aB^a$ are the squared magnitudes of the electric and the magnetic fields respectively, while $h_{ab}=g_{ab}+u_au_b$ is the symmetric tensor that projects orthogonal to the $u_a$-field (with $h_{ab}u^b=0$ -- see Appendix~\ref{AppA1}).\footnote{It is worth comparing Eq.~(\ref{eqn:T-EM1}) to the stress-energy tensor of a general imperfect fluid. Decomposed relative to the $u_a$ 4-velocity field, the latter reads
\begin{equation}
T^{(m)}_{ab}= \rho u_au_b+ ph_{ab}+ 2q_{(a}u_{b)}+ \pi_{ab}\,,  \label{mTab}
\end{equation}
with $\rho=T_{ab}u^au^b$ and $p=T_{ab}h^{ab}/3$ representing the energy density and the (isotropic) pressure of the matter, while $q_a=-h_a{}^bT_{bc}u^c$ and $\pi_{ab}=h_{\langle a}{}^ch_{b\rangle}{}^dT_{cd}$ are the associated energy-flux vector and viscosity tensor respectively (with $q_au^a=0$, $\pi_{ab}=\pi_{(ab)}$ and $\pi_{ab}u^b=0=\pi_a{}^a$).} Expression (\ref{eqn:T-EM1}) allows for an imperfect-fluid description of the electromagnetic field, where $\rho^{(em)}=(E^2+B^2)/2$ is the energy density and $p^{(em)}=(E^2+B^2)/6$ is the isotropic pressure. These are supplemented by an effective energy-flux, given by the Poynting vector $\mathcal{Q}_a=\epsilon_{abc}E^bB^c$, and by the electromagnetic viscosity tensor $\Pi_{ab}=-E_{\langle a}E_{b\rangle}-B_{\langle a}B_{b\rangle}$ (with $\mathcal{Q}_au^a=0=\Pi_{ab}u^b$ by construction).\footnote{Round and square brackets indicate symmetrisation and antisymmetrisation as usual. Angular brackets, on the other hand, denote the symmetric and traceless part of spacelike tensors. In particular, $E_{\langle a}E_{b\rangle}=E_{(a}E_{b)}-(E^2/3)h_{ab}$. Note that angled brackets are also used to denote the orthogonally projected part of vectors. For instance, $\dot{\mathcal{A}}_{\langle a\rangle}= h_a{}^b\dot{\mathcal{A}}_b$ (see Eq.~(\ref{el-field-pot}) in \S~\ref{ssVSPs} next).}\\

\subsection{Vector and scalar potentials}\label{ssVSPs}
The Faraday tensor can be also expressed in terms of the electromagnetic 4-potential, the existence of which is implied by the set of homogeneous Maxwell's equations (see (\ref{eqn:Maxw-eqns1}b) in \S~\ref{ssMEs} below). More specifically, we have:
\begin{equation}
F_{ab}= \nabla_aA_b- \nabla_bA_a= \partial_aA_b- \partial_bA_a\,,  \label{Fab2}
\end{equation}
where the second equality follows from the symmetry of the connection in Riemannian spaces. Relative to the $u_a$-field of the timelike observers, the 4-potential splits into its temporal and spatial parts according to the decomposition:
\begin{equation}
A_a= \Phi u_a+ \mathcal{A}_a\,,  \label{Aa}
\end{equation}
with $\Phi=-A_au^a$ being the scalar potential and $\mathcal{A}_a=h_a{}^bA_b$ representing the 3-vector potential (so that $\mathcal{A}_au^a=0$). Setting the divergence of the above to zero leads to the expression:
\begin{equation}
\dot{\Phi}+ \Theta\Phi+{\rm D}^{a}\mathcal{A}_{a}+\dot{u}^{a}\mathcal{A}_{a}=0\,, \label{Lorenz-gauge}
\end{equation}
which reproduces the familiar Lorenz-gauge condition (i.e.~$\nabla^{a}A_{a}=0$) in 1+3 covariant form. Note that $\dot{\Phi}=u^a\nabla_a\Phi$ is the time-derivative of the scalar potential, relative to the $u_a$-field. Also, the variables $\dot{u}_a=u^b\nabla_bu_a$ (with $\dot{u}_au^a=0$ by construction) and $\Theta={\rm D}^au_a=\nabla^au_a$ are (irreducible) kinematic quantities, respectively representing the 4-acceleration vector and the volume scalar associated with the 4-velocity field (see Appendices~\ref{AppA1} and~\ref{AppA2} for more details).\\
Combining the definitions of the electric and the magnetic fields (see Eq.~(\ref{Fab1}) in \S~\ref{ssEMVs} above) with decomposition (\ref{Aa}), leads to:
\begin{equation}
E_a= -\dot{\mathcal{A}}_{\langle a\rangle}- \frac{1}{3}\,\Theta\mathcal{A}_a- \left(\sigma_{ab}-\omega_{ab}\right)\mathcal{A}^b- {\rm D}_{a}\Phi- \Phi\dot{u}_a  \label{el-field-pot}
\end{equation}
and
\begin{equation}
B_a= \text{curl}\mathcal{A}_a- 2\Phi\omega_a\,,
\label{magn-field-pot}
\end{equation}
recalling that $\dot{\mathcal{A}}_{\langle a\rangle}= h_a{}^b\dot{\mathcal{A}}_b$ (see footnote~5). Also, $\sigma_{ab}={\rm D}_{\langle b}u_{a\rangle}$ and $\omega_{ab}={\rm D}_{[b}u_{a]}$ are the shear and the vorticity tensors respectively (with $\sigma_{ab}u^b=0=\omega_{ab}u^b$ by default). These quantities, together with the 4-acceleration vector and the volume scalar defined previously, completely determine the kinematics of the observers' worldlines (see also Appendix~\ref{AppA2}). In addition, we have ${\rm curl}\mathcal{A}_a= \epsilon_{abc}{\rm D}^b\mathcal{A}^c$ and $\omega_a=\epsilon_{abc}\omega^{bc}/2$ (with $\omega_au^a=0$) by construction. Relations (\ref{el-field-pot}) and (\ref{magn-field-pot}) express both components of the Maxwell field in terms of the scalar and the 3-vector potentials. Note the kinematic terms on the right-hand side of these expressions, which are induced by the relative motion of neighbouring observers.\footnote{Very similar, though not entirely 1+3 covariant, expressions for the electric and magnetic fields in terms of the potentials can be found in~\cite{TMcD}. Also, written in the static Minkowski space, where $\dot{u}_{a}=0=\Theta= \sigma_{ab}=\omega_{ab}$ by default, Eqs.~\eqref{el-field-pot} and~\eqref{magn-field-pot} recast into the more familiar expressions: (e.g.~see~\cite{J})
\begin{equation}
E_{a}= -\partial_{t}\mathcal{A}_{a}- \partial_{a}\Phi \hspace{15mm} {\rm and} \hspace{15mm}
B_{a}= \text{curl}\mathcal{A}_{a}\,,  \label{Minkowski}
\end{equation}
respectively (with $\text{curl}\mathcal{A}_{a}= \epsilon_{abc}\partial^{b}\mathcal{A}^{c}$ in this case).}\\

\section{Maxwell equations and wave formulae}\label{sMEWF}
Starting from Maxwell's equations one can extract a set of generalised wave formulae. Written in their full form, the latter monitor the evolution of the electric and magnetic field components in an arbitrary Riemannian spacetime~\cite{T1}. In the current section, we will derive the analogous wave-like equations for the electromagnetic vector and scalar potentials.\\

\subsection{Maxwell equations}\label{ssMEs}
The behaviour of electromagnetic fields is determined by Maxwell equations. Written in terms of the Faraday tensor, they take the compact covariant form:
\begin{equation}
\nabla^{b}F_{ab}= J_{a}\hspace{15mm} \text{and} \hspace{15mm} \nabla_{[c}F_{ab]}= 0\,,
\label{eqn:Maxw-eqns1}
\end{equation}
where $J_a=\mu u_a+\mathcal{J}_a$ is the electric 4-current (with $\mu\equiv-J_au^a$ and $\mathcal{J}_a\equiv h_a{}^bJ_b$ representing the electric charge and the associated 3-current respectively). The former set of equations comes from the Maxwellian action (see eq.~\eqref{eq-Einstein-Maxwell-action}), whilst the latter from the Ricci identities for the Maxwell field (recall eq.~\eqref{eqn:Ricci-id-4pot}), both considered in a Riemannian framework. Also, note that constraint (\ref{eqn:Maxw-eqns1}b) implies the existence of the electromagnetic 4-potential ($A_{a}$) seen in eq.~(\ref{Fab2}). Finally, a $1+3$ decomposition of the charge conservation condition $\nabla_{a}J^{a}=0$, leads to:
\begin{equation}
    \dot{\mu}=-\Theta\mu+{\rm D}_{a}\mathcal{J}^{a}-\dot{u}_{a}\mathcal{J}^{a}\,,
    \label{eqn:1+3-charge-conserv}
\end{equation}
where the definitions for $\mu$ and $\mathcal{J}^{a}$ are as stated above. Similarly, applying a $1+3$ decomposition of Maxwell's formulae with the aid of~\eqref{Fab1}, we arrive at the following propagation equations for the electric and magnetic components, namely (refer to appendix~\ref{AppB-Maxwell-eqs})
\begin{equation}
\dot{E}_{\langle a\rangle}= -\frac{2}{3}\Theta E_a+ \left(\sigma_{ab}+\epsilon_{abc}\omega^c\right)E^b+ \epsilon_{abc}\dot{u}^bB^c+ {\rm curl}B_a- \mathcal{J}_a
\label{eqn:el-field-prop}
\end{equation}
and
\begin{equation}
\dot{B}_{\langle a\rangle}= -\frac{2}{3}\Theta B_a+ \left(\sigma_{ab}+\epsilon_{abc}\omega^c\right)B^b- \epsilon_{abc}\dot{u}^bE^c- {\rm curl}E_a\,;
\label{eqn:magn-field-prop}
\end{equation}
as well as to the constraints:
\begin{equation}
{\rm D}^aE_a= \mu- 2\omega^aB_a \hspace{15mm} {\rm and} \hspace{15mm} {\rm D}^aB_a= 2\omega^aE_a\,.
\label{magn-div}
\end{equation}
All the kinematic terms seen on the right-hand side of (\ref{eqn:el-field-prop})-(\ref{magn-div}b) are induced by the relative motion of neighbouring observers (associated with the velocity field $u^{a}$) and vanish in a static Minkowski-like spacetime. Recall that analogous relative-motion effects were also observed in eqs.~(\ref{el-field-pot}) and (\ref{magn-field-pot}) in~\S~\ref{ssVSPs} earlier. It is worth noting that expressions \eqref{eqn:el-field-prop}, \eqref{eqn:magn-field-prop}, (\ref{magn-div}a) and (\ref{magn-div}b) consist of 1+3 covariant versions of Amp\`ere's, Faraday's, Coulomb's and Gauss's laws respectively.\\

\subsection{Wave equations for the potentials}\label{ssWEPs}
The wave formulae of both electromagnetic potentials follow from the sourceful set of Maxwell's equations (see expression (\ref{eqn:Maxw-eqns1}a) in \S~\ref{ssMEs}). More specifically, the wave formula for the vector potential follows from Amp\`ere's law, whereas that of its scalar counterpart derives from Coulomb's law (see eqs.~(\ref{eqn:el-field-prop}) and (\ref{magn-div}a) respectively). Next, we will provide the relevant expressions and refer the reader to Appendix~\ref{AppB} for technical details. Note that both formulae hold in an arbitrary Riemannian spacetime filled with a general imperfect fluid.

Substituting expressions \eqref{el-field-pot} and \eqref{magn-field-pot} into eq.~(\ref{eqn:el-field-prop}) and then imposing the Lorenz-gauge condition (see constraint (\ref{Lorenz-gauge}) in \S~\ref{ssVSPs}, as well as Appendix~\ref{AppB1} for additional details, auxiliary relations and intermediate steps), we arrive at:
\begin{eqnarray}
\ddot{\mathcal{A}}_{\langle a\rangle}- {\rm D}^{2}\mathcal{A}_{a}&=& -\Theta\dot{\mathcal{A}}_{\langle a\rangle}+\frac{1}{3}\left[\frac{1}{2}\,\kappa\left(\rho+3p\right) -\frac{1}{3}\,\Theta^2+ 4\sigma^2-\frac{1}{3}\,\dot{u}_b\dot{u}^b\right]\mathcal{A}_a- \mathcal{R}_{ba}\mathcal{A}^b+ E_{ab}\mathcal{A}^b \nonumber\\ &&\left[\frac{1}{3}\,\Theta(\sigma_{ab}+\omega_{ab}) -\frac{1}{2}\,\kappa\pi_{ab}+2\sigma_{c\langle a}(\sigma^c{}_{b\rangle}-\omega^c{}_{b\rangle}) -2\sigma_{c[a}\omega^c{}_{b]}-\frac{1}{3}\,\dot{u}_{\langle a}\dot{u}_{b\rangle}\right]\mathcal{A}^b \nonumber\\ && -\frac{7}{3}\,\dot{\Phi}\dot{u}_a-\left[\ddot{u}_{\langle a\rangle} +\Theta\dot{u}_a-(\sigma_{ab}+3\omega_{ab})\dot{u}^b-{\rm D}_a\Theta -2{\rm curl}\omega_a\right]\Phi \nonumber\\ &&+\dot{u}^b\left({\rm D}_{\langle b}\mathcal{A}_{a\rangle}+{\rm D}_{[b}\mathcal{A}_{a]}\right)+\frac{2}{3}\,\Theta{\rm D}_a\Phi+ 2(\sigma_{ab}+\omega_{ab}){\rm D}^b\Phi+ \mathcal{J}_a\,,  \label{eqn:vec-pot}
\end{eqnarray}
where ${\rm D}^2={\rm D}^a{\rm D}_a$ is the spatial covariant Laplacian operator. The above is a wave-like formula with extra terms, which reflect the fact that the host spacetime is not static, it contains matter and has non-Euclidean geometry. The latter, namely gravity, is represented by the $3$-Ricci tensor of the spatial sections and by the electric Weyl tensor ($\mathcal{R}_{ab}$ and $E_{ab}$ respectively -- see Appendix~\ref{AppA3} for details). The explicit presence of the 3-Ricci tensor and of the electric Weyl tensor on the right-hand side of (\ref{eqn:vec-pot}) ensures that spatial curvature and the Weyl field can drive fluctuations in the vector potential. Both terms are the direct result of the gravito-electromagnetic coupling reflected in Ricci identities (see Eqs.~(\ref{eqn:Ricci-id-4pot}) and (\ref{Ricci-vec-pot}) in \S~\ref{ssec:Ricci-Maxwell}).

Proceeding in an analogous way, one also obtains a wave equation for the scalar potential. The latter follows from Coulomb's law (see expression (\ref{magn-div}a) in \S~\ref{ssMEs} above) after making use of eqs.~(\ref{el-field-pot}) and (\ref{magn-field-pot}). In so doing (see Appendix~\ref{AppB2} for further details), one arrives at:
\begin{eqnarray}
\ddot{\Phi}-{\rm D}^{2}\Phi&=& -\frac{5}{3}\,\Theta\dot{\Phi}+ \dot{u}^{a}{\rm D}_{a}\Phi+ \left[\frac{1}{2}\,\kappa\left(\rho+3p\right) -\frac{1}{3}\,\Theta^2+2\left(\sigma^2+\omega^2\right) -\dot{u}^{a}\dot{u}_{a}\right]\Phi \nonumber\\
&&+\left[{\rm D}_{a}\Theta-\frac{4}{3}\,\Theta\dot{u}_a +2\text{curl}\omega_a-2\kappa q_{a} +\sigma_{ab}\dot{u}^b+3\epsilon_{abc}\dot{u}^b\omega^c -\ddot{u}_a\right]\mathcal{A}^a \nonumber\\
&&+2\sigma_{ab}{\rm D}^b\mathcal{A}^a- 2\omega_a{\rm curl}\mathcal{A}^a- 2\dot{u}_a\dot{\mathcal{A}}^a+ \mu\,,
\label{eqn:scalar-pot}
\end{eqnarray}
which (like eq.~(\ref{eqn:vec-pot}) above) shows wave-propagation at the speed of light. Comparing the above to expression (\ref{eqn:vec-pot}), one immediately notices the complete absence of explicit curvature terms on the right-hand side of (\ref{eqn:scalar-pot}). This difference, which was largely anticipated given the scalar nature of the related potential, implies that spacetime curvature can only indirectly affect the evolution of $\Phi$.\footnote{An alternative way of extracting the wave-like equations \eqref{eqn:vec-pot} and \eqref{eqn:scalar-pot} is by substituting (\ref{Fab2}) into Maxwell's formulae (see Eqs.~(\ref{eqn:Maxw-eqns1}) in \S~\ref{ssMEs}) and then projecting the resulting expression along and orthogonal to the $u_{a}$-field.}

Unlike the sourceful of Maxwell's formulae, the sourceless (see eq.~(\ref{eqn:Maxw-eqns1}b) in \S~\ref{ssMEs}) is trivially satisfied by the electromagnetic potential. Recall that expression (\ref{eqn:Maxw-eqns1}b) is the one that allows for the presence of the potential in the first place. As a result, substituting relations (\ref{el-field-pot}) and (\ref{magn-field-pot}) into Faraday's and Gauss' laws (see Eqs.~(\ref{eqn:magn-field-prop}) and (\ref{magn-div}b) in \S~\ref{ssMEs}) does not provide any additional propagation or constraint equations, but instead leads to trivial identities.

Finally, before closing this section, we should note that the matter terms seen on the right-hand side of (\ref{eqn:vec-pot}) and (\ref{eqn:scalar-pot}) correspond to the total (effective) fluid. Put another way, the energy density ($\rho$), the pressure ($p$), the energy flux ($q_a$) and the viscosity ($\pi_{ab}$) contain the involved contributions of the electromagnetic field (see \S~\ref{ssEMVs} previously) as well.\\

\section{Application to cosmology and astrophysics}\label{sACAs}
The wave equations of the previous section hold in a general Riemannian spacetime containing an imperfect fluid in the presence of electromagnetic field. As a result, they  can be linearised around almost any background model and applied to a variety of astrophysical and cosmological environments. In what follows, we will consider two characteristic applications.\\

\subsection{Electromagnetic potentials in FRW spacetimes}\label{ssEMPFRWSs}
The high symmetry of the Friedmann models, namely their spatial homogeneity and isotropy ensure that they cannot naturally accommodate electromagnetic fields. Therefore, in order to study the Maxwell field in FRW-like environments, one needs to introduce it as a perturbation.\footnote{One may also introduce a sufficiently random electromagnetic field, which does not affect the isotropy of the Friedmann host but only adds to the total energy of the matter sources. In other words, one may assume that $\langle E_a\rangle=0=\langle B_a\rangle$ on average, whereas $\langle E^2\rangle$, $\langle B^2\rangle\neq0$ (e.g.~see~\cite{LKNS} and references therein). This approach does not serve the purposes of our study, however, given that the effects we are primarily interested in stem from the vector nature of the Maxwell field.} Proceeding along these lines, let us consider a FRW universe filled with a single perfect fluid and then perturb it by allowing for the presence of a source-free electromagnetic field. Then, the wave formula of the vector potential (see Eq.~(\ref{eqn:vec-pot}) in \S~\ref{ssWEPs} above) linearises to:
\begin{equation}
\ddot{\mathcal{A}}_a- {\rm D}^2\mathcal{A}_a= -\Theta\dot{\mathcal{A}}_a- \frac{1}{9}\,\Theta^2\mathcal{A}_a+ \frac{1}{6}\,\kappa\left(\rho+3p\right)\mathcal{A}_a- \mathcal{R}_{ab}\mathcal{A}^b\,,  \label{eqn:vec-pot-FRW}
\end{equation}
given that $\mathcal{J}_a=0$ in the absence of charges. Note that $\mathcal{R}_{ab}=(2K/a^{2})h_{ab}$ in a Friedmann universe, with $K=0,\pm1$ being the $3$-curvature index. Starting from the above, recalling that $\Theta=3H$ in FRW cosmologies and employing the background Raychaudhuri equation, namely
\begin{equation}
\dot{H}= -H^2- \frac{1}{6}\,\kappa\left(\rho+3p\right)\,,  \label{Ray}
\end{equation}
the linear relation \eqref{eqn:vec-pot-FRW} recasts into:
\begin{equation}
\ddot{\mathcal{A}}_a- {\rm D}^2\mathcal{A}_a= -3H\dot{\mathcal{A}}_a- \left(2H^2+\dot{H}+\frac{2K}{a^2}\right)\mathcal{A}_{a}\,, \label{eqn:vec-pot-FRW2}
\end{equation}
where $a=a(t)$ is the cosmological scale factor (with $H=\dot{a}/a$ -- see also Appendix~\ref{AppA2}). The above is a wave-like differential equation, with extra terms due to the expansion and gravity and with time-dependent coefficients. After a simple Fourier decomposition, eq.~(\ref{eqn:vec-pot-FRW2}) leads to the following expression:
\begin{equation}
\ddot{\mathcal{A}}_{(n)}+ \left(\frac{n}{a}\right)^2\mathcal{A}_{(n)}= -3H\dot{\mathcal{A}}_{(n)}- \left(2H^2+\dot{H}+\frac{2K}{a^2}\right)\mathcal{A}_{(n)}\,,  \label{hcAa1}
\end{equation}
for the $n$-th harmonic mode of the vector potential.\footnote{We employ the familiar Fourier expansion $\mathcal{A}_a= \sum_{n}\mathcal{A}_{(n)}\mathcal{Q}^{(n)}_a$, in terms of the vector harmonics $\mathcal{Q}^{(n)}_a$, so that ${\rm D}_{a}\mathcal{A}_{(n)}=0= \dot{\mathcal{Q}}^{(n)}_a$ and ${\rm D}^{2}\mathcal{Q}^{(n)}_a= -(n/a)^2\mathcal{Q}^{(n)}_a$. Note that $n$ is the Laplacian eigenvalue, which coincides with the comoving wavenumber of the mode when the background FRW universe is spatially flat. In that case, as well as in Friedmann models with hyperbolic spatial sections (i.e.~when $K=0,-1$) the eigenvalue is continuous with $n^2>0$. When $K=+1$, on the other hand, the eigenvalue is discrete with $n^2\geq3$ (e.g.~see~\cite{TCM,EMM}).}\\
Our next step is to recast (\ref{hcAa1}) with respect to conformal time ($\eta= \int(dt/a)=\int da/(\dot{a}a)$). In so doing, we arrive at:
\begin{equation}
\mathcal{A''}_{(n)}+ n^{2}\mathcal{A}_{(n)}= -2\left(\frac{a'}{a}\right)\mathcal{A'}_{(n)}- \left(\frac{a''}{a}+2K\right)\mathcal{A}_{(n)}\,,  \label{hcAa2}
\end{equation}
with the primes indicating differentiation in terms of conformal time. Finally, after introducing the rescaled potential $\mathfrak{A}_{(n)}=a\mathcal{A}_{(n)}$, the above takes the compact form:
\begin{equation}
\mathfrak{A}''_{(n)}+ \left(2K+n^{2}\right)\mathfrak{A}_{(n)}= 0\,,  \label{hcAa3}
\end{equation}
where $K=0,\pm1$ depending on the geometry of the background spatial hypersurfaces. This wave equation agrees with the one obtained in~\cite{K} (compare to eq.~(17) there), provided the latter is applied to source-free electromagnetic fields, or to environments of very low (essentially zero) electrical conductivity.\\
Assuming FRW backgrounds with Euclidean or spherical spatial geometry, namely setting $K=0,+1$, eq.~(\ref{hcAa3}) leads to the following oscillatory solution for the vector potential,
\begin{equation}
\mathcal{A}_{(n)}= \frac{1}{a}\left[\mathcal{C}_1\cos\left(\sqrt{n^2+2K}\,\eta\right) +\mathcal{C}_2\sin\left(\sqrt{n^2+2K}\,\eta\right)\right]\,,  \label{Aa0+1}
\end{equation}
with the integration constants ($\mathcal{C}_1$ and $\mathcal{C}_2$) determined by the initial conditions. Therefore, in flat and closed Friedmann models, the vector potential oscillates with amplitude that decays as $\mathcal{A}_{(n)}\propto1/a$ on all scales. Recall that, in a flat FRW universe, the wavenumber is continuous with $n^2>0$, while $\eta>0$ as well. When dealing with closed FRW models, on the other hand, $n$ is discrete with $n^2\geq3$. Also, in that case, conformal time satisfies the constraint $\eta\in[0,2\pi/(1+3w)]$, where $w=p/\rho$ is the barotropic index of matter.

In spatially open Friedmann universes, with $K=-1$, the coefficient $2K+n^2$ of the second term on the left-hand side of (\ref{hcAa3}) is positive only when $n^2>2$. On the associated scales, the vector potential oscillates with decreasing amplitude in line with solution (\ref{Aa0+1}). However, on longer wavelengths (those with $0<n^2<2$), the solution of eq.~(\ref{hcAa3}) reads:
\begin{eqnarray}
\mathcal{A}_{(n)}&=& \frac{1}{a}\left[\mathcal{C}_1\cosh\left(\sqrt{|n^2+2K|}\,\eta\right) +\mathcal{C}_2\sinh\left(\sqrt{|n^2+2K|}\,\eta\right)\right] \nonumber\\ &=&\frac{1}{a}\left(\mathcal{C}_3e^{\eta\sqrt{2-n^2}} +\mathcal{C}_4e^{-\eta\sqrt{2-n^2}}\right)\,,  \label{Aa-1}
\end{eqnarray}
since $K=-1$. In Friedmann models with hyperbolic spatial geometry, the scale factor evolves as $a\propto\sinh(\beta\eta)^{1/\beta}$, where $\beta=(1+3w)/2$ and $\eta>0$. Therefore, when the universe is dominated by conventional matter with $\beta>0$, the late-time evolution of the scale factor is $a\propto e^{\eta}$. In such an environment, the dominant mode of solution (\ref{Aa-1}) evolves according to the power law:
\begin{equation}
\mathcal{A}_{(n)}\propto a^{\sqrt{2-n^2}-1}\,,  \label{amplAa}
\end{equation}
with $0<n^2<2$. Consequently, as long as $1<n^2<2$ the vector potential keeps decaying, though at a rate slower than $\mathcal{A}\propto1/a$. However, on longer wavelengths (those with $0<n^2<1$), the amplitude of the vector potential starts increasing. In fact, at the infinite wavelength limit (i.e.~for $n\rightarrow0$,) the vector potential grows as $\mathcal{A}_{(0)}\propto a^{\sqrt{2}-1}$. Therefore, in perturbed FRW cosmologies with open spatial sections, and on sufficiently large scales, the decay of the electromagnetic vector potential is reversed solely due to curvature effects.

Solutions (\ref{Aa0+1}) and (\ref{Aa-1}) are in full agreement with the ones describing the linear evolution of electric and magnetic fields in Friedmann models (see~\cite{T1} for details). When the FRW background is flat or closed (i.e.~for $K=0,+1$), for example, the magnetic field obeys the oscillatory solution:
\begin{equation}
B_{(n)}= \frac{1}{a^2}\left[\mathcal{C}_1\cos\left(\sqrt{n^2+2K}\,\eta\right) +\mathcal{C}_2\sin\left(\sqrt{n^2+2K}\,\eta\right)\right]\,, \label{eqn:B-amplitude1}
\end{equation}
on all scales. The above result also holds in perturbed Friedmann cosmologies with open spatial sections, as long as $n^2>2$. Otherwise, namely on longer wavelengths with $0<n^2<2$, we have:
\begin{equation}
B_{(n)}= \frac{1}{a^2}\left[\mathcal{C}_1\cosh\left(\eta\sqrt{2-n^2}\right) +\mathcal{C}_2\sinh\left(\eta\sqrt{2-n^2}\right)\right]\,, \label{eqn:B-amplitude2}
\end{equation}
since $K=-1$.\footnote{The agreement between the sets (\ref{Aa0+1}), (\ref{eqn:B-amplitude1}) and (\ref{Aa-1}), (\ref{eqn:B-amplitude2}) becomes intuitively plausible once we recall that $B_a= \epsilon_{abc}{\rm D}^b\mathcal{A}^c$ to linear order on FRW backgrounds (see Eq.~(\ref{magn-field-pot}) in \S~\ref{ssVSPs} earlier).} This solution exhibits exponential behaviour closely analogous to that of the vector potential seen in (\ref{Aa-1}). More specifically, in open Friedmann models with conventional matter, the dominant magnetic mode of (\ref{eqn:B-amplitude2}) obeys the power law $B_{(n)}\propto a^{\sqrt{2-n^2}-2}$, as long as $0<n^2<2$~\cite{BT,T2}. Again, the reason for the qualitative change in the magnetic evolution is the negative curvature of the universe's spatial sections.

Following (\ref{amplAa}) and (\ref{eqn:B-amplitude2}), in Friedmann universes with hyperbolic spatial geometry, both the vector potential and the magnetic field are superadiabatically amplified, a term originally coined in gravitationally-wave studies~\cite{G}.\footnote{The reader is referred to~\cite{dGML} for a comparison of graviton production in closed and open Friedmann models.} It should be noted that, in our case, the superadiabatic amplification occurs despite the conformal invariance of the Maxwell field, which still holds. This happens because, in contrast to the flat FRW spacetime which is globally conformal to the Minkowski space, the conformal flatness of its curved counterparts is only local (e.g.~see~\cite{S}-\cite{IKM}). As a result, in the latter type of models, the electromagnetic wave equation acquires extra curvature-related terms and the familiar adiabatic decay law is not a priori guaranteed. Instead, on spatially open FRW backgrounds, the Maxwell field can be superadiabatically amplified (see also~\cite{BT} for further discussion).\\

\subsection{Gravitational-wave effects on electromagnetic 
signals}\label{ssG-WEEMSs}
Studies on the interaction between gravitational and electromagnetic waves have a long history, with most of the available treatments involving the electric and the magnetic fields directly (e.g.~see~\cite{CCG} and~\cite{Co}-\cite{BH}). In what follows, we will provide an alternative approach that involves the potentials of the Maxwell field.

\subsubsection{The Weyl-Maxwell coupling in Minkowski 
space}\label{sssW-MCMS}
Provided that the gravito-electromagnetic interaction takes place in the low-density interstellar space, away from massive compact stars, we may assume that the host environment is described by the Minkowski spacetime. There, we may also treat both the electromagnetic and the gravitational waves as test fields propagating in an otherwise empty and static space. In such an environment, the wave formulae of the vector and the scalar potentials (see Eqs.~(\ref{eqn:vec-pot}) and (\ref{eqn:scalar-pot}) in \S~\ref{ssWEPs}) linearise to
\begin{equation}
\ddot{\mathcal{A}}_a- {\rm D}^2\mathcal{A}_a= 0 \hspace{15mm} {\rm and} \hspace{15mm} \ddot{\Phi}- {\rm D}^2\Phi= 0\,,  \label{Mwp}
\end{equation}
respectively. The above accept simple plane-wave solutions of the form:
\begin{equation}
\mathcal{A}_{(n)}= \mathcal{C}\sin(nt+\mathcal{\theta}_\mathcal{C}) \hspace{15mm} {\rm and} \hspace{15mm} \Phi_{(n)}= \mathcal{D}\sin(nt+\mathcal{\theta}_\mathcal{D})\,, \label{eqn:orig-scalar-pot}
\end{equation}
with $\mathcal{A}_{(n)}$ representing the $n$-th harmonic modes of the vector potential and $\Phi_{(n)}$ the one of its scalar counterpart.\footnote{Solution (\ref{eqn:orig-scalar-pot}a) follows after introducing the harmonic splitting $\mathcal{A}_a= \sum_n\mathcal{A}_{(n)}\mathcal{Q}_a^{(n)}$ (see also footnote~9), while in (\ref{eqn:orig-scalar-pot}b) we have assumed that $\Phi=\sum_n\Phi_{(n)}\mathcal{Q}^{(n)}$. In the latter case, $\mathcal{Q}^{(n)}$ are standard scalar harmonic functions, with ${\rm D}^2\mathcal{Q}^{(n)}=-n^2\mathcal{Q}^{(n)}$. Also, $\dot{\mathcal{Q}}^{(n)}=0={\rm D}_a\Phi_{(n)}$ by construction.} Given the flatness of the Minkowski background, $n$ is the physical wavenumber of the mode, with $n^2=n_an^a$ and $n_a$ representing the corresponding eigenvector. Also, $\mathcal{C}$, $\mathcal{D}$ and $\mathcal{\theta}_{\mathcal{C}}$, $\mathcal{\theta}_{\mathcal{D}}$ are the associated amplitudes and phase constants, to be determined by the initial conditions.

Within the framework of the 1+3 covariant approach, gravitational radiation is described by the electric and the magnetic components of the Weyl field (see Appendix~\ref{AppA3}). Also, isolating linear gravitational waves requires imposing a number of constraints to guarantee that only the pure-tensor part of the free gravitational field is accounted for~\cite{TCM,EMM}. In practice, this means ensuring that ${\rm D}^bE_{ab}=0={\rm D}^bH_{ab}$ and that only the transverse component of these traceless tensors survives. Given the absence of matter and the static nature of the Minkowski space, this is achieved by demanding that $\omega_a=0=\dot{u}_a$ to first order. These translate into the following linear relations:
\begin{equation}
\dot{\sigma}_{ab}= -E_{ab}\,, \hspace{8mm} \dot{E}_{ab}=\text{curl}H_{ab} \hspace{8mm} {\rm and} \hspace{8mm} H_{ab}= {\rm curl}\sigma_{ab}\,,  \label{MsGW}
\end{equation}
between the Weyl tensors and the shear (see Eqs.~(\ref{shear-prop}) and (\ref{kcon3}) in Appendix~\ref{AppA2}). Therefore, in our environment, the linear evolution of both $E_{ab}$ and $H_{ab}$ is determined by shear perturbations and more specifically by the transverse (i.e.~the pure tensor -- ${\rm D}^b\sigma_{ab}=0$) part of the shear. The latter satisfies the wave equation~\cite{T3,KT}:
\begin{equation}
\ddot{\sigma}_{ab}- {\rm D}^2\sigma_{ab}=0\,.  \label{Mwsh}
\end{equation}
Note that in deriving the above, we have taken into account that:
\begin{equation}
   \text{curl}H_{ab}=\text{curl}\left(\text{curl}\sigma_{ab}\right)=-{\rm D^{2}}\sigma_{ab}+\frac{1}{2}\mathcal{R}_{ac}\sigma^{c}{}_{b}+\frac{1}{2}\mathcal{R}_{dbac}\sigma^{cd}
    \label{double-curl-sigma}
\end{equation}
which to first-order reduces to
\begin{equation}
    \text{curl}H_{ab}=-{\rm D^{2}}\sigma_{ab}\,.
    \label{double-curl-1st}
\end{equation}
Now considering the second temporal derivative and employing the harmonic decomposition of the shear tensor\footnote{We assume a monochromatic solution for simplicity, so that: $\sigma_{ab}= \sigma_{(k)}\mathcal{Q}^{(k)}_{ab}$, where $\dot{\mathcal{Q}}^{(k)}_{ab}=0={\rm D}_a\sigma_{(k)}$, and $\mathcal{Q}^{(k)}_{ab}$ are pure-tensor harmonics that satisfy the constraints $\mathcal{Q}_{ab}^{(k)}=\mathcal{Q}_{(ab)}^{(k)}$, ${\rm D}^b\mathcal{Q}_{ab}^{(k)}=0$ and ${\rm D}^2\mathcal{Q}_{ab}^{(k)}=-k^2\mathcal{Q}_{ab}^{(k)}$, ($k^{2}=k^{a}k_{a}$ with $k^{a}$ being the wave 3-vector)}, equation~\eqref{Mwsh} transforms into:
\begin{equation}
    \ddot{\sigma}_{(k)}+k^{2}\sigma_{(k)}=0\,.
    \label{shear-wave-k-mode}
\end{equation}
The latter accepts the following harmonic solution:
\begin{equation}
\sigma_{(k)}= \mathcal{G}\sin(kt+\mathcal{\theta}_\mathcal{G})\,. \label{eqn:orig-shear}
\end{equation}
In the above $k$ is the physical wavenumber of the mode, $\mathcal{G}$ is the amplitude of the gravitational wave and $\mathcal{\theta}_\mathcal{G}$ is the associated phase. Solution~\eqref{eqn:orig-shear} represents the amplitude of a monochromatic, transverse, gravitational plane-wave solution.\\
Solutions (\ref{eqn:orig-scalar-pot}) describe linear electromagnetic waves propagating on a Minkowski background in terms of the associated potentials, while solution (\ref{eqn:orig-shear}) does the same for gravitational radiation in terms of the corresponding shear perturbations. The interaction between these two sources is monitored by the wave formulae:
\begin{equation}
\ddot{\tilde{\mathcal{A}}}_a- {\rm D}^2\tilde{\mathcal{A}}_a= 2\sigma_{ab}{\rm D}^b\Phi \hspace{15mm} {\rm and} \hspace{15mm}
\ddot{\tilde{\Phi}}- {\rm D}^2\tilde{\Phi}= 2\sigma_{ab}{\rm D}^{a}\mathcal{A}^b\,,   \label{eqn:scalar-pot-coupl}
\end{equation}
at the second perturbative level (see eqs.~\eqref{eqn:vec-pot} and~\eqref{eqn:scalar-pot} in \S~\ref{ssWEPs}). Note that $\mathcal{A}_{a}$ and $\Phi$ represent the (linear) potentials prior to the gravito-electromagnetic interaction, while their `tilded' counterparts (i.e.~$\tilde{\mathcal{A}}_{a}$ and $\tilde{\Phi}$) are the (second order) potentials that emerged from the interaction. Also, in deriving \eqref{eqn:scalar-pot-coupl}, we have taken into account that, on our Minkowski background, the Gauss-Codacci equation (see (\ref{GC}) in Appendix~\ref{AppA3}) linearises to $\mathcal{R}_{ab}=E_{ab}$. It is also worth noting that the wave formulae (\ref{eqn:scalar-pot-coupl}a) and (\ref{eqn:scalar-pot-coupl}b) account for the ``backreaction'' of the scalar potential upon the its vector counterpart and vice-versa. Including these effects allows us to extend the analysis of~\cite{T3,KT}, where the analogous backreaction of the electric upon the magnetic component (and vice-versa) was bypassed.

We proceed to analyse the coupling between the Weyl and the Maxwell fields, by harmonically decomposing the gravitationally induced potentials. In other words, we set:
\begin{equation}
\tilde{\mathcal{A}}_a= \tilde{\mathcal{A}}_{(\ell)}\tilde{\mathcal{Q}}^{(\ell)}_a \hspace{10mm} {\rm and} \hspace{10mm} \tilde{\Phi}= \tilde{\Phi}_{(\ell)}\tilde{\mathcal{Q}}^{(\ell)}\,,  \label{hsplit2}
\end{equation}
where $\ell$ is the physical wavenumber of the induced modes (with $\ell^2=\ell_a\ell^a$ and $\ell_a$ being the associated wavevector).\footnote{The vector and scalar harmonics seen in  (\ref{hsplit2}) are $\tilde{\mathcal{Q}}^{(\ell)}_a= \mathcal{Q}^{(n)}\mathcal{Q}^{(k)}_{ab}n^b$ and $\tilde{\mathcal{Q}}^{(\ell)}= \mathcal{Q}^{(k)}_{ab}\mathcal{Q}^a_{(n)}n^b$ by construction, where $n_a$ is the wavevector of the potentials. Note that, since $\dot{\mathcal{Q}}^{(n)}=0=\dot{\mathcal{Q}}^{(k)}_{ab}$, it follows that $\dot{\tilde{\mathcal{Q}}}^{(\ell)}_a=0= \dot{\tilde{\mathcal{Q}}}^{(\ell)}$ as well. In addition, recalling that ${\rm D}^2\mathcal{Q}^{(n)}=-n^2\mathcal{Q}^{(n)}$ and that ${\rm D}^2\mathcal{Q}^{(k)}_{ab}=-k^2\mathcal{Q}^{(n)}_{ab}$, one can show that ${\rm D}^2\tilde{\mathcal{Q}}^{(\ell)}_a= -\ell^2\tilde{\mathcal{Q}}^{(\ell)}_a$ and that ${\rm D}^2\tilde{\mathcal{Q}}^{(\ell)}= -\ell^2\tilde{\mathcal{Q}}^{(\ell)}$, with $\ell$ satisfying conditions (\ref{l-k-n}).} These are related to the wavevectors and the wavenumbers of the initially interacting sources via the expressions:
\begin{equation}
\ell_a= k_a+ n_a \hspace{10mm} {\rm and} \hspace{10mm} \ell^2= n^2+ k^2+ 2nk\cos\phi\,,  \label{l-k-n}
\end{equation}
with $0\leq\phi\leq\pi$ representing the interaction angle of the original linear waves.

\subsubsection{Weyl-Maxwell resonances}\label{sssWMRs}
Substituting decompositions (\ref{hsplit2}), together with those of the initially interacting electromagnetic and gravitational signals (see footnotes~9 and~10) back into the second-order formulae (\ref{eqn:scalar-pot-coupl}a) and (\ref{eqn:scalar-pot-coupl}b), the latter take the form:
\begin{equation}
\ddot{\tilde{\mathcal{A}}}_{(\ell)}+ \ell^{2}\tilde{\mathcal{A}}_{(\ell)}= 2\sigma_{(k)}\Phi_{(n)} \hspace{15mm} {\rm and} \hspace{15mm} \ddot{\tilde{\Phi}}_{(\ell)}+ \ell^2\tilde{\Phi}_{(\ell)}= 2\sigma_{(k)}\mathcal{A}_{(n)}\,,  \label{hwtAtF}
\end{equation}
respectively.\footnote{The phase factor ${\rm e}^{\imath\pi/2}$ in the 3-gradient of the potential has been `absorbed' into the associated wavevector.} Employing the linear solutions (\ref{eqn:orig-scalar-pot}) and (\ref{eqn:orig-shear}), the first of the above differential equations recasts as:
\begin{equation}
\ddot{\tilde{\mathcal{A}}}_{(\ell)}+ \ell^2\tilde{\mathcal{A}}_{(\ell)}= \mathcal{E}\left\{\cos[(k-n)t+\theta_{\mathcal{E}_1}] -\cos[(k+n)t+\theta_{\mathcal{E}_2}]\right\}\,,  \label{ddcA}
\end{equation}
while the latter reads:
\begin{equation}
\ddot{\Phi}_{(\ell)}+ \ell^2\Phi_{(\ell)}= \mathcal{M}\left\{\cos[(k-n)t+\theta_{\mathcal{M}_1}] -\cos[(k+n)t+\theta_{\mathcal{M}_2}]\right\}\,.  \label{ddPhi}
\end{equation}
Here, $\mathcal{E}=\mathcal{G}\mathcal{D}$ and  $\mathcal{M}= \mathcal{G}\mathcal{C}$ are the amplitudes of the gravitationally induced potential waves, while $\theta_{\mathcal{E}_{1,2}}= \theta_\mathcal{G}\mp\theta_\mathcal{D}$ and $\theta_{\mathcal{M}_{1,2}}=\theta_{\mathcal{G}}\mp \theta_{\mathcal{C}}$, $\theta_{J_{2}}\equiv \theta_{F}+\theta_{D}$ are the associated phases (all fixed at the onset of the gravito-electromagnetic interaction). According to (\ref{ddcA}) and (\ref{ddPhi}), the induced electromagnetic signal is driven by the superposition of two waves, with effective wave numbers $m_{1,2}=k\mp n$. Solving eqs.~(\ref{ddcA}) and (\ref{ddPhi}) leads to:
\begin{equation}
\tilde{\mathcal{A}}_{(\ell)}= \mathfrak{D}\sin(\ell t+\vartheta)+ \mathcal{K}_1\cos[m_1t+\theta_{\mathcal{E}_1}]- \mathcal{K}_2\cos[m_2t+\theta_{\mathcal{E}_2}]  \label{cA}
\end{equation}
and
\begin{equation}
\Phi_{(\ell)}= \mathfrak{D}\sin(\ell t+\vartheta)+ \mathcal{L}_1\cos[m_1t+\theta_{\mathcal{M}_1}]- \mathcal{L}_2\cos[m_2t+\theta_{\mathcal{M}_2}]\,,  \label{Phi}
\end{equation}
respectively. Note that $\mathfrak{D}$, $\vartheta$, $\mathcal{K}_{1,2}$ and $\mathcal{L}_{1,2}$ are constants determined at the onset of the Weyl-Maxwell interactions, with the latter two given by:
\begin{equation}
\mathcal{K}_{1,2}= \frac{\mathcal{E}}{\ell^2-m_{1,2}^2} \hspace{15mm} {\rm and} \hspace{15mm} \mathcal{L}_{1,2}= \frac{\mathcal{M}}{\ell^{2}-m_{1,2}^2}\,.  \label{cKcL}
\end{equation}
Accordingly, the gravito-electromagnetic coupling leads to resonances when $\ell\rightarrow m_{1,2}=k\mp n$. In particular, when $\ell\rightarrow m_1=k-n$, relation~\eqref{l-k-n} implies that the two original waves propagate in opposite directions (i.e. $\phi\rightarrow\pi$). When $\ell\rightarrow m_2=k+n$, on the other hand, the original waves propagate along the same direction (i.e. $\phi\rightarrow0$). Note that these results are in agreement with those obtained after employing the electromagnetic fields instead of their potentials~\cite{T3,KT}.

Despite the appearances, the resonances identified in this section do not generally suggest an arbitrarily strong enhancement of the emerging electromagnetic wave. Instead, and in analogy with forced harmonic oscillations in classical mechanics, the aforementioned resonances imply linear (in time) growth for the amplitude of the electromagnetic signal. Typically, this requires the `smooth' transition between the potentials prior and after the interaction, namely it follows naturally after imposing the conditions $\tilde{\mathcal{A}}_{(\ell)}=\mathcal{A}_{(n)}$ and $\tilde{\Phi}_{(\ell)}=\Phi_{(n)}$ at the onset of the Weyl-Maxwell coupling. We refer the reader to \S~III in~\cite{KT} for a thorough discussion of the gravito-electromagnetic case, as well as to~\cite{LL} for the presentation of the mechanical analogue.\\

\subsection{Electromagnetic radiation effects on gravitational signals}\label{ssec:EM-rad-effects-on-GWs}

We now examine the inverse problem, namely the waveform and the resonances arising from the effect of electromagnetic radiation on a gravitational wave signal. In contrast to the previous case, we work now with the $E$ and $B$ components of the Maxwell field, instead of the potentials. As it will become clear in the following, potentials do not facilitate calculations in this case.

\subsubsection{The Weyl-Maxwell coupling in Minkowski space} \label{sssec:Grav-EM-waves-Minkowski}

Our background space assumption remains the same as in the inverse problem (recall eq.~\eqref{Mwsh}). In other words, away from massive compact stars and in low-density interstellar/intergalactic environments, where the gravitational field is expected to be weak, we consider the propagation of gravitational and electromagnetic radiation on a nearly empty, static and irrotational ($\rho\rightarrow 0=\Theta$, $\mu=0=\mathcal{J}_{a}$ and $\omega_{ab}=0$), perturbed Minkowski background. Given the above, Euler's equation implies that:
\begin{equation}
    \dot{u}_{a}=\lim_{P \to 0}\frac{{\rm D}_{a}P}{P}=0 \quad \text{or a constant}\,,
    \label{4-accel-approx}
\end{equation}
according to de l' H\^{o}pital's rule (note that $D_{a}P=\partial_{a}P$, where $\partial_{a}$ represents the ordinary $3$-D gradient operator). 

\subsubsection{Linear equations}\label{sssec:Linear-eqs}

On the one hand, the (first order) shear wave equation is the same with that used in the inverse problem. On the other hand, Maxwell's equations in terms of $E^{a}$ and $B^{a}$, namely~\eqref{eqn:el-field-prop}, \eqref{eqn:magn-field-prop} and~\eqref{magn-div}, reduce to:
\begin{equation}
    \dot{E}_{a}=\text{curl}B_{a}\,,
    \label{dot-E-appr}
\end{equation}
\begin{equation}
    \dot{B}_{a}=-\text{curl}E_{a}
    \label{dot-B-appr}
\end{equation}
and
\begin{equation}
    {\rm D}^{a}E_{a}=0={\rm D}^{a}B_{a}\,.
    \label{E-B-div-appr}
\end{equation}
As we will see in the following, the zero divergence conditions for the electromagnetic fields imply the existence of transverse wave solutions. On taking the dot derivative of~\eqref{dot-E-appr} and~\eqref{dot-B-appr}, we arrive at the linear wave equations:
\begin{equation}
    \ddot{E}_{a}-{\rm D^{2}}E_{a}=0 \hspace{15mm} {\rm and} \hspace{15mm} \ddot{B}_{a}-{\rm D^{2}}B_{a}=0\,,
    \label{E-B-wave-eqs-appr}
\end{equation}
where we have taken into account that to linear order, we have~\cite{T3}:
\begin{equation}
    \ddot{E}_{a}\approx \text{curl}\dot{B}_{a}\approx -\text{curl}(\text{curl}E_{a})\approx +{\rm D^{2}}E_{a}\,.
    \label{double-curl-E-appr}
\end{equation}
A similar relation holds for the magnetic field (for details refer to the Appendix). Subsequently, the harmonic splitting of the fields \footnote{In analogy with the Maxwell potentials, we now harmonically decompose the $E$ and $B$ components: $E_{a}= E_{(n)}\mathcal{Q}^{(\epsilon)(n)}_{a}$ and $B_{a}= B_{(n)}\mathcal{Q}^{(\beta)(n)}_{a}$, where $\dot{\mathcal{Q}}^{(\epsilon,\beta)(n)}_{a}=0={\rm D}_{a}E_{(n)}={\rm D}_{a}B_{(n)}$, and $\mathcal{Q}^{(\epsilon,\beta)(n)}_{a}$ are pure-tensor harmonics that satisfy the constraints ${\rm D}^{a}\mathcal{Q}_{a}^{(\epsilon,\beta)(n)}=0$ and ${\rm D}^2\mathcal{Q}_{a}^{(\epsilon,\beta)(n)}=-n^2\mathcal{Q}_{a}^{(\epsilon,\beta)(n)}$ ($n^{2}=n^{a}n_{a}$ with $n^{a}$ being the electromagnetic wave 3-vector).} transforms equations~\eqref{E-B-wave-eqs-appr} into:
\begin{equation}
    \ddot{E}_{a}+n^{2}E_{a}=0 \hspace{15mm} {\rm and} \hspace{15mm} \ddot{B}_{a}+n^{2}B_{a}=0\,.
    \label{E-B-wave-eqs-appr2}
\end{equation}
The above accept the following monochromatic, transverse, plane-wave solutions:
\begin{equation}
    E_{(n)}=C\sin(nt+\theta_{C}) \hspace{15mm} {\rm and} \hspace{15mm} B_{(n)}=D\sin(nt+\theta_{D})\,,
    \label{E-B-sol}
\end{equation}
where $C$, $D$ and $\theta_{C}$, $\theta_{D}$ are the associated amplitude and phase constants respectively.

\subsubsection{Second order (resonant) waveform}\label{ssec:2nd-order-GWs}

 We have considered so far electromagnetic and gravitational waves as linear perturbations (represented by the Maxwell and Weyl fields, namely $E^{a}$, $B^{a}$, and $E_{ab}$, $H_{ab}$, the latter pair reduced to $\sigma_{ab}$) propagating on a Minkowski background. In the following subsections, we allow for higher order terms and examine the arising gravitational wave form; firstly, up to second order and subsequently up to third order terms, involving the Maxwell-Weyl coupling. Being particularly interested in pointing out the electromagnetic amplification of gravitational radiation, we work under the assumption that the electromagnetic density is significantly greater than the gravitational, i.e. $\mathbf{E^{2}\sim B^{2}\gg \sigma^{2}}$. In practice, the aforementioned approximation excludes second and third order shear terms (i.e. $\sigma^{2}$ and $\sigma^{3}$) from our discussion.\\
In the first place, considering up to second-order perturbations in reference to the Minkowski background, the shear wave equation reads:
\begin{eqnarray}
    &&\ddot{\sigma}^{(2)}_{ab}-{\rm D^{2}}\sigma^{(2)}_{ab}=4\dot{\sigma}_{c\langle a}\sigma^{c}{}_{b\rangle}-2E_{\langle a}\text{curl}B_{b\rangle}+2B_{\langle a}\text{curl}E_{b\rangle}\hspace{5mm}\text{or approximately}\nonumber\\
    &&\ddot{\sigma}^{(2)}_{ab}-{\rm D^{2}}\sigma^{(2)}_{ab}\approx -2E_{\langle a}\text{curl}B_{b\rangle}+2B_{\langle a}\text{curl}E_{b\rangle}\,.
    \label{shear-wave2-EM-terms}
\end{eqnarray}
The above describes the propagation of a gravitational wave in the presence of electromagnetic field. No Weyl-Maxwell coupling takes place at this perturbative level. With the aid of the previously introduced harmonic decomposition of the gravitational and the electromagnetic fields, as well as of the new definition (see the Appendix for some relevant details) $\mathcal{Q}^{(m)}_{ab}\equiv\mathcal{Q}^{(\epsilon)(n)}_{\langle a}\mathcal{Q}^{(\epsilon)(n)}_{b\rangle}+\mathcal{Q}^{(\beta)(n)}_{\langle a}\mathcal{Q}^{(\beta)(n)}_{b\rangle}=-\hat{n}_{\langle a}\hat{n}_{b\rangle}e^{2in^{c}x_{c}}=s^{(2)}_{ab}e^{im^{c}x_{c}}$, so that $s^{(2)}_{ab}=-\hat{n}_{\langle a}\hat{n}_{b\rangle}e^{i(2n^{c}-m^{c})x_{c}}$ ($\hat{n}^{a}$ is the unit electromagnetic wave vector), we have:
\begin{equation}
    s_{ab}^{(2)}=-\hat{n}_{\langle a}\hat{n}_{b\rangle}\hspace{8mm}\text{and}\hspace{8mm} m_{a}=2n_{a}\,,
    \label{tilde-s}
\end{equation}
where $s^{(2)}_{ab}$ represents the unit shear tensor field of second order and $m_{a}$ the associated wave vector of $\sigma^{(2)}_{ab}=\sigma^{(2)}_{(m)}\mathcal{Q}^{(2)}_{ab}$ with $\mathcal{Q}^{(2)}_{ab}=\mathcal{Q}^{(m)}_{ab}=s^{(2)}_{ab}e^{im^{c}x_{c}}$; eventually, equation~\eqref{shear-wave2-EM-terms} recasts into the following (temporal) relation for shear's $m$-th mode:
\begin{equation}
\ddot{\sigma}^{(2)}_{(m)}+m^{2}\sigma^{(2)}_{(m)}=2E_{(n)}B_{(n)}=M_{1}\cos(2nt+\theta_{C}+\theta_{D})+M_{2}\,.
    \label{wave-eq-shear-2nd-mth-mode}
\end{equation}
In the above, $M_{1}\equiv-CD$ and $M_{2}\equiv CD\cos(\theta_{C}-\theta_{D})$ are constants which come from using the plane wave solutions~\eqref{E-B-sol} for the electromagnetic field. Equation~\eqref{wave-eq-shear-2nd-mth-mode} is solved directly giving:
\begin{equation}
    \sigma^{(2)}_{(m)}=M\sin(mt+\theta_{M})+\frac{M_{1}}{m^{2}-4n^{2}}\cos(2nt+\theta_{C}+\theta_{D})+\frac{M_{2}}{4n^{2}}
    \label{shear-sol2a}
\end{equation}
Note that the harmonic decomposition implies that $m_{a}=2n_{a}$, namely the wave vector of the modulated gravitational wave has the direction of the electromagnetic wave vector and twice as much norm. Therefore, the second term on the right hand side of~\eqref{shear-sol2a} is indeterminate. Aiming to deal with this indeterminacy by making appeal to de l' H\^{o}pital's rule, we redefine appropriately the constants $M$ and $\theta_{M}$ so that the equation in question is rewritten as:
\begin{equation}
    \sigma^{(2)}_{(m)}=M\cos(mt+\theta_{M})+\frac{M_{1}}{m^{2}-4n^{2}}\left[\cos(2nt+\theta_{C}+\theta_{D})-\cos(mt+\theta_{C}+\theta_{D})\right]+\frac{M_{2}}{4n^{2}}\,.
    \label{shear-sol2b}
\end{equation}
Finally, on calculating the limit $m\rightarrow 2n$ of the above relation we arrive at the solution:
\begin{equation}
    \sigma^{(2)}_{(m=2n)}=M\cos(2nt+\theta_{M})+\frac{M_{1}}{4n}t\sin(2nt+\theta_{C}+\theta_{D})+\frac{M_{2}}{4n^{2}}\,,
    \label{shear-sol3}
\end{equation}
where the observed linear growth in the second term reflects the familiar resonant behavior. In other words, the general solution of~\eqref{wave-eq-shear-2nd-mth-mode} is a pure resonance. Note that alternatively we could have arrived directly at~\eqref{shear-sol3} by setting $m=2n$ in~\eqref{wave-eq-shear-2nd-mth-mode} and solving.\\

\subsection{Third order waveform} \label{ssec:GravitoEM-coupling-Minkowski}

In the present subsection, we move on to consider the impact that an electromagnetic signal has on a gravitational wave by considering the perturbed wave equation for the shear up to third order in reference to the Minkowski background. It turns out that the Maxwell-Weyl coupling manifests itself at the third (actually the highest) perturbative level. It is worth noting that when considering the inverse phenomenon, namely gravitational wave effects on electromagnetic radiation, the aforementioned coupling appears at the second perturbative level. The difference in question implies more powerful resonant solutions for the case we examine.

Our presentation proceeds as following: First, we construct the full shear wave equation; second, we apply the approximation $\mathbf{E^{2}\sim B^{2}\gg \sigma^{2}}$ (note that this does \textbf{not} generally imply that $E\sim B\gg\sigma$) and solve the associated (harmonically decomposed) equation; third, we derive the resonant solutions.\\
To begin with, let us see how the divergence-free conditions, imposed for isolating gravitational waves, are written by keeping up to third order terms. In particular, equations~\eqref{kcon12}, \eqref{div-Eab} and~\eqref{div-Hab} reduce to:
\begin{equation}
    {\rm D}^{b}\sigma_{ab}=0=q^{\text{(em)}}_{a}=\epsilon_{abc}E^{b}B^{c}\,,
    \label{shear-div2}
\end{equation}
\begin{equation}
    {\rm D}^{b}E_{ab}=0=\frac{1}{6}{\rm D}_{a}(E^{2}+B^{2})+\epsilon_{abc}\sigma^{b}{}_{d}H^{cd}
    \label{div-Eab-2}
\end{equation}
and
\begin{equation}
    {\rm D}^{b}H_{ab}=0=\epsilon_{abc}\sigma^{b}{}_{d}(E^{cd}+\frac{1}{2}\pi_{\text{(em)}}^{cd})\,.
    \label{div-Hab-2}
\end{equation}
In addition, the divergence of the electromagnetic anisotropic pressure leads to the constraint:
\begin{equation}
    {\rm D}^{b}\pi^{\text{(em)}}_{ab}=0=-E_{b}{\rm D}^{b}E_{a}-B_{b}{\rm D}^{b}B_{a}+\frac{1}{3}{\rm D}_{a}(E^{2}+B^{2})\,.
    \label{div-pi-2}
\end{equation}
Having stated the above, we proceed to our principal task. Namely, we derive the wave equation for the shear by taking the dot derivative of~\eqref{shear-prop} under the assumptions defining our spacetime model. In the first place, we have:
\begin{equation}
    \ddot{\sigma}_{\langle ab\rangle}=-2\dot{\sigma}_{c\langle a}\sigma^{c}{}_{b\rangle}-\dot{E}_{\langle ab\rangle}+\frac{1}{2}\dot{\pi}^{\text{(em)}}_{\langle ab\rangle}\,,
    \label{wave-eq-shear-3rd-1}
\end{equation}
where
\begin{equation}
    \dot{E}_{\langle ab\rangle}=-\frac{1}{3}(\rho^{\text{(em)}}+P^{\text{(em)}})\sigma_{ab}+\text{curl}H_{ab}-\frac{1}{2}\dot{\pi}^{\text{(em)}}_{ab}+3\sigma_{\langle a}{}^{c}E_{b\rangle c}-\frac{1}{2}\sigma_{\langle a}{}^{c}\pi^{\text{(em)}}_{b\rangle c}\,,
    \label{dot-Eab2}
\end{equation}
in accordance with~\eqref{dot-Eab}. From the last two equations it is apparent that the coupled terms are of third order, i.e. of second order regarding their electromagnetic part (refer to subsection~\ref{ssEMVs}) and of first order regarding their gravitational (shear) part. Recalling the definition of the electromagnetic anisotropic pressure tensor, $\pi^{\text{(em)}}_{ab}=-E_{\langle a}E_{b\rangle}-B_{\langle a}B_{b\rangle}$, we calculate its dot derivative with the aid of~\eqref{eqn:el-field-prop} and~\eqref{eqn:magn-field-prop}:
\begin{equation}
    \dot{\pi}^{\text{(em)}}_{ab}=-2\left(\sigma_{c\langle a}E_{b\rangle}E^{c}+\sigma_{c\langle a}B_{b\rangle}B^{c}\right)-2\left(E_{\langle a}\text{curl}B_{b\rangle}-B_{\langle a}\text{curl}E_{b\rangle}\right)\,.
    \label{dot-pi}
\end{equation}
By making use of~\eqref{dot-Eab2} and~\eqref{dot-pi}; substituting $\text{curl}H_{ab}$ and $E_{ab}$ with their equivalents from~\eqref{curl-Hab} (see the Appendix) and \eqref{shear-prop} respectively; and recalling the definitions for $\rho_{\text{(em)}}$ and $\pi_{\text{(em)}}$, relation~\eqref{wave-eq-shear-3rd-1} ultimately recasts into:
\begin{eqnarray}
    \ddot{\sigma}_{\langle ab\rangle}-{\rm D}^{2}\sigma_{ab}&=&4\dot{\sigma}_{c\langle a}\sigma^{c}{}_{b\rangle}+6\left(\sigma_{c\langle a}\sigma^{c}{}_{|d|}\sigma^{d}{}_{b\rangle}-\sigma^{2}\sigma_{ab}\right)+2\sigma_{c\langle a}\left(E_{b\rangle}E^{c}+B_{b\rangle}B^{c}\right)\nonumber\\ 
    &&-\frac{3}{2}(E^{2}+B^{2})\sigma_{ab}-2\left(E_{\langle a}\text{curl}B_{b\rangle}-B_{\langle a}\text{curl}E_{b\rangle}\right)\,.
    \label{wave-eq-shear-3rd-3}
\end{eqnarray}
We observe that the right-hand side of the last equation consists of the following kinds of source terms\footnote{Note that the term $\dot{\sigma}_{c\langle a}\sigma^{c}{}_{b\rangle}$ includes both a coupled and a pure shear contribution.}: three coupled/Maxwell-Weyl (first, third and fourth); three pure shear (first and second); two pure electromagnetic (fifth). Subsequently, adopting our above stated approximation (electromagnetic energy density much greater than the gravitational), and neglecting the self-coupling of the Maxwell field (e.g. first order $E^{a}$ with second order $B^{a}$), our wave equation now reads:
\begin{equation}
    \ddot{\sigma}_{\langle ab\rangle}-{\rm D}^{2}\sigma_{ab}\approx 4\dot{\sigma}_{c\langle a}\sigma^{c}{}_{b\rangle}+2\sigma_{c\langle a}\left(E_{b\rangle}E^{c}+B_{b\rangle}B^{c}\right)-\frac{3}{2}(E^{2}+B^{2})\sigma_{ab}\,.
    \label{eqn:approx-wave-eq}
\end{equation}
The next step consists of harmonically decomposing the above according to the notation used in section~\ref{ssec:2nd-order-GWs} as well as that $\sigma^{(3)}_{ab}\equiv\sigma^{(3)}_{(l)}e^{il^{c}x_{c}}s^{(3)}_{ab}$, where $l_{a}$ and $s^{(3)}_{ab}$ denote respectively the wave vector and the unit tensor of the third order shear field. In this case, it turns out that there are two individual tensor components, so that:
\begin{equation}
s^{(3)}_{ab}=s^{(1)}_{ab}+s^{(1)}_{c\langle a}\left(\epsilon^{c}\epsilon_{b\rangle}+\beta^{c}\beta_{b\rangle}\right)\,,\hspace{4mm} l_{a}=k_{a}+2n_{a}\hspace{4mm}\text{and}\hspace{4mm} l^{2}=l^{a}l_{a}=k^{2}+4n^{2}+4kn\cos{\phi}\,,
\label{eqn:s3-l-φ}
\end{equation}
with $\phi$ being the angle between the original gravitational and electromagnetic waves. Note that the form of the wave vector $l_{a}$ reflects the $2:1$ ratio of the Maxwell-Weyl coupling. Therefore, for the $l$-th mode of the final wave form we arrive at the following two equations (corresponding to the tensor components $s^{(1)}_{c\langle a}\left(\epsilon^{c}\epsilon_{b\rangle}+\beta^{c}\beta_{b\rangle}\right)$ and $s^{(1)}_{ab}$ respectively):
\begin{equation}
\ddot{\sigma}^{(3)}_{(l)}+l^{2}\sigma^{(3)}_{(l)}=4\left(\dot{\sigma}^{(2)}_{(m=2n)}\sigma^{(1)}_{(k)}+\dot{\sigma}^{(1)}_{(k)}\sigma^{(2)}_{(m=2n)}\right)+2\left(E^{2}_{(n)}+B^{2}_{(n)}\right)\sigma_{(k)}
\label{eqn:l-mode1}
\end{equation}
and
\begin{equation}
\ddot{\sigma}^{(3)}_{(l)}+l^{2}\sigma^{(3)}_{(l)}=-\frac{8}{3}\left(\dot{\sigma}^{(2)}_{(m=2n)}\sigma^{(1)}_{(k)}+\dot{\sigma}^{(1)}_{(k)}\sigma^{(2)}_{(m=2n)}\right)-\frac{3}{2}\left(E^{2}_{(n)}+B^{2}_{(n)}\right)\sigma_{(k)}\,.
\label{eqn:l-mode2}
\end{equation}
Now substituting the associated harmonic modes from eqs~\eqref{E-B-sol},~\eqref{eqn:orig-shear} and~\eqref{shear-sol3}, into the above, we eventually find out the following linear, second order differential equations:
\begin{equation}
\ddot{\sigma}^{(3)}_{(l)}+l^{2}\sigma^{(3)}_{(l)}=A_{11}\sin{(kt)}+\left(A_{3}t+A_{12}\right)\sin{\left[(k+2n)t\right]}+\left(A_{4}t+A_{13}\right)\sin{\left[(k-2n)t\right]}
\label{eqn:l-mode3}
\end{equation}
and
\begin{equation}
\ddot{\sigma}^{(3)}_{(l)}+l^{2}\sigma^{(3)}_{(l)}=A_{8}\sin{(kt)}+\left(-\frac{2}{3}A_{3}t+A_{9}\right)\sin{\left[(k+2n)t\right]}+\left(-\frac{2}{3}A_{4}t+A_{10}\right)\sin{\left[(k-2n)t\right]}\,,
\label{eqn:l-mode4}
\end{equation}
where the various $A_{i}$ constants can be expressed in terms of the initial amplitudes $\mathcal{G}$, $C$ and $D$ (see appendix), and where we have neglected to write (for simplicity\footnote{Besides, for our purposes, namely for determining the resonant solutions and the electromagnetic trace on gravitational waves, phase constants are irrelevant.}) all the phase constants in the trigonometric terms. In fact, if we want to express $A_{i}$-s in terms of the initial wave amplitudes $C$, $D$, $\mathcal{G}$, it is necessary to keep the phase constants non-equal to zero. Note that eqs~\eqref{eqn:l-mode3} and~\eqref{eqn:l-mode4} differ only in their amplitudes' magnitude. The general solutions of the aforementioned equations are respectively:
\begin{eqnarray}
    \sigma^{(3)}_{(l)}&=&L\sin{(lt)}+\frac{A_{11}}{l^{2}-k^{2}}\sin{(kt)}+\left[\frac{A_{3}t+A_{12}}{l^{2}-(k+2n)^{2}}\right]\sin{\left[(k+2n)t\right]}\nonumber\\
    &&\frac{-2(k+2n)A_{3}}{\left[l^{2}-(k+2n)^{2}\right]^{2}}\cos{\left[(k+2n)t\right]}+\left[\frac{A_{4}t+A_{13}}{l^{2}-(k-2n)^{2}}\right]\sin{\left[(k-2n)t\right]}\nonumber\\
    &&\frac{-(k-2n)A_{4}}{\left[l^{2}-(k-2n)^{2}\right]^{2}}\cos{\left[(k-2n)t\right]}
    \label{eqn:l-mode-sol1}
\end{eqnarray}
and
\begin{eqnarray}
    \sigma^{(3)}_{(l)}&=&L\sin{(lt)}+\frac{A_{8}}{l^{2}-k^{2}}\sin{(kt)}+\left[\frac{-2A_{3}t+A_{9}}{3\left[l^{2}-(k+2n)^{2}\right]}\right]\sin{\left[(k+2n)t\right]}\nonumber\\
    &&\frac{4(k+2n)A_{3}}{\left[l^{2}-(k+2n)^{2}\right]^{2}}\cos{\left[(k+2n)t\right]}+\left[\frac{-2A_{4}t+A_{10}}{l^{2}-(k-2n)^{2}}\right]\sin{\left[(k-2n)t\right]}\nonumber\\
    &&\frac{2(k-2n)A_{4}}{\left[l^{2}-(k-2n)^{2}\right]^{2}}\cos{\left[(k-2n)t\right]}
    \label{eqn:l-mode-sol2}\,.
\end{eqnarray}
Three different modes are observed in the above solutions, corresponding to wave numbers $l$ (given by~\eqref{eqn:s3-l-φ}) and $k\pm 2n$. As it is apparent, the waveform's amplitude increases linearly with time in the general case. In the following subsection, we isolate the resonant solutions from the above expression.

\subsubsection{Weyl-Maxwell resonances}\label{ssec:resonant-sol}

It is evident that there are three indeterminate cases predicted by~\eqref{eqn:l-mode-sol1} and~\eqref{eqn:l-mode-sol2}, i.e. $l\rightarrow k\pm 2n$ and $l\rightarrow k$, associated with resonances. The first two cases occur when $\phi=0$ or $\phi=\pi$, namely when the original interacting waves are parallel or antiparallel respectively. As for the third case, it corresponds to $n=k$ and $\phi=\pi$, namely equal original wave numbers and opposite propagation directions.\\
Making use of de l' H\^{o}pital's rule, in analogy with section~\ref{ssec:2nd-order-GWs}, we determine the individual resonant solutions. In particular, from~\eqref{eqn:l-mode-sol1} (similar solutions obviously hold for~\eqref{eqn:l-mode-sol1}) we have:
\begin{eqnarray}
    \sigma^{(3)}_{l=k+2n}&=&L_{1}\cos{\left[(k+2n)t\right]}+\frac{A_{11}}{4n(k+n)}\sin{(kt)}-\frac{A_{3}t^{2}+2A_{12}t}{4(k+n)}\cos{\left[(k+2n)t\right]}\nonumber\\
    &&+\frac{A_{4}t+A_{13}}{8kn}\sin{\left[(k-2n)t\right]}-\frac{(k-2n)A_{4}}{\left(8kn\right)^{2}}\cos{\left[(k-2n)t\right]}\,,
    \label{eqn:res-l1}
\end{eqnarray}

\begin{eqnarray}
    \sigma^{(3)}_{l=k-2n}&=&L_{2}\cos{\left[(k-2n)t\right]}+\frac{A_{11}}{4n(n-k)}\sin{(kt)}-\frac{A_{3}t+A_{12}}{4kn}\sin{\left[(k+2n)t\right]}\nonumber\\
    &&-\frac{3A_{4}t^{2}+4A_{13}t}{8(k-2n)}\cos{\left[(k-2n)t\right]}-\frac{(k+n)A_{3}}{32}\cos{\left[(k+2n)t\right]}
    \label{eqn:res-l2}
\end{eqnarray}
and
\begin{eqnarray}
    \sigma^{(3)}_{l=k}&=&\left(L_{3}+\frac{A_{11}}{2k}t\right)\cos{(kt)}-\frac{A_{3}t+A_{12}}{4n(k+n)}\sin{\left[(k+2n)t\right]}\nonumber\\
    &&-\frac{(k+2n)A_{3}}{8n^{2}(k+n)^{2}}\cos{\left[(k+2n)t\right]}+\frac{A_{4}t+A_{13}}{4n(k-n)}\sin{\left[(k-2n)t\right]}\nonumber\\
    &&-\frac{(k-2n)A_{4}}{16n^{2}(k-n)^{2}}\cos{\left[(k-2n)t\right]}\,.
    \label{eqn:res-l3}
\end{eqnarray}
The resonant solutions for the emerging waveform follow, as appears above, an overall parabolic/quadratic increase with time; a significantly higher rate of amplification in comparison to the linear one, experienced by the gravitationally enhanced electromagnetic signal (i.e. the inverse waveform).\\

\section{Concluding remarks}\label{sD}
Electromagnetic fields appear everywhere in the universe, either in the form of `individual' electric and magnetic fields, or as traveling electromagnetic radiation. A special feature of the Maxwell field, which separates it from the other known energy sources, is its vector nature. The latter ensures a purely geometrical coupling between electromagnetism and spacetime curvature that is manifested through the Ricci identities and goes beyond the standard interplay between matter and geometry introduced by Einstein's equations. As a result, the evolution of electric and magnetic fields, as well as the propagation of electromagnetic signals, are affected by the curvature of the host spacetime via both of the aforementioned relations.

Most of the available studies employ, as well  as target, the electric and magnetic components directly. Here, we provide an alternative (fully general relativistic) treatment, which uses the 1+3 covariant formalism and involves the electromagnetic vector and scalar potentials. Although the latter may not be directly measurable physical entities, their existence is theoretically allowed by the form of Maxwell's equations. In addition, the temporal and spatial gradients of the vector and scalar potentials give rise to the actual electric and magnetic fields. Therefore, depending on the nature of the problem in hand, one is in principle free to choose either description when analysing electromagnetic phenomena. Given that an 1+3 covariant treatment of electromagnetic fields in curved spacetimes was already given in~\cite{T1}, we have provided here a supplementary study involving the scalar and the vector potentials of the Maxwell field. Moreover, we have included in the discussion the resonances induced by electromagnetic radiation on a gravitational waveform.

We began by introducing a family of observers, which facilitated the 1+3 splitting of the spacetime into a temporal direction (along the observers' 4-velocity vector) and 3-dimensional spatial hypersurfaces orthogonal to it. This in turn allowed us to decompose the electromagnetic 4-potential into its timelike and spacelike parts, respectively represented by the associated scalar and vector potentials. The latter were shown to satisfy wave-like equations, which were directly derived from Maxwell's formulae and contained driving terms reflecting the nature and the material content of the host spacetime. Given that the electromagnetic potential trivially satisfies one of Maxwell's equations, both of the aforementioned wave formulae were derived from the other. More specifically, Faraday's law leads to the wave equation of the vector potential and Coulomb's law to that of its scalar counterpart. In the case of the vector potential, some of the aforementioned driving terms were due to the nonzero spacetime curvature. We found, in particular, that both the spatial and the Weyl parts of the curvature can affect the evolution of the vector potential, through the latter's purely geometrical coupling to the spacetime geometry (mediated by the Ricci identities). No such coupling holds for the scalar potential, which explains why there were no direct spacetime curvature effects in the wave equation of the latter.

Since our principal aim was to study the Maxwell field in curved spacetimes, we applied the wave formula of the vector potential to a Friedmann model with non-Euclidean spatial geometry. Confining to the linear regime of an almost-FRW universe, we found that in spatially closed models, the potential oscillates with an amplitude that decays inversely proportional to the cosmological scale factor, just like it does in spatially flat Friedmann models. The only effect of the positive curvature, was to increase the frequency of the oscillation. On the other hand, the hyperbolic spatial geometry of the open FRW universes modified the evolution of the vector potential in a more `dramatic' way. There, the model's negative curvature changed the standard oscillatory behaviour to a power-law evolution. Not surprisingly, this qualitative change was found to occur on sufficiently large scales, where the effects of the non-Euclidean geometry become more prominent. Exactly analogous curvature effects were also observed during the evolution of source-free electric and magnetic fields in perturbed Friedmann models~\cite{BT,T2}.

We then turned to astrophysical environments and employed the electromagnetic potentials as well as the $E$ and $B$ components to investigate the coupling between the Maxwell and the Weyl fields in the low-density interstellar space. In practice, this meant studying the interaction between propagating gravitational and electromagnetic waves on a Minkowski background at the second and third perturbative order. Given that gravity-wave (i.e.~pure tensor) perturbations are monitored by shear distortions, we included the driving effects of the latter into the wave formulae of the scalar and the vector potentials. Conversely, we also studied the driving effects of the electric and magnetic Maxwell components into the wave formula of the shear. Our results showed that the gravitationally induced electromagnetic potentials as well as the electromagnetically enhanced gravitational signal perform forced oscillations, driven by the coupling between the originally interacting waves. This immediately opens the possibility of resonances, which in our case occur when the initial electromagnetic and gravitational waves propagate along the same, or in the opposite, direction. In most realistic situations, the aforementioned gravito-electromagnetic resonances lead to the linear amplification of the emerging electromagnetic signal. Exactly analogous resonances and amplification effects were reported in the studies of~\cite{T3,KT}, which employed the electric and the magnetic components of the Maxwell field, instead of the potentials. On the other hand, electromagnetic radiation results into the quadratic/parabolic resonant amplification of the emerging gravitational signal. We finally note that, in the present analysis, we also accounted for the backreaction effects between the scalar and the vector potentials, while those of~\cite{T3,KT} bypassed similar backreaction effects between the electric and the magnetic fields. This underlines the considerable technical simplification that one can achieve by involving the electromagnetic potentials instead of the actual electric and magnetic fields.

The compete agreement between our results and those of the previous more conventional studies, together with the technical advantages the use of the potentials seems to bring in, suggest that the formalism introduced and developed here could prove particularly useful when probing the behaviour of electromagnetism in technically demanding astrophysical and cosmological environments. Here, we considered the highly symmetric Minkowski and FRW backgrounds. In principle, however, our analysis can be also applied to, say, the vicinity and perhaps the interior of massive compact stars, or to the very early stages of the universe's evolution and to the study of the Cosmic Microwave Background (CMB).\\

\begin{subappendices}

\section{1+3 Covariant approach}\label{AppA}
In the present section we outline the basic principles of the \textit{1+3 covariant approach} (refer to the extensive reviews of~\cite{EMM} and~\cite{TCM}), we introduce the kinematic quantities and subsequently provide the background for the description of a charged, conducting fluid. The covariant approach to relativity, as described in the following, differs from the more familiar \textit{metric based approach} in that the evolution equations, as well as the relevant constraints satisfied by the individual components of all spacetime quantities, are derived from the Bianchi and the Ricci identities, instead of the metric. Therefore, due to their geometric generality, the covariant formulae can be readily adapted to a wider spectrum of applications.\\

\subsection{Spacetime splitting}\label{AppA1}
In the context of the 1+3 covariant approach to general relativity, the 4-dimensional spacetime decomposes into a temporal direction and a 3-dimensional space orthogonal to it~\cite{TCM,EMM}. This splitting is achieved by introducing a family of (fundamental) observers, moving along their timelike worldlines. These have parametric equations of the form $x^{a}=x^{a}(\tau)$, where $\tau$ is the observer's proper time.\footnote{Throughout this study, Latin indices vary between 0 and 3 and we have set the velocity of light to unity.} The tangent vector to these worldlines is the observer's 4-velocity (with $u^{a}={\rm d}x^{a}/{\rm d}\tau$ and $u^au_a=-1$) and defines their temporal direction. Then, assuming that $g_{ab}$ is the spacetime metric, the symmetric tensor $h_{ab}=g_{ab}+u_au_b$, with $h_{ab}u^{b}=0$, $h_a{}^a=3$, $h_a{}^ch_{cb}=h_{ab}$ by construction, projects orthogonal to the $u_a$-field and into the observer's 3-D rest-space.

On using the $u_a$-field and the associated projection tensor $h_{ab}$, one can decompose every spacetime vector and tensor, every operator and every equation into their temporal and spatial components. For instance, the 4-vector $V_a$ decomposes as:
\begin{equation}
V_a= \mathcal{V}u_a+ \mathcal{V}_a\,,  \label{Vsplit}
\end{equation}
where $\mathcal{V}=-V_au^a$ is the timelike part parallel to $u_a$ and $\mathcal{V}_a=h_a{}^bV_a$ is its spacelike counterpart orthogonal to $u_a$. Similarly, the symmetric second-rank tensor $T_{ab}$ splits as:
\begin{equation}
T_{ab}= tu_au_b+ \frac{1}{3}\left(T+t\right)h_{ab}+ 2u_{(a}t_{b)}+ t_{ab}\,,  \label{Tsplit}
\end{equation}
with $T=T_a{}^a$, $t=T_{ab}u^au^b$, $t_a=-h_a{}^bT_{bc}u^c$ and $t_{ab}=h_{\langle a}{}^ch_{b\rangle}{}^dT_{cd}$.\footnote{Recall that round brackets denote symmetrisation, square ones antisymmetrisation and angular brackets describe the symmetric traceless part of orthogonally projected second-rank tensors (e.g.~$T_{\langle ab\rangle}=T_{(ab)}-(1/3)T_c{}^ch_{ab}$).} The above decomposition follows from the expression $T_{ab}=g_{ac}g_{bd}T^{cd}= (h_{ac}-u_au_c)(h_{bd}-u_bu_d)T^{cd}$ and its most familiar application is on the energy-momentum tensor of a general imperfect fluid (e.g.~see~\cite{TCM,EMM}). An additional useful splitting is that of the 4-D Levi-Civita tensor ($\eta_{abcd}=\eta_{[abcd]}$). Relative to the $u_a$-field, the latter decomposes according to:
\begin{equation}
\eta_{abcd}= 2u_{[a}\epsilon_{b]cd}- 2\epsilon_{ab[c}u_{d]}
\end{equation}
where $\epsilon_{abc}=\epsilon_{[abc]}=\eta_{abcd}u^d$ is the Levi-Civita tensor of the 3-D spatial hypersurfaces. Then, $\epsilon_{abc}u^c=0$ and  $\epsilon_{abc}\epsilon^{def}= 3!h_{[a}{}^dh_b{}^eh_{c]}^f$ by construction.\\
Once the time-direction and the orthogonal 3-space have been introduced, one needs to define temporal and spatial differentiation. For a general tensor field $T_{ab\cdots}{}^{cd\cdots}$, the time and the 3-space derivatives are respectively given by
\begin{equation}
\dot{T}_{ab\cdots}{}^{cd\cdots}= u^{e}\nabla_{e}T_{ab\cdots}{}^{cd\cdots} \hspace{10mm} {\rm and} \hspace{10mm} {\rm D}_{e}T_{ab\cdots}{}^{cd\cdots}= h_e{}^sh_a{}^fh_b{}^ph_q{}^ch_r{}^d\cdots \nabla_{s}T_{fp\cdots}{}^{qr\cdots}\,,  \label{tempspatder}
\end{equation}
with $\nabla_a$ representing the 4-D covariant derivative operator. It follows that ${\rm D}_ah_{bc}=0={\rm D}_d\epsilon_{abc}$ and that $\dot{\epsilon}_{abc}=3u_{[a}\epsilon_{bc]d}\dot{u}^d$, with $\dot{u}_a$ being the 4-acceleration (see Appendix~\ref{AppA2} next).\\

\subsection{Covariant kinematics}\label{AppA2}
All the information regarding the kinematic evolution of the 4-velocity field is encoded in its covariant gradient. The latter decomposes into the irreducible kinematic variables of the motion according to:
\begin{equation}
\nabla_bu_a= \sigma_{ab}+ \omega_{ab}+ \frac{1}{3}\,\Theta h_{ab}- \dot{u}_au_b\,,
\end{equation}
with $\sigma_{ab}\equiv {\rm D}_{\langle b}u_{a\rangle}$, $\omega_{ab}\equiv {\rm D}_{[b}u_{a]}$, $\Theta\equiv \nabla^{a}u_{a}={\rm D}^{a}u_{a}$ and $\dot{u}_{a}\equiv u^{b}\nabla_{b}u_{a}$ respectively representing the shear and the vorticity tensors, the volume expansion/contraction scalar and the 4-acceleration vector. Overall, ${\rm D}_{b}u_{a}=\sigma_{ab}+\omega_{ab}+(1/3)\Theta h_{ab}$ describes the relative motion of neighbouring observers. In detail, the shear monitors distortion in the shape of a moving fluid element and nonzero vorticity implies rotation. The volume scalar, on the other hand, determines the expansion/contraction of the fluid (when it is positive/negative). Finally, a nonzero 4-acceleration reveals the presence of non-gravitational forces, which in turn ensures non-geodesic motion. Note that the vorticity tensor leads to the vector $\omega_a=\epsilon_{abc}\omega^{bc}/2$, which determines the rotation axis. Also, the volume scalar is typically used to define a representative length-scale ($a$), so that $\dot{a}/a=\Theta/3$. In cosmological studies, $a$ is identified with the scale factor of the universe, which the volume scalar and the Hubble parameter ($H$) are related by $\Theta/3=H$.

The evolution of the volume scalar, the shear and the vorticity is monitored by a set of three propagation equations, supplemented by an equal number of constraints. These are obtained after applying the Ricci identity (see (\ref{eqn:Ricc-Electr}) in \S~\ref{ssec:Ricci-Maxwell}) to the 4-velocity field. More specifically, the timelike component of the resulting expression leads to the Raychaudhuri equation:
\begin{equation}
\dot{\Theta}= -\frac{1}{3}\,\Theta^2- \frac{1}{2}\,\kappa(\rho+3p)- 2\left(\sigma^2-\omega^2\right)+ {\rm D}^a\dot{u}_a+ \dot{u}^a\dot{u}_a\,,  \label{eqn:Ray}
\end{equation}
to the shear evolution formula:
\begin{equation}
\dot{\sigma}_{\langle ab\rangle}= -\frac{2}{3}\,\Theta\sigma_{ab}- \sigma_{c\langle a}\sigma^c{}_{b\rangle}- \omega_{\langle a}\omega_{b\rangle}+ {\rm D}_{\langle a}\dot{u}_{b\rangle}+ \dot{u}_{\langle a}\dot{u}_{b\rangle}- E_{ab}+ \frac{1}{2}\,\kappa\pi_{ab}  \label{shear-prop}
\end{equation}
and to the propagation equation of the vorticity tensor,
\begin{equation}
\dot{\omega}_{\langle ab\rangle}= -\frac{2}{3}\,\Theta\omega_{ab}+ {\rm D}_{[b}\dot{u}_{a]}- 2\sigma_{c[a}\omega^c{}_{b]}\,,  \label{dotvorten}
\end{equation}
where $\dot{\sigma}_{\langle ab\rangle}= h_a{}^ch_b{}^d\dot{\sigma}_{cd}$ and $\dot{\omega}_{\langle ab\rangle}= h_a{}^ch_b{}^d\dot{\omega}_{cd}$ by construction. Note that, recalling the $\omega_{ab}=\epsilon_{abc}\omega^c$, one could replace the above with the evolution formula of the vorticity vector,
\begin{equation}
\dot{\omega}_{\langle a\rangle}= -\frac{2}{3}\,\Theta\omega_a- \frac{1}{2}\,{\rm curl}\dot{u}_a+\sigma_{ab}\omega^b\,.
\end{equation}

The propagation formulae of the irreducible kinematic variables are supplemented by three constraints. These are obtained from the spatial part of the aforementioned Ricci identities and they are given by:
\begin{equation}
{\rm D}^b\sigma_{ab}= \frac{2}{3}\,{\rm D}_a\Theta+ {\rm curl}\omega_a+ 2\epsilon_{abc}\dot{u}^b\omega^c- \kappa q_a\,, \hspace{15mm} {\rm D}^a\omega_a= \dot{u}^a\omega_a  \label{kcon12}
\end{equation}
and
\begin{equation}
H_{ab}= {\rm curl}\sigma_{ab}+ {\rm D}_{\langle a}\omega_{b\rangle}+ 2\dot{u}_{\langle a}\omega_{b\rangle}\,.  \label{kcon3}
\end{equation}
Note that $\sigma^2=\sigma_{ab}\sigma^{ab}/2$ and $\omega^2=\omega_{ab}\omega^{ab}/2=\omega_a\omega^a$ are the (square) magnitudes of the shear and the vorticity respectively, while $E_{ab}$ and $H_{ab}$ are the electric and the magnetic components of the Weyl tensor (see Appendix~\ref{AppA3}). Finally, ${\rm curl}\sigma_{ab}=\epsilon_{cd\langle a}{\rm D}^{c}\sigma_{b\rangle}{}^{d}$ by construction.\\

\subsection{Proving Maxwell equations in terms of E and B fields}\label{AppB-Maxwell-eqs}

In the present Appendix subsection we provide proofs for the covariant Maxwell equations, written in terms of the $E$ and $B$ components. Note for reference that eq.~\eqref{Fab1} plays the key role in our following calculations.\\
Let us begin with decomposing the equation set~(\ref{eqn:Maxw-eqns1}a). Substituting~\eqref{Fab1} into the latter and applying the product rule for differentiation, we receive:
\begin{equation}
    (\nabla^{b}u_{a})E_{b}+u_{a}\nabla^{b}E_{b}-\Theta E_{a}-\dot{E}_{a}+(\nabla^{b}\epsilon_{abc})B^{c}+\epsilon_{abc}\nabla^{b}B^{c}=J_{a}\,.
    \label{eqn:M1}
\end{equation}
Projecting the above orthogonal to $u^{a}$ and deploying the decomposition of the velocity gradient, yields:
\begin{equation}
\left(\sigma_{ab}+\omega_{ab}+\frac{1}{3}\Theta h_{ab}\right)E^{b}-\Theta E_{a}-\dot{E}_{\langle a\rangle}+\text{curl}B_{a}-\epsilon_{abc}\dot{u}^{b}B^{c}-\mathcal{J}_{a}=0\,.
\label{eqn:M2}
\end{equation}
Note that in deriving the above, we have taken into account the following auxiliary relations (recall $\epsilon_{abc}\equiv\eta_{abcd}u^{d}$ and $\dot{E}_{\langle a\rangle}\equiv h_{a}{}^{b}\dot{E}_{b}$):
\begin{equation}
\left(\nabla^{b}\epsilon_{abc}\right)B^{c}=\eta_{abcd}\left(\nabla^{b}u^{d}\right)B^{c}=-\eta_{abcd}u^{b}\dot{u}^{d}B^{c}=\epsilon_{abc}\dot{u}^{b}B^{c}
\label{eqn:M3}
\end{equation}
as well as
\begin{equation}
    \nabla^{a}E_{a}={\rm D}^{a}E_{a}+E_{a}\dot{u}^{a}\,.
    \label{eqn:M4}
\end{equation}
Finally, rearranging terms in eq.~\eqref{eqn:M2}, we arrive at a covariant expression (involving kinematic effects) of Amp\`{e}re's law:
\begin{equation}
\dot{E}_{\langle a\rangle}=-\frac{2}{3}\Theta E_{a}+\left(\sigma_{ab}+\epsilon_{abc}\omega^{c}\right)E^{b}+\epsilon_{abc}\dot{u}^{b}B^{c}+\text{curl}B_{a}-\mathcal{J}_{a}\,.
\label{eqn:M5}
\end{equation}
Now we get back to eq.~(\ref{eqn:Maxw-eqns1}a) which we project along $u^{a}$, so that (recall the definitions for $E$ and $B$ fields):
\begin{equation}
    -\nabla^{b}E_{b}-\left(\nabla^{b}u^{a}\right)F_{ab}=-\mu\hspace{2mm}\rightarrow\hspace{2mm} {\rm D}^{a}E_{a}+E_{a}\dot{u}^{a}-\omega^{ab}F_{ab}-E_{a}\dot{u}^{a}=-\mu\,.
    \label{eqn:M6}
\end{equation}
The above eventually recasts into the following form of Gauss law:
\begin{equation}
    {\rm D}^{a}E_{a}+2\omega^{a}B_{a}=\mu\,.
    \label{eqn:M7}
\end{equation}
Turning subsequently our attention to the other set of Maxwell equations (i.e. see~(\ref{eqn:Maxw-eqns1}b)). In particular, we start by projecting the equation in question along $u^{a}$, and applying Leibniz's product rule:
\begin{equation}
    \dot{F}_{ab}+2\nabla_{[a}E_{b]}+2\nabla_{[a}u_{|c|}F^{c}{}_{b]}=0\,.
     \label{eqn:M8}
\end{equation}
In the next step, we make a projection with the $3$-D Levi-Civita pseudotensor $\epsilon_{d}{}^{ab}$, which leads to:
\begin{equation}
\dot{\left(\epsilon_{dab}F^{ab}\right)}-\dot{\epsilon_{dab}}F^{ab}+2\text{curl}E_{d}-2({\rm D}_{b}u_{d})B^{b}+\Theta B_{d}=0\,.
\label{eqn:M9}
\end{equation}
In deducing the above, we have made use of $\text{curl}E_{a}\equiv\epsilon_{abc}{\rm D}^{b}E^{c}$ and
\begin{eqnarray}
    2\epsilon_{d}{}^{ab}{\nabla}_{[a}u_{|c|}F^{c}{}_{b]}&=&2\epsilon_{d}{}^{ab} {\rm D}_{a}u_{c}F^{c}{}_{b}=-2\epsilon_{d}{}^{ab}\epsilon_{ceb}{\rm D}_{a}u^{c}B^{e}=-2\left(h_{dc}h^{a}{}_{e}-h_{de}h^{a}{}_{c}\right){\rm D}_{a}u^{c}B^{e}\nonumber\\
    &&-2({\rm D}_{b}u_{d})B^{b}+\Theta B_{d}\,.
    \label{eqn:M10}
\end{eqnarray}
Therefore, recalling that $\dot{\epsilon}_{abc}=3u_{[a}\epsilon_{bc]d}\dot{u}^{d}$, we get:
\begin{equation}
    \dot{\epsilon_{dab}}F^{ab}=\frac{1}{2}\left(u_{d}\epsilon_{abe}+u_{b}\epsilon_{dae}+u_{a}\epsilon_{bde}\right)F^{ab}\dot{u}^{e}=\frac{1}{2}u_{d}\epsilon_{abc}F^{ab}\dot{u}^{c}+\epsilon_{abc}\dot{u}^{b}E^{c}\,.
    \label{eqn:M11}
\end{equation}
Finally, multiplying eq.~\eqref{eqn:M9} by $h_{a}{}^{d}$ and deploying eqs~\eqref{eqn:M10} and~\eqref{eqn:M11}, the former recasts into the covariant version of Faraday's formula:
\begin{equation}
\dot{B}_{\langle a\rangle}=\left(\sigma_{ab}+\epsilon_{abc}\omega^{c}-\frac{2}{3}\Theta h_{ab}\right)B^{b}+\epsilon_{abc}\dot{u}^{b}E^{c}-\text{curl}E_{a}\,,
\label{eqn:M12}
\end{equation}
where recall that $\nabla_{a}u_{b}={\rm D}_{a}u_{b}-u_{a}\dot{u}_{b}$. On the other hand, getting back to~(\ref{eqn:Maxw-eqns1}b)) and considering the scalar equation coming from projection with $\epsilon^{abc}$, we have:
\begin{equation}
3\epsilon^{abc}\nabla_{a}F_{bc}=0\hspace{2mm}\rightarrow\hspace{2mm} 2\epsilon^{abc}\left(\nabla_{a}u_{b}\right)E_{c}+\epsilon^{abc}\left(\nabla_{a}\epsilon_{bcd}\right)B^{d}+\epsilon^{abc}\epsilon_{bcd}\nabla_{a}B^{d}=0\,,
\label{eqn:M13}
\end{equation}
which successively simplifies to:
\begin{equation}
-2\epsilon^{abc}\omega_{ab}E_{c}+2{\rm D}^{a}B_{a}=0
\label{eqn:M14}
\end{equation}
and ultimately to the following form of Gauss's law for the magnetic flux:
\begin{equation}
{\rm D}^{a}B_{a}-2\omega^{a}E_{a}=0\,.
\label{eqn:M15}
\end{equation}
\\

\subsection{Spatial and Weyl curvature}\label{AppA3}
The Riemann tensor ($R_{abcd}$), which determines the curvature of the 4-D spacetime satisfies the symmetries $R_{abcd}=R_{cdab}$, $R_{abcd}=R_{[ab][cd]}$ and $R_{a[bcd]}=0$. Also, the trace of $R_{abcd}$ leads to the symmetric Ricci tensor via the contraction $R_{ab}=R^c{}_{acb}$. The latter, together with the Ricci scalar $R=R^a{}_a$, determine the local gravitational field due to the presence of matter by means of Einstein's equations (see expression (\ref{EFE}) in \S~\ref{ssec:Ricci-Maxwell}).\\
The (intrinsic) curvature of the 3-D hypersurfaces orthogonal to the the observers' 4-velocity is determined by the associated 3-Riemann tensor (see eqs.~(\ref{eqn:Ricc-Electr}b) in~\S~\ref{ssec:Ricci-Maxwell}), given by:
\begin{equation}
\mathcal{R}_{abcd}= h_a{}^eh_b{}^fh_c{}^qh_d{}^sR_{efqs}- {\rm D}_cu_a{\rm D}_du_b+ {\rm D}_du_a{\rm D}_cu_b\,,  \label{3Riemann}
\end{equation}
with the 4-velocity gradient ${\rm D}_bu_a=(\Theta/3)h_{ab}+ \sigma_{ab}+\omega_{ab}$ describing the extrinsic curvature. When there is no vorticity (i.e.~for $\omega_{ab}=0$), the 3-Riemann tensor shares all the symmetries of its 4-D counterpart. In the opposite case, we have $\mathcal{R}_{abcd}=\mathcal{R}_{[ab][cd]}$ only. Then, the corresponding 3-Ricci tensor $\mathcal{R}_{ab}= \mathcal{R}^c{}_{acb}$ satisfies the Gauss-Codacci equation,
\begin{eqnarray}
\mathcal{R}_{ab}&=& \frac{2}{3}\left(\kappa\rho-\frac{1}{3}\,\Theta^2+\sigma^2-\omega^2\right)h_{ab}- E_{ab}+ \frac{1}{2}\,\kappa\pi_{ab}- \frac{1}{3}\,\Theta(\sigma_{ab}+\omega_{ab}) \nonumber\\ &&+\sigma_{c\langle a}\sigma^c{}_{b\rangle}- \omega_{c\langle a}\omega^c{}_{b\rangle}+ 2\sigma_{c[a}\omega^c{}_{b]}\,.  \label{GC}
\end{eqnarray}
It follows that, in contrast to its 4-D counterpart, $\mathcal{R}_{ab}$ is no longer symmetric. Instead, in rotating spacetimes, the 3-Ricci tensor has an antisymmetric part that is given by:
\begin{equation}
\mathcal{R}_{[ab]}= -\frac{1}{3}\,\Theta\omega_{ab}+ 2\sigma_{c[a}\omega^c{}_{b]}\,.  \label{cR[ab]}
\end{equation}
Finally, the trace of (\ref{GC}) leads to the 3-Ricci scalar $\mathcal{R}=\mathcal{R}^a{}_a=2[\rho-(\Theta^2/3)+\sigma^2 -\omega^2]$, which measures the mean curvature of the 3-D spatial sections.\\
The long-range gravitational field, namely tidal forces and gravity waves are monitored by the Weyl curvature tensor $\mathcal{C}_{abcd}$ (with $\mathcal{C}_{abcd}=\mathcal{C}_{cdab}$, $\mathcal{C}_{abcd}=\mathcal{C}_{[ab][cd]}$, $\mathcal{C}_{a[bcd]}=0$ and  $\mathcal{C}^{c}{}_{bcd}=0$), which satisfies the relation:
\begin{equation}
\mathcal{C}_{abcd}= R_{abcd}- \frac{1}{2}\left(g_{ac}R_{bd}+g_{bd}R_{ac} -g_{bc}R_{ad}-g_{ad}R_{bc}\right)+ \frac{1}{6}\,R\left(g_{ac}g_{bd}-g_{ad}g_{bc}\right)\,,  \label{Weyl}
\end{equation}
share all the symmetries of the Riemann tensor and it is also trace-free. In addition, relative to the $u_a$-field, the Weyl tensor splits into an electric and a magnetic component given by:
\begin{equation}
E_{ab}= \mathcal{C}_{acbd}u^cu^d \hspace{10mm} {\rm and} \hspace{10mm}
H_{ab}= \frac{1}{2}\,\epsilon_a{}^{cd}\mathcal{C}_{cdbe}u^e\,,  \label{EMWeyl}
\end{equation}
both of which are symmetric, traceless and ``live'' in the observers rest-space (i.e.~$E_{ab}u^b=0=H_{ab}u^b$).It is worth noting that the electric component is a generalisation of the Newtonian tidal tensor while the magnetic one has no Newtonian counterpart. Employing the above, the Weyl curvature tensor decomposes as:
\begin{equation}
\mathcal{C}_{ab}{}^{cd}= 4\left(u_{[a}u^{[c}+h_{[a}{}^{[c}\right)E_{b]}{}^{d]}+
2\epsilon_{abe}u^{[c}H^{d]e}+ 2u_{[a}H_{b]e}\epsilon^{cde}\,, \label{Weylsplit}
\end{equation}
relative to the $u_a$-field, or alternatively as:
\begin{equation}
    \mathcal{C}_{abcd}=(g_{abqp}g_{cdsr}-\eta_{abqp}\eta_{cdsr})u^{q}u^{s}E^{pr}-(\eta_{abqp}g_{cdsr}+g_{abqp}\eta_{cdsr})u^{q}u^{s}H^{pr}\,,
  \label{Weylsplit1}  
\end{equation}
where $g_{abcd}=g_{ac}g_{bd}-g_{ad}g_{bc}$. Regarding the dynamical description of long range gravity, as encoded by the Weyl tensor, it is achieved via the Bianchi identity in the following form:
\begin{equation}
    \nabla^{d}\mathcal{C}_{abcd}=\nabla_{[b}R_{a]c}+\frac{1}{6}g_{c[b}\nabla_{a]}R\,.
    \label{Bianchi-Weyl}
\end{equation}
The latter relation actually comes from the contraction of the Bianchi identity, as satisfied in principle by the Riemann tensor,
\begin{equation}
    \nabla_{e}R_{abcd}+\nabla_{d}R_{abec}+\nabla_{c}R_{abde}=0\, ,
    \label{Bianchi-Riemann}
\end{equation}
on taking into account eq.~\eqref{Weyl}. With the aid of~\eqref{EMWeyl} and on projecting appropriately along and orthogonal to $u_{a}$, the Bianchi identity~\eqref{Bianchi-Weyl} splits into the following dynamical equations:
\begin{eqnarray}
    \dot{E}_{\langle ab\rangle}&=&-\Theta E_{ab}-\frac{1}{2}(\rho+P)\sigma_{ab}+\text{curl}H_{ab}-\frac{1}{2}\dot{\pi}_{ab}-\frac{1}{6}\Theta\pi_{ab}-\frac{1}{2}D_{\langle a}q_{b\rangle}-\dot{u}_{\langle a}q_{b\rangle} \nonumber\\
    &&+3\sigma_{\langle a}{}^{c}\left(E_{b\rangle c}-\frac{1}{6}\pi_{b\rangle c}\right)+\epsilon_{cd\langle a}\left[2\dot{u}^{c}H_{b\rangle}{}^{d}-\omega^{c}\left(E_{b\rangle}{}^{d}+\frac{1}{2}\pi_{b\rangle}{}^{d} \right) \right]
    \label{dot-Eab}
\end{eqnarray}
and
\begin{eqnarray}
 \dot{H}_{\langle ab\rangle}=-\Theta H_{ab}-\text{curl}E_{ab}+\frac{1}{2}\text{curl}\pi_{ab}+3\sigma_{\langle a}{}{c}H_{b\rangle c}-\frac{3}{2}\omega_{\langle a}q_{b\rangle}-\epsilon_{cd\langle a}\left(2\dot{u}^{c}E_{b\rangle}{}^{d}-\frac{1}{2}\sigma^{c}{}_{b\rangle}q^{d}+\omega^{c}H_{b\rangle}{}^{d} \right)\,,
 \label{dot-Hab}
\end{eqnarray}
which describe the propagation of the long range gravitational field as governed by the matter distribution, as well as the constraints:
\begin{equation}
    {\rm D}^{b}E_{ab}=\frac{1}{3}{\rm D}_{a}\rho-\frac{1}{2}{\rm D}^{b}\pi_{ab}-\frac{1}{3}\Theta q_{a}+\frac{1}{2}\sigma_{ab}q^{b}-3H_{ab}q^{b}+\epsilon_{abc}\left(\sigma^{b}{}_{d}H^{cd}-\frac{3}{2}\omega^{b}q^{c}\right)
    \label{div-Eab}
\end{equation}
and
\begin{equation}
    {\rm D}^{b}H_{ab}=(\rho+P)\omega_{a}-\frac{1}{2}\text{curl}q_{a}+3E_{ab}\omega^{b}-\frac{1}{2}\pi_{ab}\omega^{b}-\epsilon_{abc}\sigma^{b}{}_{d}\left(E^{cd}+\frac{1}{2}\pi^{cd}\right)\,.
    \label{div-Hab}
\end{equation}
It is worth noting the close analogy between equations~\eqref{dot-Eab}-\eqref{div-Hab} and Maxwell equations (see subsection~\ref{ssMEs}). This resemblance, which has been thought as a possible sign of a closer underlying connection between the electromagnetic and the gravitational fields, has been the subject of debate for many decades.\\
Also, the transverse degrees of freedom in the components of the Weyl tensor provide a covariant description of gravitational waves. This means that in the transverse, traceless gauge, the Weyl field has to be divergence-free (i.e. ${\rm D}^{b}E_{ab}=0={\rm D}^{b}H_{ab}$) to linear order. The same condition has to be satisfied by all the orthogonally projected, transverse, traceless, second-rank tensors (e.g. the shear and the anisotropic pressure) involved in the problem in question.\\

\section{Deriving the wave equations for the potentials}\label{AppB}
This part of the Appendix provides guidance and some of the key steps leading to the wave formulae of the vector and the scalar electromagnetic potentials given in \S~\ref{ssWEPs}.\\

\subsection{The wave formula for the vector potential}\label{AppB1}
The wave formula for the vector potential (see Eq.~(\ref{eqn:vec-pot}) in \S~\ref{ssWEPs}) follows after combining expressions (\ref{el-field-pot}) and (\ref{magn-field-pot}) with Ampere's law (\ref{eqn:el-field-prop}). In particular, taking the time derivative of (\ref{el-field-pot}), using Raychaudhuri's equation (\ref{Ray}), together with the propagation formulae of the shear and the vorticity tensors (see (\ref{shear-prop}) and (\ref{dotvorten}) respectively), one arrives at
\begin{eqnarray}
\dot{E}_{\langle a\rangle}&=& -\ddot{\mathcal{A}}_{\langle a\rangle}- \frac{1}{3}\,\Theta\dot{\mathcal{A}}_{\langle a\rangle}- (\sigma_{ab}-\omega_{ab})\dot{\mathcal{A}}^b+ \frac{1}{3}\left[\frac{1}{3}\,\Theta^2+\frac{1}{2}\,\kappa(\rho+3p) +2\left(\sigma^2-\omega^2\right)-{\rm D}_b\dot{u}^b\right] \mathcal{A}^b \nonumber\\ &&+\left[\frac{2}{3}\,\Theta(\sigma_{ab}-\omega_{ab}) +\sigma_{c\langle a}\sigma^c{}_{b\rangle}+\omega_{\langle a}\omega_{b\rangle}-2\sigma_{c[a}\omega^c{}_{b]}- \frac{1}{2}\,\kappa\pi_{ab}- {\rm D}_{\langle b}\dot{u}_{a\rangle} +{\rm D}_{[b}\dot{u}_{a]}\right]\mathcal{A}^b \nonumber\\ &&+E_{ab}\mathcal{A}^b- 2\dot{\Phi}\dot{u}_a- \Phi\ddot{u}_{\langle a\rangle}- {\rm D}_a\dot{\Phi}+ \frac{1}{3}\,\Theta{\rm D}_a\Phi+ (\sigma_{ab}-\omega_{ab}){\rm D}^b\Phi\,.  \label{AppB11}
\end{eqnarray}
Note that in deriving the above, which expresses the left-hand side of Eq.~(\ref{eqn:vec-pot}) in terms of the electromagnetic vector and scalar potentials, we have also used the auxiliary relation
\begin{equation}
h_a{}^b\left({\rm D}_a\Phi\right)^{\cdot}= \dot{\Phi}\dot{u}_a+ {\rm D}_a\dot{\Phi}- \frac{1}{3}\,\Theta{\rm D}_a\Phi- (\sigma_{ab}-\omega_{ab}){\rm D}^b\Phi\,,  \label{AppB12}
\end{equation}
monitoring the commutation between the spatial and the temporal derivatives of $\Phi$.

The terms on the right-hand side of Ampere's law are also expressed in terms of the aforementioned potentials by means of (\ref{el-field-pot}) and (\ref{magn-field-pot}). The most involved derivation is that of ${\rm curl}B_a$, since it requires the use of the 3-Ricci identities (see expression (\ref{eqn:Ricc-Electr}b) in \S~\ref{ssec:Ricci-Maxwell}). In so doing and after applying the Lorenz-gauge condition (see Eq.~(\ref{Lorenz-gauge}) in \S~\ref{ssVSPs}) twice, we obtain
\begin{eqnarray}
{\rm curl}B_a&=& -{\rm D}^2\mathcal{A}_a+ \mathcal{R}_{ba}\mathcal{A}^b+ 2\omega_{ab}\dot{\mathcal{A}}^b- \frac{1}{3}\left({\rm D}_b\dot{u}^b -\frac{1}{3}\,\dot{u}_b\dot{u}^b\right)\mathcal{A}_a+ \frac{1}{3}\,\dot{u}_{\langle a}\dot{u}_{b\rangle}\mathcal{A}^b \nonumber\\ &&-\mathcal{A}^b\left({\rm D}_{\langle b}\dot{u}_{a\rangle}-{\rm D}_{[b}\dot{u}_{a]}\right)- \dot{u}^b\left({\rm D}_{\langle b}\mathcal{A}_{a\rangle}-{\rm D}_{[b}\mathcal{A}_{a]}\right)+ \frac{1}{3}\,\dot{\Phi}\dot{u}_a \nonumber\\ &&+\left(\frac{1}{3}\,\Theta\dot{u}_a-{\rm D}_a\Theta- 2{\rm curl}\omega_a\right)\Phi-{\rm D}_a\dot{\Phi}- \Theta{\rm D}_a\Phi- 2\omega_{ab}{\rm D}^b\Phi\,,  \label{AppB13}
\end{eqnarray}
where $\mathcal{R}_{ab}$ satisfies the Gauss-Codacci equation (see expression (\ref{GC}) in Appendix~\ref{AppA3}). Using the auxiliary relations given above and following the recommended steps, one may recast Ampere's law into the wave-like formula (\ref{eqn:vec-pot}), governing the evolution of the vector potential in an arbitrary Riemannian spacetime.\\

\subsection{The wave formula for the scalar potential}\label{AppB2}
The wave formula for the scalar potential (see Eq.~(\ref{eqn:scalar-pot}) in \S~\ref{ssWEPs}) is obtained after substituting expressions (\ref{el-field-pot}) and (\ref{magn-field-pot}), into Coulomb's law (see (\ref{magn-div}a) in \S~\ref{ssMEs}). To begin with, taking the spatial divergence of (\ref{el-field-pot}), we initially obtain
\begin{eqnarray}
{\rm D}^aE_a&=& {\rm D}^a\dot{\mathcal{A}}_{\langle a\rangle}- {\rm D}^2\Phi- \dot{u}_a{\rm D}^a\Phi- \Phi{\rm D}^a\dot{u}_a- \frac{1}{3}\mathcal{A}^a{\rm D}_a\Theta- \mathcal{A}^b{\rm D}^a\left(\sigma_{ba}+\omega_{ba}\right) \nonumber\\ && -\frac{1}{3}\,\Theta{\rm D}_a\mathcal{A}^a- \left(\sigma_{ba}+\omega_{ba}\right){\rm D}^a\mathcal{A}^b\,.  \label{AppB21}
\end{eqnarray}
Employing the Ricci identities (see Eq.~(\ref{eqn:Ricci-id-4pot}) in \S~\ref{ssec:Ricci-Maxwell}) and using the symmetries of the Riemann tensor (see \S~\ref{AppA3} previously), the first term on the right-hand side of the above reads
\begin{eqnarray}
{\rm D}^a\dot{\mathcal{A}}_{\langle a\rangle}&=& \frac{1}{3}\,\Theta{\rm D}^a\mathcal{A}_a+ \left({\rm D}^a\mathcal{A}_a\right)^{\cdot}- \frac{2}{3}\,\Theta\dot{u}_a\mathcal{A}^a- \dot{u}_a\dot{\mathcal{A}}^a+ R_{ab}u^a\mathcal{A}^b+ \left(\sigma_{ab}+\omega_{ab}\right)\mathcal{A}^a\dot{u}^b \nonumber\\ &&+\left(\sigma_{ba}+\omega_{ba}\right){\rm D}^a\mathcal{A}^b\,,  \label{AppB22}
\end{eqnarray}
with $R_{ab}=R^c{}_{acb}$ representing the 4-D Ricci tensor. Substituting this result back into Eq.~(\ref{AppB21}), adopting the Lorenz-gauge (i.e.~imposing condition (\ref{Lorenz-gauge}) in \S~\ref{ssVSPs}), employing Raychaudhuri's formula (see Eq.~(\ref{Ray}) in Appendix~\ref{AppA2}), using constraint (\ref{kcon12}a), while also taking into account that $R_{ab}u^a\mathcal{A}^b= T_{ab}u^a\mathcal{A}^b=-\kappa q_a\mathcal{A}^a$ (see footnote~4 in \S~\ref{ssEMVs}) and keeping in mind that $\omega_{abc}=\epsilon_{abc}\omega^c$ (see Appendix~\ref{AppA2}), we arrive at
\begin{eqnarray}
{\rm D}^aE_a&=& \ddot{\Phi}- {\rm D}^2\Phi+ \frac{5}{3}\,\Theta\dot{\Phi}- \left[\frac{1}{2}\,\kappa(\rho+3p)- \frac{1}{3}\,\Theta^2+2\left(\sigma^2-\omega^2\right) -\dot{u}_a\dot{u}^a\right]\Phi- \dot{u}^a{\rm D}_a\Phi \nonumber\\ &&-\left[{\rm D}_a\Theta-\frac{4}{3}\,\Theta\dot{u}_a+2{\rm curl}\omega_a-2\kappa q_a+(\sigma_{ab}+3\omega_{ab})\dot{u}^b -\ddot{u}_a\right]\mathcal{A}^a+ 2\dot{u}_a\dot{\mathcal{A}}^a \nonumber\\ &&-2\sigma_{ab}{\rm D}^b\mathcal{A}^a\,.  \label{AppB23}
\end{eqnarray}
At the same time, the right-hand side of expression (\ref{magn-field-pot}) gives $2\omega^aB_a=2\omega^a{\rm curl}\mathcal{A}_a-4\omega^2\Phi$. Combining the latter with (\ref{AppB23}), we can finally recast Eq.~(\ref{magn-div}a) into the wave-formula (\ref{eqn:scalar-pot}) of the scalar potential.\\

\section{Wave equation for the shear field}\label{AppC}

In the present section, we include some auxiliary calculations related with the curl of the magnetic Weyl tensor as well as the harmonic decomposition of the shear wave equation.\\

\subsection{Curl of the magnetic Weyl component}\label{AppC1}

In deriving the wave equation for the shear in the main text, we need to calculate the curl of the magnetic Weyl tensor. In detail, deploying equation~\eqref{kcon3}, the term in question can be written as:
\begin{equation}
    \text{curl}H_{ab}=\text{curl}(\text{curl}\sigma_{ab})=\epsilon_{c}{}^{d}_{\langle a}\epsilon_{ef\langle d}{\rm D}^{c}{\rm D}^{e}\sigma^{f}{}_{b\rangle\rangle}\,,
\end{equation}
where
\begin{equation}
    \epsilon_{cda}\epsilon^{efb}{\rm D}^{c}{\rm D}_{e}\sigma_{f}{}^{d}=3!h_{[c}{}^{e}h_{d}{}^{f}h_{a]}{}^{b}{\rm D}^{c}{\rm D_{e}}\sigma_{f}{}^{d}={\rm D}^{c}{\rm D}_{a}\sigma_{bc}-{\rm D}^{2}\sigma_{ab}\,.
\end{equation}
Subsequently, making use of the Ricci identities in the form of~\eqref{eqn:3-Ricci-2nd-rank-tensor}, the first term on the right hand side of the above becomes:
\begin{equation}
    {\rm D}^{c}{\rm D}_{a}\sigma_{bc}={\rm D}_{a}{\rm D}^{c}\sigma_{bc}-2\omega^{c}{}_{a}\dot{\sigma}_{\langle bc\rangle}+\mathcal{R}_{ebac}\sigma^{ce}+\mathcal{R}_{ea}\sigma_{b}{}^{e}\,.
\end{equation}
Aiming to isolate gravitational waves in the context of the spacetime model adopted in subsection~\ref{sssW-MCMS}, we find out that the aforementioned term transforms into:
\begin{equation}
    \text{curl}H_{ab}=\mathcal{R}_{e\langle ba\rangle c}\sigma^{ce}+\mathcal{R}_{e\langle a}\sigma_{b\rangle}{}^{e}-{\rm D}^{2}\sigma_{ab}\,,
\end{equation}
or equivalently into:
\begin{equation}
    \text{curl}H_{ab}=\rho^{\text{(em)}}\sigma_{ab}+\frac{3}{2}\sigma_{c\langle a}\pi^{\text{(em)}c}{}_{b\rangle}+3E_{c\langle a}\sigma^{c}{}_{b\rangle}+2\sigma^{2}\sigma_{ab}-{\rm D}^{2}\sigma_{ab}\,.
    \label{curl-Hab}
\end{equation}
Note the Weyl-Maxwell coupling manifested in the first two terms. The remaining ones, are purely Weyl, taking into account that the component $E_{ab}$ (like its magnetic counterpart $H_{ab}$) reduces to the shear field $\sigma_{ab}$.\\

\subsection{Harmonic decomposition}\label{AppC2}

Let us focus our attention on the harmonic decomposition of the terms $-2\text{curl}B_{\langle a}E_{b\rangle}$ and $2\text{curl}E_{\langle a}B_{b\rangle}$, which appear in the right hand side of~\eqref{shear-wave2-EM-terms}. To begin with, we have:
\begin{equation}
    -\left(\text{curl}B_{a}\right)E_{b}=-\left(\epsilon_{acd}{\rm D}^{c}B^{d}\right)E_{b}=-i\epsilon_{acd}k^{c}\mathcal{Q}^{(\beta)d}\mathcal{Q}^{(\epsilon)}_{b}=i\mathcal{Q}^{(\epsilon)}_{a}\mathcal{Q}^{(\epsilon)}_{b}\,,
\end{equation}
where we have taken into account that $\mathcal{Q}^{(\epsilon)}_{a}\equiv e^{in^{c}x_{c}}\epsilon_{a}$ and $\mathcal{Q}^{(\beta)}_{a}\equiv e^{in^{c}x_{c}}\beta_{a}$ ($\epsilon_{a}$ and $\beta_{a}$ are unit vectors along the directions of $E_{a}$ and $B_{a}$ fields while $x_{a}$ is a 3-D position vector). Finally, we have considered that for a plane electromagnetic wave, the relations: $\beta_{a}=\epsilon_{abc}\hat{n}^{b}\epsilon^{c}$, $\epsilon_{a}=\epsilon_{abc}\beta^{b}\hat{n}^{c}$ hold\footnote{With $\hat{n}^{a}$ we denote the unit electromagnetic wave vector.}. In the same way, we figure out that:
\begin{equation}
\left(\text{curl}E_{a}\right)B_{b}=i\mathcal{Q}^{(\beta)}_{a}\mathcal{Q}^{(\beta)}_{b}\,.
\end{equation}
Note that in the main text, we incorporate the $90$ degrees phase shift introduced by the imaginary unit $i=e^{i\pi/2}$, into the definition of the wave vector $n^{a}$. Overall, we have:
\begin{equation}
    \mathcal{Q}^{(\epsilon)}_{\langle a}\mathcal{Q}^{(\epsilon)}_{b\rangle}+\mathcal{Q}^{(\beta)}_{\langle a}\mathcal{Q}^{(\beta)}_{b\rangle}=e^{2in^{c}x_{c}}(\epsilon_{\langle a}\epsilon_{b\rangle}+\beta_{\langle a}\beta_{b\rangle})=-\hat{n}_{\langle a}\hat{n}_{b\rangle}e^{2in^{c}x_{c}}\,,
\end{equation}
where we have made use of the intermediate relation:
\begin{equation}
    \beta_{\langle a}\beta_{b\rangle}=-\epsilon_{\langle a}\epsilon_{b\rangle}-\hat{n}_{\langle a}\hat{n}_{b\rangle}\,.
\end{equation}
The last expression actually comes from:
\begin{equation}
    \beta_{a}\beta_{b}=-\epsilon_{a}\epsilon_{b}-\hat{n}_{ a}\hat{n}_{b}+\frac{1}{3}h_{ab}\,,
\end{equation}
which can be proved by recalling that $\epsilon_{abc}\epsilon^{def}= 3!h_{[a}{}^dh_b{}^eh_{c]}^f$.

\end{subappendices}

\newpage

\include{chap:2}

\chapter{Gravito-electromagnetic equivalence in metric affine framework}\label{chap:2}

We revisit the relativistic coupling between gravity and electromagnetism, putting particularly into question the status of the latter; whether it behaves as a source or as a form of gravity. Considering a metric-affine framework and a simple action principle, we find out that a component of gravity, the so-called homothetic curvature field, satisfies both sets of Maxwell equations. Therefore, we arrive at a gravito-electromagnetic equivalence analogous to the mass-energy equivalence. We raise and discuss some crucial questions implied by the aforementioned finding concerning the geometric nature of electromagnetism~\cite{M-GRAV-EM}.\\
The results of the present chapter have essentially been derived in past works within different formulations (e.g. see~\cite{OVEH}-\cite{TW} and the references in~\cite{KPS}). Those were unknown to the author at the time of writing. Our novelty consists of the particular perspective and formulation under which we envisage electromagnetism in a metric affine framework.

\section{Introduction}\label{sec:Introduction}

There are two kinds of well-known (fundamental) macroscopic field quantities introduced to causally describe the motion of matter on large (macroscopic) scales. These are the gravitational and electromagnetic fields, which are conventionally described by General Relativity and Maxwellian Electromagnetism. Due to the wide presence of electromagnetic fields in astrophysical and cosmological environments, we frequently need to consider the parallel presence, coupling or coexistence of gravity and electromagnetism on large-scales. In practice, our conventional perspective consists of envisaging electromagnetic fields (in analogy with matter fields) as sources of gravitation, and therefore generalising the laws of electrodynamics to curved spacetimes (we talk about electrodynamics in curved spacetime). However, unlike (ordinary) matter fields\footnote{There are in fact two explicitly known forms of matter (taking into account the mass-energy equivalence), `ordinary' matter and electromagnetic fields. These can be described by scalar and vector/tensor fields respectively.}, electromagnetic ones possess a geometric nature which allows for their double coupling with spacetime curvature, not only (indirectly) via Einstein's equations but (directly) through the so-called Ricci identities as well\footnote{Apart from the so-called \textit{Einstein-Maxwell coupling}, the \textit{Weyl-Maxwell coupling} (long-range curvature and electromagnetic field) has also been studied within the literature~\cite{MB}.}~\cite{T11} (see also the introduction to Part~I).\\
Overall, it seems to us that the aforementioned special coupling, described through~\eqref{eqn:Ricci-id-4pot}, makes the status of electromagnetic fields essentially different from that of a classical (scalar field) source of gravitation. The aforementioned observation along with another, consisting of the mathematical similarity between the Faraday tensor $F_{ab}=2\partial_{[a}A_{b]}$ and the so-called \textit{homothetic curvature} tensor field $\hat{R}_{ab}=\partial_{[a}Q_{b]}$ (associated with length changes-see the introduction to Chapter~$2$), motivated us to investigate whether electromagnetic fields could be envisaged as a form of spacetime curvature. \\

\subsection{Metric affine framework}\label{ssec:Metric-affine}

Let us briefly present the metric-affine framework~\cite{I}, within which, the above mentioned field $\hat{R}_{ab}$ exists. To begin with, the transition from relativistic to metric affine spacetime requires raising two constraints of Riemannian geometry. On the one hand, we allow for an antisymmetric connection part, $S_{ab}{}^{c}\equiv\Gamma^{c}{}_{[ab]}$ (i.e. the torsion tensor); on the other hand, for a non-vanishing covariant derivative of the metric tensor, $Q_{abc}\equiv-\nabla_{a}g_{bc}\neq 0$ (note that $Q_{a}=g^{bc}Q_{abc}=Q_{ac}{}^{c}$ and $q_{a}=g^{bc}Q_{cba}=Q^{c}{}_{ca}$ are the non-metricity vectors). The former is associated with the impossibility to form infinitesimal parallelograms via parallel transport of a vector upon the direction of another; the latter implies the vector length change during parallel transport. Within a metric-affine geometry, the Ricci tensor has also an antisymmetric part, containing contributions from both torsion and non-metricity. Homothetic curvature $\hat{R}_{ab}=\partial_{[a}Q_{b]}$ is just a component of that antisymmetric part. In the particular case of torsionless spacetime, one has $R_{[ab]}=\hat{R}_{ab}$. While Riemann curvature (or \textit{direction curvature}) is responsible for changes in the direction of parallelly transported vectors along a closed curve, homothetic curvature (or \textit{length curvature}) is associated with changes in vectors' length. It is worth noting that within the literature, the spacetime property of vectors' length change has been argued that it leads to the so-called \textit{second clock effect}, the exclusion of existence of sharp spectral lines, and therefore to a non-physical theory. In particular, the aforementioned problem dates back to Weyl's gauge theory of gravity and Einstein's associated objections (for some historical information refer to e.g.~\cite{HG}; for a modern approach to Weyl's theory see e.g.~\cite{SLFDR}). Interestingly however, it has been recently shown~\cite{HL1},~\cite{HL2} that under appropriate redefinition of proper time and the covariant derivative, the second clock effect does not actually arise in gravity theories with non-metricity.

Up to this point our aim may have already become clear. We will examine whether $\hat{R}_{ab}$ satisfies Maxwell equations, and whether there is a correspondence between homothetic curvature and the Maxwell field. In particular, it is the goal of this Chapter to present the observation that there is indeed a (metric-affine) curvature component field which actually turns out to present an equivalence with the Maxwell field. In face of this finding, we put into question our conventional perspective regarding the way we envisage macroscopic electromagnetic fields and their relation to gravity.\\

\section{Homogeneous (metric affine) Maxwell equations and the implication for gravito-electromagnetic equivalence}\label{sec:Homog-Maxwell-eqs}

Let us start from the expression $\nabla_{[a}F_{bc]}=(1/3)(\nabla_{a}F_{bc}+\nabla_{c}F_{ab}+\nabla_{b}F_{ca})$, within a Riemannian framework. According to the homogeneous Maxwell equations, it has to be equal to zero. Taking thus into account that the Faraday tensor comes from a potential 4-vector, we follow the operations:
\begin{eqnarray}
    \nabla_{[a}F_{bc]}&=&\frac{1}{3!}\left[2\nabla_{[a}\nabla_{b]}A_{c}+2\nabla_{[c}\nabla_{a]}A_{b}+2\nabla_{[b}\nabla_{c]}A_{a}\right]=\frac{1}{3}\left(R_{abcd}+R_{cabd}+R_{bcad}\right)A^{d} \nonumber\\&&
    =R_{[abc]d}A^{d}=0\,.
    \label{eq:Maxwell-Ricci}
\end{eqnarray} 
In other words, we have recalled that if a second-rank antisymmetric tensor field can be written as the gradient of a 4-vector field, then the homogeneous Maxwell equations are a consequence of two geometric properties of the Riemannian spacetime\footnote{Besides, the homogeneous Maxwell equations can be derived theoretically in Minkowski spacetime~\cite{LMEN1} through variation of the action $\mathcal{S}=\int{\left(-\sum{m_{i}}\sqrt{\eta_{ab}\dot{x}_{(i)}^{a}\dot{x}_{(i)}^{b}}-\frac{1}{4}F_{cd}F^{cd}-\sum{e_{i}}A_{a}\dot{x}_{(i)}^{a}\right)\,d\tau}$ with respect to the particles' coordinates $x^{a}(\tau)$ ($\tau$ is the particle's proper--time, its world--line parameter). Subsequently, the homogeneous Maxwell equations are generalised to curved (Riemannian) spacetime via the so--called minimal substitution rule.}; these are the Ricci identities in the form of~\eqref{eqn:Ricci-id-4pot} and the first Bianchi identities (i.e. $R_{[abc]d}=0$).\footnote{For an arbitrary vector field $A_{a}$ the aforementioned properties imply that $\nabla_{[a}\nabla_{b}A_{c]}=0$.} Inversely, if the homogeneous Maxwell equations are satisfied, the second-rank antisymmetric tensor field can be written as the gradient of a 4-vector field in Riemannian spacetime. Therefore, it is clear that $\nabla_{[a}\hat{R}_{bc]}=0$, within the geometric framework in question. It is worth noting that the above well-known conclusion can be generalised to (non-Riemannian) geometries which possess non-metricity\footnote{It can be shown that both the Ricci and the first Bianchi identities maintain their Riemannian form when the relativistic background is modified by the additional non-metricity requirement. In fact, non-metricity is incorporated into the Riemann tensor.} (e.g. see eqs. (1.152) and (1.158) in~\cite{I}, corresponding to the metric-affine version of the Ricci identities and of the first Bianchi identities respectively). Nevertheless, in a general metric affine geometry, possessing torsion as well, the homogeneous Maxwell equations cease to be valid (once again see eqs. (1.152) and (1.158) of~\cite{I}, in combination with~\eqref{eq:Maxwell-Ricci}). In this case, homothetic curvature satisfies the following generalised version of Bianchi identities (known as \textit{Weitzenbock identities}-see eq (1.169) in~\cite{I}):
\begin{equation}
    \nabla_{[a}\hat{R}_{bc]}=2\hat{R}_{d[a}S_{bc]}{}^{d}\,.
    \label{eqn:Weitzenbock}
\end{equation}
Observe that in the absence of torsion, $\hat{R}_{ab}$ satisfies the homogeneous set of Maxwell equations (i.e. $\nabla_{[a}\hat{R}_{bc]}=0$). Besides, it is known that Einstein-Hilbert action\footnote{In fact, there is a generalised action (known under the name \textit{quadratic theory}~\cite{I2}), containing the Einstein-Hilbert, which has as a consequence the property $S_{ab}{}^{c}=-(2/3)S_{[b}\delta_{a]}{}^{c}$.} implies that $S_{ab}{}^{c}=-(2/3)S_{[b}\delta_{a]}{}^{c}$ (with $S_{a}\equiv S_{ab}{}^{b}$ being one of the torsion vectors)-see~\cite{I}. Given the aforementioned property, let us stress out the observation that homothetic curvature satisfies (recall eq.~\eqref{eqn:Weitzenbock}) the following homogeneous set of Maxwell-like equations, namely 
\begin{equation}
    \textcolor{red}{\hat{\nabla}_{[a}\hat{R}_{bc]}=0}\,,\hspace{5mm} \text{where}\hspace{5mm} \hat{\nabla}_{a}=\nabla_{a}-\frac{4}{3}S_{a}\,,\hspace{5mm}\text{for}\hspace{5mm} S_{ab}{}^{c}=-\frac{2}{3}S_{[b}\delta_{a]}{}^{c}\,.
    \label{eqn:Homogeneous-Maxwell}
\end{equation}
We note once again that the above turns out to hold for a generalised action (quadratic theory~\cite{I2}), a part of which is the Einstein-Hilbert. A possible correspondence between the Faraday tensor and the Maxwell potential with the homothetic curvature and the non-metricity vector is apparent. In particular, let us focus on the correspondence $A_{a}\rightarrow Q_{a}$ and $F_{ab}\rightarrow\hat{R}_{ab}$. Taking into account that in geometrised units, $A_{a}$ and $g_{ab}$ are dimensionless, a coupling constant $k$ of length dimension is needed so that dimensional equivalence is established, i.e. 
\begin{equation}
A_{a}=kQ_{a}\hspace{5mm} \text{and} \hspace{5mm} F_{ab}=k\hat{R}_{ab}\,,
\label{eq:grav-em-equivalence}
\end{equation}
where $Q_{a}$ obviously has inverse length dimension. Thus, a potential equivalence between the homogeneous Maxwell equations and~\eqref{eqn:Homogeneous-Maxwell} makes its appearance via the correspondence: $F_{ab}\rightarrow k\hat{R}_{ab}$ and $\nabla\rightarrow\hat{\nabla}$. The question is: \textit{Is there an action reproducing both Einstein and Maxwell field equations, and satisfying the condition $S_{ab}{}^{c}=-\frac{2}{3}S_{[b}\delta_{a]}{}^{c}$ (appearing in~\eqref{eqn:Homogeneous-Maxwell}) as well?} On finding such an action, the above assumed equivalence will be established.\\

\section{Inhomogeneous Maxwell equations: From electrodynamics in curved spacetime to metric affine (gravitational) equivalent of Maxwellian electrodynamics}\label{sec:Inhomog-Maxwell-eqs}

In contrast to the homogeneous set of Maxwell equations, which springs from a purely geometric principle, the inhomogeneous one is known to be a consequence of an action principle (involving the electromagnetic field's strength and its coupling with matter.\\

\subsection{Maxwellian action in curved (relativistic) spacetime}\label{ssec:Maxwell-action-curved-spacetime}

Before answering the question stated in the end of the previous subsection, let us recall that the action for electrodynamics in curved (Riemannian-relativistic) spacetime, reads (e.g. see~\cite{LL1} and~\cite{DFC1}):
 \begin{equation}
\mathcal{S}_{CEM}=\int{\left(R_{ab}g^{ab}+\mathcal{L}_{\text{m}}-\frac{1}{4}F_{ac}F_{bd}g^{ab}g^{cd}-A_{a}J_{b}g^{ab}\right)\sqrt{-g}\,d^{4}x}\,,
     \label{eq-Einstein-Maxwell-action}
 \end{equation}
where $J^{a}$ is the current 4-vector, $\mathcal{L}_{\text{m}}$ is the Lagrangian density of matter and $g$ the determinant of the metric tensor. In the aforementioned combined action, the electromagnetic field couples with the metric tensor of the gravitational field to form the scalar (Lorentz invariant) inner products $F_{ab}F^{ab}=F_{ac}F_{bd}g^{ab}g^{cd}$ and $A^{a}J_{a}=A_{a}J_{b}g^{ab}$. Note that in the above, there are two fundamental fields, the spacetime geometry or gravitation, and the Maxwell gauge potential. In this context, the metric tensor acts as a mediator between fields-sources of gravity-with geometric nature (vectors, tensors) and their energy content (i.e. Lagrangian densities). On the one hand, variations with respect to the potential $A_{a}$ lead to Maxwell equations of the form:
\begin{equation}
{\rm D}_{b}F^{ba}\equiv\frac{1}{\sqrt{-g}}\nabla_{b}\left(\sqrt{-g}F^{ba}\right) =J^{a}\,,\hspace{5mm}\text{where}\hspace{5mm} \frac{1}{\sqrt{-g}}\nabla_{a}\sqrt{-g}=-\frac{1}{2}Q_{a}\,.
\label{eqn:inhomog-Maxwell}
\end{equation}
The above formula reduces to $\nabla_{b}F^{ba}=J^{a}$ in Riemannian spacetime, where $Q^{a}$ vanishes.
Variations with respect to the metric field, on the other hand, lead to Einstein's equations~\eqref{EFE} and the energy-momentum tensors for the matter and Maxwell fields.\\

\subsection{Metric affine (gravitational) equivalent of the Maxwellian action and field equations}\label{ssec:Metric-affine-equivalent-of-Maxwell-action}

We have seen that General Relativity accommodates separate field equations for gravity and electromagnetism, which are derived by a common combined (or `coupled') action. Let us now return to our question regarding the search for an action reproducing inhomogeneous Maxwell-like equations for $\hat{R}_{ab}$, under the condition: $S_{ab}{}^{c}=-(2/3)S_{[b}\delta_{a]}{}^{c}$ (so that the homogeneous set~\eqref{eqn:Homogeneous-Maxwell} is also satisfied). Motivated by~\eqref{eq-Einstein-Maxwell-action}, the simplest action we can imagine, consists of the Einstein-Hilbert and a gravitational analogue of the Maxwellian-electromagnetic action-based on the correspondence~\eqref{eq:grav-em-equivalence}. Besides, our action (aside from the term $Q_{a}J^{a}$), is a particular case of a general model, known as \textit{quadratic theory} (e.g. see~\cite{I2},~\cite{HM} and~\cite{OVEH}). We consider the following:
\begin{equation}
    \mathcal{S}_{GEM}=\int{\left(R+\mathcal{L}_{\text{(m)}}-\frac{k^2}{4}\hat{R}_{ab}\hat{R}^{ab}-\frac{k}{2}Q_{a}J^{a}\right)\sqrt{-g}\,d^{4}x}\,,
     \label{eq-Einstein-Maxwell-action2}
\end{equation}
where $Q_{a}J^{a}$ represents a coupling between charged currents and the non-metricity vector (in analogy with the coupling $A_{a}J^{a}$ between matter and electromagnetic fields\footnote{The electric charge can be envisaged as a kind of coupling constant between matter and electromagnetic fields.}). Within the spirit of our work, unlike electromagnetic fields, we do not envisage matter (and therefore the current $J^{a}$) as a geometric quantity\footnote{Our consideration, regarding the non-geometric origin of matter, differs from the historical effort for gravito-electromagnetic unification in a metric-affine framework, started by Eddington and developed by Einstein~\cite{HG}}. Therefore, the term $Q_{a}J^{a}$ expresses a coupling between charged matter and an element of metric affine curvature. It is worth noting that no new-unknown fields are introduced, just a gravitational analogue of the classical electromagnetic action. Moreover, all action terms are invariant under general coordinate transformations (in contrast to e.g.~\cite{KPS}. Note that in the aforementioned paper the homogeneous set of Maxwell equations is not satisfied). In eq.~\eqref{eq-Einstein-Maxwell-action2} both $Q_{a}$ and therefore $\hat{R}_{ab}$ depend on the metric tensor as well as on the connection (for details see~\cite{I}). Also, we shall keep in mind that the metric appears in the Lagrangian inner products and scalars (i.e. $\hat{R}_{ab}\hat{R}^{ab}=\hat{R}_{ac}\hat{R}_{bd}~g^{ab}g^{cd}$, $Q_{a}J^{a}=Q_{a}J_{b}~g^{ab}$ and $R=R_{ab}~g^{ab}$).\\
First of all, let us consider metric variations of~\eqref{eq-Einstein-Maxwell-action2}. Taking into account the auxiliary relations in the appendix~\ref{appA11}, we arrive at Einstein field equations (of the form~\eqref{EFE}) with stress-energy tensor $T_{ab}=T^{\text{(m)}}_{ab}-(k^{2}/4)\hat{R}_{cd}\hat{R}^{cd}g_{ab}-k^{2}\hat{R}_{ac}\hat{R}^{c}{}_{b}$. Note that $R_{ab}$ and $R$ contain now contributions from torsion and non-metricity, while $T^{\text{(m)}}_{ab}$ refers to the energy-momentum tensor for matter.\\
 Regarding variations with respect to the connection (see the appendix), we receive the following field equations:
 \begin{equation}
     \frac{1}{2}Q_{c}g^{ab}-Q_{c}{}^{ab}-\frac{1}{2}Q^{a}\delta^{b}{}_{c}+q^{a}\delta^{b}{}_{c}+2S_{c}g^{ab}-S^{a}\delta^{b}{}_{c}+g^{ad}S_{dc}{}^{b}+k^2\delta^{b}{}_{c}{\rm D}_{d}\hat{R}^{da}-kJ^{a}\delta^{b}{}_{c}=0\,,
     \label{eq:Gamma-FEQS}
 \end{equation}
 where ${\rm D}_{a}\equiv(1/\sqrt{-g})\nabla_{a}(\sqrt{-g}...)$. Note that the first four terms represent the so-called Palatini tensor. Moreover, all the first seven terms originate from the Einstein-Hilbert action, allowing for non-vanishing torsion and non-metricity (see chapter 2 of~\cite{I}).
 Subsequently, taking the three traces of~\eqref{eq:Gamma-FEQS}, leads to the relations:
 \begin{eqnarray}
 -\frac{3}{2}Q^{a}+3q^{a}+&4k^2{\rm D}_{b}\hat{R}^{ba}&-4kJ^{a}-4S^{a}=0\,, \hspace{10mm}
 \frac{1}{2}Q_{a}+q_{a}+k^2{\rm D}_{b}\hat{R}^{b}{}_{a}-kJ_{a}+4S_{a}=0 \nonumber \\
 &&\text{and}\hspace{5mm}\textcolor{red}{k{\rm D}_{b}\hat{R}^{ba}=J^{a}}~~(\text{with}~\textcolor{red}{{\rm D}_{a}J^{a}=0})\,.
 \label{eq:CFEQS-traces}
 \end{eqnarray}
 Note that eq.~(\ref{eq:CFEQS-traces}c) represents the inhomogeneous set of Maxwell equations. Within the same (metric-affine) framework, action~\eqref{eq-Einstein-Maxwell-action} would lead to the same equations for the Faraday field, namely $\boldsymbol{{\rm D}}_{\boldsymbol{b}}\boldsymbol{F}^{\boldsymbol{ba}}=\boldsymbol{J}^{\boldsymbol{a}}$. Let us point out that eq.~(\ref{eq:CFEQS-traces}c) is essentially a consequence of two basic mathematical properties and one physical property. In detail, the two mathematical properties are: firstly, the similar mathematical construction between the homothetic curvature $\hat{R}_{ab}$ and the Faraday $F_{ab}$ tensor field (i.e. written as the gradient of a vector field); secondly, the linear dependence of the non-metricity vector $Q_{a}$ on the connection, so that $\delta_{\Gamma}Q_{a}=2\delta_{a}{}^{d}\delta_{c}{}^{b}\delta\Gamma^{c}{}_{bd}$. The aforementioned physical property is associated with the action~\eqref{eq-Einstein-Maxwell-action2} itself.\\

\subsection{Constraints}\label{ssec:Constraints}
 
 We observe that charge conservation is expressed in the form ${\rm D}_{a}J^{a}=0~(\Leftrightarrow \nabla_{a}J^{a}=(1/2)Q_{a}J^{a})$. Moreover, taking the nabla divergence of~(\ref{eq:CFEQS-traces}c), we come up with the constraint:
 \begin{equation}
 \nabla_{a}J^{a}=\frac{k}{2}\left(\hat{R}_{ab}\hat{R}^{ab}+Q_{a}\nabla_{b}\hat{R}^{ba}\right)\hspace{5mm} \text{or}\hspace{5mm} k\hat{R}_{ab}\hat{R}^{ab}=Q_{a}\left(J^{a}-k\nabla_{b}\hat{R}^{ba}\right)\,.
 \label{eqn:constraint}
\end{equation}
 In other words, we have figured out that the last two terms in the action~\eqref{eq-Einstein-Maxwell-action2} are actually related with each other through the above expression. \\
 Subsequently, considering various combinations of the three traces in~\eqref{eq:CFEQS-traces} with the initial field equations~\eqref{eq:Gamma-FEQS} (this involves some lengthy but straightforward algebra)\footnote{Note that due to the non-metricity requirement, raising indices is no-longer a trivial operation. For instance, raising indices in~(\ref{eq:CFEQS-traces}b) leads to:
 \begin{equation}
     \frac{1}{2}Q^{a}+q^{a}-kQ_{b}{}^{ca}\hat{R}^{b}{}_{c}+\frac{k^{2}}{2}Q_{b}\hat{R}^{ba}+4S^{a}=0\,.\nonumber
 \end{equation}
 }, we eventually arrive at the constraints:
 \begin{equation}
     Q^{a}=4q^{a}=-\frac{16}{3}S^{a}\,.
     \label{eqn:constraints2}
 \end{equation}
 Namely, within the framework of the action~\eqref{eq-Einstein-Maxwell-action2}, the non-metricity and torsion vectors are linearly dependent, so that they all together correspond to only one degree of freedom. The same thing generally happens when considering only the Einstein-Hilbert action (e.g. see~\cite{I}). In particular, it is well-known that Einstein-Hilbert action does not reproduce general relativity. Instead, it leads to Einstein's field equations along with an additional degree of freedom expressed by~\eqref{eqn:constraints2}. As a consequence of the latter, relation~\eqref{eq:grav-em-equivalence} recasts into:
 \begin{equation}
     A_{a}=kQ_{a}=4kq_{a}-\frac{16}{3}kS_{a}\hspace{5mm}\text{and}\hspace{5mm} F_{ab}=k\hat{R}_{ab}=4kq_{ab}=-\frac{16}{3}kS_{ab}\,,
     \label{eq:gravito-electrom-equiv2}
 \end{equation}
 where $q_{ab}\equiv \partial_{[a}q_{b]}$ and $S_{ab}\equiv \partial_{[a}S_{b]}$. In other words, the vectorial degree of freedom expressed by~\eqref{eqn:constraints2} and allowed by the Einstein-Hilbert action, provides a gravitational equivalent for the Maxwell field. Furthermore, following some lengthy operations, involving eqs.~\eqref{eq:Gamma-FEQS} and~\eqref{eq:CFEQS-traces} (see Chapter~$2$ of~\cite{I}), it can be shown that the torsion and non-metricity tensors are related with the associated vectors via
 \begin{equation}
     S_{ab}{}^{c}=-\frac{2}{3}S_{[b}\delta_{a]}{}^{c}\hspace{5mm}\text{and}\hspace{5mm} Q_{abc}=\frac{1}{4}Q_{a}g_{bc}\,.
     \label{eqn:tensor-vector-S-Q-constraints}
 \end{equation}
 The above constraints hold exactly the same for action~\eqref{eq-Einstein-Maxwell-action2}, given that eq~\eqref{eq:Gamma-FEQS} reduces to Einstein-Hilbert $\Gamma$-field equations under~(\ref{eq:CFEQS-traces}c). Therefore, the homogeneous set of Maxwell equations in the form of~\eqref{eqn:Homogeneous-Maxwell}, is also satisfied by the $\hat{R}_{ab}$ field in the case of the action we examine.\\

\section{Closing remarks-Questions for further research}\label{sec:Discussion}

 Although the present work was initially motivated by the problem of classical gravito-electromagnetic unification, our study points out more a potential equivalence between the Maxwell field and a metric affine component of the gravitational field (i.e. homothetic or length curvature), analogous to mass-energy equivalence. If someone would like to place the present effort within the unified theories context, then it would belong somewhere between the lines of Weyl and Eddington-Einstein. It shares some similarities with both the aforementioned approaches but it essentially differs from both. In particular, envisaging electromagnetism as a component of metric-affine gravity, dates back to the efforts of Weyl, Eddington and Einstein~\cite{HG},~\cite{LOR} (refer to the aforementioned reviews for any information concerning past efforts and failures of unification). Despite that unifying theories are widely regarded by the modern scientific community as a vain dream (presumably because of a long history of failures), history of physics tends to favour an antidiametrically opposite point of view. Let us recall for instance, the many new paths opened through unification of electricity and magnetism, as well as of electromagnetic and weak interactions, in the distant and recent past.
 
 Overall, we have shown that the antisymmetric part of the Ricci tensor, namely the homothetic curvature, satisfies all of Maxwell equations. This finding points out the fundamental question: \textit{Is it possible to exist two different kinds of fields both satisfying Maxwell equations and describing different things? If not, should electromagnetism be envisaged as a form, instead of a source, of gravity on large scales? Alternatively, are electromagnetic fields equivalent to gravitational fields, and which is the equivalence relation?}. Our work shows that there must be such an equivalence, taking the form of~\eqref{eq:grav-em-equivalence}, so that the Maxwell field can be calculated from a given metric. The aforementioned relation implies that a given electromagnetic field has a gravitational equivalent determined via the conversion constant $k$. It is worth noting that there is a remarkable analogy between gravito-electromagnetic (eq.~\eqref{eq:grav-em-equivalence}) and mass-energy equivalence, i.e. $E=mc^2$ ($k$ is the counterpart of $c^{2}$). Presenting and supporting the idea of a potential \textit{gravito-electromagnetic equivalence} is essentially  the contribution of the present piece of work. Therefore, two crucial questions arise.
 
 Firstly, which is the nature of the conversion constant $k$ and how can it be determined? Let us make a \textit{conjecture}. On the one hand, we observe that the action term $J_{a}Q^{a}$, introduced in~\eqref{eq-Einstein-Maxwell-action2}, establishes a coupling between matter and non-metricity, mediated by the electric charge. On the other hand, within classical electrodynamics, the electric charge is known to act as a coupling constant between matter and electromagnetic field (see $J_{a}A^{a}$ in~\eqref{eq-Einstein-Maxwell-action}). The aforementioned double coupling potentially implies an equivalence relation between non-metricity and the Maxwell field, where the electric charge plays the role of the coupling constant. Besides, we take into account that the electric charge has length dimension in geometrised units. Therefore and in other words, we state the following question: Could the coupling constant $k$ (with length geometrised dimension) be identified as the total electric charge of a given charged distribution? If this is the case, it would appear that the electric charge behaves on large-scales as a quantity which translates a given electromagnetic field into its gravitational equivalent. Furthermore, according to~\eqref{eq:grav-em-equivalence}, with $k\rightarrow\mathcal{Q}$, opposite charges correspond to homothetic curvature of opposite sign. Could the macroscopic interaction between a positive and a negative charge distribution be envisaged as a consequence of an `interaction' between opposite kinds of homothetic curvature?
 
 Secondly, how are the properties of the Maxwell field ($A_{a}=kQ_{a}=(16/3)kS_{a}$, via~\eqref{eqn:constraints2}) reconciled with the geometric significance/properties of non-metricity and torsion? The aforementioned properties are respectively the change to a vector's magnitude under its parallel transport along a given curve, and the impossibility to form a closed (small) parallelogram under parallel transport of one vector to the direction of another~\cite{I}.\\
 Addressing the above exposed questions/problems is left to the future.\\ 

\begin{subappendices}

\section{Metric and connection variations}\label{appA11}

In deriving the field equations within the main text, we make use of the following relations for metric and connection variations~\cite{I},~\cite{I2}. Concerning the former, we have:
\begin{equation}
    \delta_{g}Q_{a}=\partial_{a}\left(g_{bc}\delta g^{bc}\right)\,,\hspace{4mm} \delta_{g}\hat{R}_{ab}=\partial_{[a}\delta_{g}Q_{b]}=0\,,\hspace{4mm}\delta_{g}\sqrt{-g}=-(1/2)\sqrt{-g}g_{ab}\delta g^{ab}\,,\nonumber
\end{equation}
\begin{equation}
\delta_{g}\left(\hat{R}_{ab}\hat{R}^{ab}\right)=\delta_{g}\left(\hat{R}_{ab}\hat{R}_{cd}g^{ac}g^{bd}\right)=-2\hat{R}_{ac}\hat{R}^{c}{}_{b}\delta g^{ab}\nonumber
\end{equation}
and
\begin{equation}
    \delta_{g}\left(\hat{R}_{ab}\hat{R}^{ab}\sqrt{-g}\right)=\left(-2\hat{R}_{ac}\hat{R}^{c}{}_{b}-(1/2)\hat{R}_{cd}\hat{R}^{cd}g_{ab}\right)\sqrt{-g}\\\delta g^{ab}\,.\nonumber
\end{equation}
As for the latter, we deploy:
\begin{equation}
    \delta_{\Gamma}Q_{a}=2\delta_{a}{}^{d}\delta_{c}{}^{b}\delta\Gamma^{c}{}_{bd}\hspace{4mm}\text{and}\hspace{4mm}\delta_{\Gamma}(\hat{R}_{ab}\hat{R}^{ab})=-4\nabla_{b}\hat{R}^{ba}\delta\Gamma^{c}{}_{ca}=-4\nabla_{d}\hat{R}^{da}\delta^{b}{}_{c}\delta\Gamma^{c}{}_{ab}\,,\nonumber
\end{equation}
with $\delta_{a}{}^{b}$ being the Kronecker symbol.\\

\end{subappendices}

\newpage

\part{}

\section{Magnetised fluids in astrophysical and cosmic environments}\label{sec:magnetised-fluids}

Large-scale magnetic fields form an unambiguously existing component of the universe's energy content, which potentially contributes to cosmic dynamics and structure formation (via its effects on density inhomogeneities). Although magnetic fields are widely present in the universe, in both astrophysical and cosmological scales (i.e. compact stellar objects, interstellar medium, galaxies, galaxy clusters, intergalactic space), their origin, evolution and role have not been adequately explained.\\

\subsection{The Magnetohydrodynamics (MHD) approximation}\label{ssec:Magnetohydrodynamics}

The unambiguous detection of large-scale magnetic fields along with the parallel absence of large-scale electric fields, motivates the study of magnetised fluid models. Within the theoretical framework of magnetohydrodynamics (MHD) it is possible to isolate the magnetic component of the Maxwell field by adopting a highly conducting fluid model (e.g. see~\cite{EMM3} for the covariant description). In particular, according to Ohm's law applied in the fluid's rest frame,
\begin{equation}
    \mathcal{J}_{a}=\varsigma E_{a}\,,
    \label{Ohm}
\end{equation}
non-zero spatial currents arise for $E_{a}\rightarrow 0$ at the MHD limit (i.e. $\varsigma\rightarrow\infty$, where $\varsigma$ is the conductivity of the medium). Then, according to Alfv\'{e}n's theorem, the magnetic field lines of a highly conducting fluid behave as being frozen into the fluid; they move with it and always connect the same matter particles.\\
Under the MHD approximation, the magnetic field itself can be envisaged as a viscous fluid with energy-momentum tensor:
\begin{equation}
T^{\text{(magn)}}_{ab}=\frac{1}{2}B^{2}u_{a}u_{b}+\frac{1}{6}B^{2}h_{ab}-B_{\langle a}B_{b\rangle}\,,
\label{eqn:magnetic-energy-tensor}
\end{equation}
where $\rho_{B}=B^{2}/2$, $P_{B}=B^{2}/6$ and $\pi^{B}_{ab}=-B_{\langle a}B_{b\rangle}$ encode the magnetic energy density, isotropic pressure and anisotropic stress respectively. Similarly, Maxwell equations reduce to the following propagation equation:
\begin{equation}
    \dot{B}_{\langle a\rangle}=\left(-\frac{2}{3}\Theta h_{ab}+\sigma_{ab}+\epsilon_{abc}\omega^{c}\right)B^{b}\,,
    \label{MHD-B-prop}
\end{equation}
also known as the \textit{magnetic induction equation}, which is the MHD version of Faraday's law; and the three constraints
\begin{equation}
    \mathcal{J}_{a}=\epsilon_{abc}\dot{u}^{b}B^{c}+\text{curl}B_{a}\,,
    \label{MHD-current}
\end{equation}
\begin{equation}
    2\omega^{a}B_{a}=\mu \quad \text{and} \quad {\rm D}^{a}B_{a}=0\,,
    \label{MHD-constraints}
\end{equation}
where according to~\eqref{MHD-current} the magnetic field lines remain frozen-in with the matter, in the form of currents. In accordance with eq.~\eqref{MHD-B-prop}, the magnetic field is not sourced by currents but instead by purely kinematic effects. It is worth noting that the multiplication of~\eqref{MHD-B-prop} by $B^{a}$ leads to the following evolution formula for the magnetic energy density:
\begin{equation}
\dot{\left(B^{2}\right)}=-\frac{4}{3}\Theta B^{2}-2\sigma_{ab}\pi^{ab}_{B}\,.
\label{eqn:magn-density-evol-eq}
\end{equation}
The above shows that in the absence of anisotropy, magnetic density changes as $B^{2}\propto a^{-4}$ (with $a$ being the scale factor).\\
For an ideal fluid, the zero divergence of the total energy-momentum tensor (matter plus magnetic field), $\nabla^{b}T_{ab}=0$ splits into the continuity equation: $\dot{\rho}=-\Theta(\rho+P)$, where there is no magnetic contribution; and the equation of motion (i.e. Euler's equation):
\begin{equation}
    (\rho+P)\dot{u}_{a}=-{\rm D}_{a}P+\epsilon_{abc}\mathcal{J}^{b}B^{c}\,,
    \label{Euler-ideal-MHD}
\end{equation}
where the pressure gradients and the magnetic Lorentz force are the remaining causes of non-geodesic motion. Substituting the current from~\eqref{MHD-current} into the last term in the above relation and following the operations we arrive at:
\begin{equation}
    \epsilon_{abc}\mathcal{J}^{b}B^{c}=-B^{2}\dot{u}_{a}+\dot{u}^{b}B_{b}B_{a}-\frac{1}{2}{\rm D}_{a}B^{2}+B^{b}{\rm D}_{b}B_{a}\,.
    \label{eqn:Lorentz-force}
\end{equation}
With the aid of the last expression, Euler's equation~\eqref{Euler-ideal-MHD} recasts into:
\begin{equation}
    (\rho+P+B^{2})\dot{u}_{a}=-{\rm D}_{a}P+\dot{u}^{b}B_{b}B_{a}-\frac{1}{2}{\rm D}_{a}B^{2}+B^{b}{\rm D}_{b}B_{a}\,.
    \label{Euler's-MHD}
\end{equation}
The last two terms in the right-hand side of the above relation split the magnetic Lorentz force into its pressure and tension component respectively.\\

\subsection{Magnetic field evolution formulae under expansion/contraction}\label{ssec:Solutions-for-B}

Let us focus our interest on Faraday's formula~\eqref{MHD-B-prop} and the search of solutions for the magnetic field. In particular, let us introduce the unit spacelike vector field $n^{a}$, along the magnetic direction (i.e. $u_{a}n^{a}=0$, $n_{a}n^{a}=1$ and $B^{a}=\mathcal{B}n^{a}$)\footnote{For details regarding the kinematics on the surface normal to $n^{a}$, see the following chapter and its appendix section~\ref{appB-after}.}. Therefore, projecting Faraday's equation along the magnetic direction $n^{a}$, it transforms into:
\begin{equation}
    \dot{\mathcal{B}}=\left(\Sigma-\frac{2\Theta}{3}\right)\mathcal{B}\,,
    \label{eqn:Faraday-projection}
\end{equation}
with $\Sigma\equiv\sigma_{ab}n^{a}n^{b}$ the shear scalar component, a quantity associated with volume expansion/contraction ($\Sigma=\Theta_{ab}n^{a}n^{b}-\Theta/3$~-see the comments below)\footnote{In fact, note that $\Sigma\equiv\sigma_{ab}n^{a}n^{b}\equiv({\rm D}_{\langle b}u_{a\rangle})n^{a}n^{b}=({\rm D}_{b}u_{a})n^{a}n^{b}-(\Theta/3)h_{ab}n^{a}n^{b}=u'_{a}n^{a}-\Theta/3$, where $u'_{a}\equiv n^{b}{\rm D}_{b}u_{a}$.}. Note that in deriving the above we have taken into account that $\dot{B}_{\langle a\rangle}=h_{a}{}^{b}\left(\dot{\mathcal{B}}n_{b}+\mathcal{B}\dot{n}_{b}\right)=\dot{\mathcal{B}}n_{a}+\mathcal{B}\alpha_{a}$ where $\dot{n}_{a}=\mathcal{A}u_{a}+\alpha_{a}$ ($\mathcal{A}\equiv\dot{u}_{a}n^{a}$ and $\alpha_{a}$ its component lying on the $2$-dimensional surface normal to $n^{a}$). Considering that $\Sigma=\lambda\Theta$ ($\lambda$ being a real number), equation~\eqref{eqn:Faraday-projection} recasts into the solvable form:
\begin{equation}
\dot{\mathcal{B}}=\left(\lambda-\frac{2}{3}\right)\Theta\mathcal{B}\,.
\label{eqn:Faraday-solvable-form}
\end{equation}
From the above it is clear that only solutions with $\lambda<2/3$ predict a decrease for the magnetic field during expansion (note that $\lambda=2/3$ implies constant magnetic density under volume changes). Therefore, we envisage as physically senseful those solutions which satisfy the aforementioned constraint. Moreover, in reference to the problem of magnetised gravitational collapse, the requirement for pure contraction translates into a negative volume scalar $\Theta<0$ and a negative generalised Hubble parameter (see~\cite{EMM}), i.e. $\Theta_{ab}n^{a}n^{b}<0$. In detail, the generalised Hubble relation reads:
\begin{equation}
    \Theta_{ab}n^{a}n^{b}=\Sigma+\frac{\Theta}{3}=\left(\lambda+\frac{1}{3}\right)\Theta <0\,,\hspace{8mm}\text{namely}\hspace{8mm} \lambda>-\frac{1}{3}\,,
\end{equation}
where $\Theta_{ab}\equiv\nabla_{(b}u_{a)}\equiv\sigma_{ab}+(\Theta/3) h_{ab}$ is the expansion tensor and $n^{a}$ is an arbitrary spatial ($3$-D) and unitary vector.
Overall, substituting $\Theta=3\dot{a}/a$ ($a$ being an average scale factor) into eq.~\eqref{eqn:Faraday-solvable-form}, leads to the following law of variation for the magnetic field:
\boldmath
\begin{equation}
\mathcal{B}\propto a^{3\lambda-2}\hspace{6mm}\text{with}\hspace{6mm}-\frac{1}{3}<\lambda<\frac{2}{3}\,,
\label{eqn:magn-law-of-variation}
\end{equation}
\unboldmath
where its lower boundary is imposed by the problem of magnetised gravitational contraction. The above fundamental expression (a new result within the literature-as far as we know) is generally valid within the context of magnetohydrodynamics. Throughout this chapter, we make use of it to study magnetic fields in astrophysical and cosmological frameworks.\\
It is worth mentioning the particular condition under which eq.~\eqref{eqn:magn-law-of-variation} holds in orthogonal Bianchi models. Following~\cite{CEll}, the kinematics of the aforementioned models is constrained by:
\begin{equation}
-\frac{\Theta}{3}\leq \Theta_{ab}n^{a}n^{b} < \Theta\hspace{10mm} \text{or}\hspace{10mm} -\frac{2\Theta}{3}\leq \Sigma < \frac{2\Theta}{3}\,,
\label{eqn:Theta-tensor-condition}
\end{equation}
where $\Theta_{ab}\equiv\nabla_{(b}u_{a)}\equiv\sigma_{ab}+(\Theta/3) h_{ab}$ and $n^{a}$ is a vector tangent to the hypersurfaces of homogeneity. In deriving eq.~(\ref{eqn:Theta-tensor-condition}b), we have considered that $\Theta_{ab}n^{a}n^{b}=\Sigma+\Theta/3$ (with $\Sigma\equiv\sigma_{ab}n^{a}n^{b}$). Consequently, eq.~\eqref{eqn:magn-law-of-variation} reduces, for orthogonal Bianchi models, to:
\begin{equation}
\mathcal{B}\propto a^{3\lambda-2}\hspace{6mm}\text{with}\hspace{6mm}-\frac{2}{3}\leq\lambda<\frac{2}{3}\,.
\label{eqn:magn-law-of-variation-orthog-Bianchi}
\end{equation}
The last expression introduces a lower boundary to the rate of decrease/increase of the magnetic field during expansion or contraction respectively.

\include{chap:3}

\chapter{\bf Gravito-magnetic elasticity and the problem of magnetised gravitational collapse }\label{chap:3}

Magnetic fields are a very special form of elastic medium. Within cosmic and astrophysical environments (magnetised stars and protogalaxies) they counteract shear and rotational distortions as well as gravitational collapse. Their vector nature allows for their extraordinary coupling with spacetime curvature in the framework of general relativity. This particular coupling points out the way to study magnetic elasticity under gravitational deformation. In the context of magnetohydrodynamics, we provide a covariant description of a self-gravitating magnetised fluid; reveal the law of gravito-magnetic elasticity and derive a parametric (kinematics dependent) evolution formula for the magnetic field. Subsequently, we arrive at a non-collapse criterion for a magnetised fluid, and address the question whether the magnetic forcelines are able to prevent contraction towards a singularity. Our answer depends on the explicit value of the parameter determining our magnetic evolution formula. Finally, our argumentation peaks by suggesting a calculation for the fracture limit of magnetic fieldlines under gravitational contraction. Two illustrative applications, in a neutron star and a white dwarf, accompany the results.\\

 \section{Introduction}\label{sec:Introduction3}

 It is well known (mainly from astrophysical studies of magnetised fluids, e.g. see~\cite{Pa} and~\cite{M}, but from relativistic as well~\cite{GT}) that magnetic forcelines behave like an elastic medium under their kinematic (shear or rotational) deformation. Namely, in analogy with a spring under pressure, they develop tension stresses resisting their deflection. However, it may be less known that magnetic fieldlines present a similar behaviour under spacetime distortions~\cite{T13}-\cite{MT}. In particular, the aforementioned studies have shown that the magnetic elasticity in question is expressed through a magneto-curvature tension stress, coming from the Ricci identities. Therefore, from a relativistic point of view, magnetic fields acquire particular interest due to their direct coupling, as vectors, with spacetime curvature (see the associated sections in chapters~\ref{chap:1} and~\ref{chap:2}).

 In detail, previous independent relativistic studies have supported the following basic ideas regarding the behaviour of magnetic fields in curved spacetimes. First, magnetic fields have the impressive ability not to self-gravitate; in other words, not to contract or collapse under their own gravity independently of the latter's strength~\cite{Mel, Th}. Second, in the presence of an external gravitational field, magnetic forcelines tend to stabilise themselves by developing naturally curvature related tension stresses (monitored by Ricci identities for the magnetic field) which resist their gravitational deformation~\cite{Mel, Th}. Third, the key factor giving rise to such an unconventional behaviour in both cases is the aforementioned magnetic  elasticity~\cite{T23, KT3, TMav}. In a sense, it appears as though the elastic properties of the magnetic forcelines are transferred into the fabric of the host space, which seems to act like an elastic medium under tension.

 In accordance with the above data, magnetic fieldlines counteract gravitational implosion of a highly conducting fluid, and potentially hold it up~\cite{T23}-\cite{Ch}. The question on such a possibility primarily motivated the work of this chapter (see~\cite{Mav-R-Soc}). In reference to this problem and given the elastic behaviour of magnetic fields, one can raise another associated question concerning the existence of a possible elastic and more crucially a fracture gravito-magnetic limit (i.e. an amount of gravity under which magnetic forcelines lose their coherence-cease to exist). Moreover, if such a fracture limit exists, could we have an estimation of it for a given magnetised (collapsing) star? Crucially, could magnetic forcelines manage to disrupt gravitational collapse before reaching their fracture limit~\cite{MT}? Besides, it is worth noting that a systematic theoretical description for gravito-magnetic elasticity is elusive. In addressing all of the above questions, one needs an evolution formula for the magnetic field of a highly conductive, self-gravitating fluid. In alignment with the magnetic elasticity, we introduce an approximation leading to such a law of variation, which we subsequently apply to the aforementioned problems.\\

\section{Kinematically induced magnetic tension stresses}

To begin with, let us consider the decomposition of the magnetic 3-D gradient $\rm D_{b}B_{a}$ into its symmetric (trace-free), antisymmetric and trace part. In other words,
\begin{equation}
    {\rm D}_{b}B_{a}={\rm D}_{\langle b}B_{a\rangle}+{\rm D}_{[b}B_{a]}+\frac{1}{3}({\rm D}^{c}B_{c})h_{ab}\,,
    \label{magn-gradient}
\end{equation}
which reveals the individual tension\footnote{Actually the magnetic tension force vector refers to the directional derivative along the field itself. See the following analysis.} components triggered by, and resisting to shape (i.e. $\sigma^{(B)}_{ab}={\rm D}_{\langle b}B_{a\rangle}$), rotational (i.e. $\omega^{(B)}_{ab}={\rm D}_{[b}B_{a]}$) and volume distortions (i.e. $\Theta^{(B)}={\rm D}^{a}B_{a}$) of the magnetic forcelines respectively. Besides, at the magnetohydrodynamic limit (MHD), the tension component opposing to volume expansion/contraction (last term) vanishes (i.e. ${\rm D}^{c}B_{c}=0$ from Gauss's law). In the above, ${\rm D}_{a}=h_{a}{}^{b}\nabla_{b}$ is the projected (3-D) covariant derivative operator and $h_{ab}=g_{ab}+u_{a}u_{b}$ (with $g_{ab}$ being the spacetime metric and $u^{a}$ being a timelike $4$-velocity vector) an operator projecting upon the observer's (3-D) rest-space.  The covariant kinematics of the magnetic tension stresses is monitored by the Ricci identities for the magnetic field:
\begin{equation}
    2\nabla_{[a}\nabla_{b]}B_{c}=R_{abcd}B^{d}\,,
    \label{4-Ricci}
\end{equation}
where $R_{abcd}$ is the Riemann spacetime tensor. In particular, the timelike part of the above leads to propagation equations for the magnetic shear $\sigma^{(B)}_{ab}$ and vorticity $\omega^{(B)}_{a}=\epsilon_{abc}\omega^{bc}$. On the other hand, its spacelike part leads to divergence conditions (constraints) for the aforementioned quantities. The equations in question, appearing here for the first time-as far as we know-, could prove useful when studying the kinematics of magnetised fluids in various contexts. However, as we do not make any use of those in the present thesis, we have chosen to place them in the brief appendix~\ref{appA-prop-eqns-magn-tension-stresses}.  \\

\section{Gravitationally induced magnetic tension stresses}

In analogy with their deflection due to kinematic effects, associated with the fluid's motion, magnetic forcelines counteract their gravitational distortion. Where does the corresponding magneto-curvature tension stress come from? The answer lies in the direct coupling of magnetic fields (as vectors) with spatial curvature via the (3-D projected) Ricci identities (e.g. see~\cite{T1} or~\cite{TCM}),
\begin{equation}
     2{\rm D}_{[a}{\rm D}_{b]}B_{c}=-2\omega_{ab}\dot{B}_{\langle c\rangle}+\mathcal{R}_{dcba}B^{d}\,,
     \label{3-Ricci}
\end{equation}
where $\mathcal{R}_{abcd}$ represents the 3-D counterpart of the Riemann tensor. Note that the aforementioned coupling manifests itself at the second differentiation order.\\

\subsection{Describing the magneto-curvature tension stress}\label{ssec:magneto-curvature tension}

Let us consider the 3-gradient of the magnetic tension force vector $\tau_{a}=B^{b}{\rm D}_{b}B_{a}$ (the non-zero tension force implies that the magnetic fieldlines are not spacelike geodesics). Employing the 3-D Ricci identities~\eqref{3-Ricci} we arrive at:
\begin{equation}
    {\rm D}_{c}\tau_{a}={\rm D}_{c}B^{b}{\rm D}_{b}B_{a}+B^{b}{\rm D}_{b}{\rm D_{c}}B_{a}+2\omega_{bc}B^{b}\dot{B}_{\langle a\rangle}+\mathcal{R}_{dabc}B^{b}B^{d}\,.
\end{equation}
The first three terms involve kinematic effects through eq~\eqref{magn-gradient} whilst the last one can be envisaged as the magneto-curvature tension stress (or the gradient of the magneto-curvature tension component). If $n^{a}$ is the magnetic field direction (i.e. $B^{a}=\mathcal{B}n^{a}$), the term in question can alternatively be written as:
\begin{equation}
s_{ac}=\mathcal{R}_{dabc}B^{b}B^{d}=\mathcal{B}^{2}\mathcal{R}_{dabc}n^{b}n^{d}=-\mathcal{B}^{2}u_{ac}\,,
    \label{Magn-curv-tensor}
\end{equation}
where $u_{ac}\equiv-\mathcal{R}_{dabc}n^{b}n^{d}$ can be envisaged as a kind of strain tensor\footnote{Typically, within conventional elastic mechanics the strain tensor is defined to be a dimensionless symmetric quantity~\cite{LL3}. However, here we allow for a non-vanishing anti-symmetric part taking into account any torsional deformation. Furthermore, our strain tensor has inverse square length dimensions in geometrised units.}, describing spatial distortions of the magnetic forcelines (for a commentary on the law of magnetic elasticity under volume gravitational distortions refer to section~\ref{sec:Magnetic-elasticity-law}). Our definition for the strain tensor is metric independent and thus essentially differs from its counterpart~(4.9) in~\cite{CQ}. As for the stress tensor $s_{ab}$, it includes those forces which act against (see the following discussion on the problem of gravitational collapse) spatial curvature and tend to restore the forcelines to their initial state. Overall, the meaning of~\eqref{Magn-curv-tensor} is the following. Due to spatial curvature, the magnetic fieldlines are bent and twisted. In analogy with an elastic rod under pressure, they react via the restoring stress $s_{ab}$ which increase in proportion to the amount of deformation $u_{ab}$ (Hooke's law of elasticity) and the magnetic density. In fact, when appearing in the kinematic equations for a magnetised fluid, it turns out that the magneto-curvature tension stress depends on the ratio of the magnetic density over the total system's density (i.e. matter and magnetic fields-see eq.~\eqref{el-law2} in the following).

Let us recall that any kind of deformation can be reduced into a sum of a pure shear ($u_{\langle ab\rangle}=\mathcal{R}_{d\langle ab\rangle c}n^{c}n^{d}$), a torsional one (or twisting)\footnote{For the sake of accuracy, vorticity or rotational deformations are included in the shear/shape type of distortions as well.} ($u_{[ab]}=\mathcal{R}_{d[ab]c}n^{c}n^{d}$) and a hydrostatic compression ($(u^{c}{}_{c}/3)h_{ab}=(1/3)\mathcal{R}_{cd}n^{c}n^{d}h_{ab}$). Hence, on splitting the magneto-curvature tension stress into its symmetric trace-free ($s_{\langle ac\rangle}$), antisymmetric ($s_{[ac]}$) and trace part ($s=s^{c}{}_{c}$), we receive its associated component counteracting shape, rotational and volume changes respectively, due to gravity (see appendix~\ref{app:magneto-curvature-tension}). Of the aforementioned components we focus here on the last one. Hence, considering the trace of~\eqref{Magn-curv-tensor} and the double projection of the Gauss-Codacci formula (e.g. see eq.~(1.3.39) in~\cite{TCM3} and eq~(92) in~\cite{MT}) along the magnetic direction $n^{a}$, we deduce that
\begin{equation}
    s=s^{c}{}_{c}=\mathcal{B}^{2}\mathcal{R}_{bd}n^{b}n^{d}=\mathcal{B}^{2}\left[\frac{2}{3}\rho+\mathcal{E}+\frac{\Pi}{2}+\left(\lambda^{2}-\frac{\lambda}{3}-\frac{2}{9}\right)\Theta^{2}+\alpha_{a}\alpha^{a}\right]\,,
    \label{elast-law1}
\end{equation}
 where $\rho$ denotes the energy density of matter; $\Pi\equiv \pi_{ab}n^{a}n^{b}$ and $\mathcal{E}\equiv E_{ab}n^{a}n^{b}$ are the anisotropic stress $\pi_{ab}$ and the tidal (or electric Weyl) tensor $E_{ab}$, twice projected along the magnetic direction, respectively. In deriving the above we have employed $\Sigma=\lambda\Theta$ and eqs~\eqref{eqn:appB1} and~\eqref{eqn:Sigma-vector} from the appendix~\ref{app:magneto-curvature-tension} (see eq.~\eqref{eqn:3D-curv-along-n} as well). Also, assuming an ideal fluid model, the anisotropic stress terms in the above vanish. Then, of particular interest is that the deformation due to gravitational compression/expansion in~\eqref{elast-law1} is determined by the density  of matter and the tidal tensor projected along the magnetic fieldlines. Note that there is no magnetic input in~\eqref{elast-law1}. In fact, recalling that $\rho_{\text{(magn)}}=\mathcal{B}^{2}/2$, $\pi^{(\text{magn})}_{ab}=-B_{\langle a}B_{b\rangle}$ and therefore $\Pi^{\text{(magn)}}\equiv\pi^{\text{(magn)}}_{ab}n^{a}n^{b}=-(2/3)\mathcal{B}^{2}$, it turns out that the magnetic anisotropic stress exactly cancels the magnetic energy density contribution (see also~\cite{TMav}). The aforementioned conclusion becomes straightforward if we assume the simple case of a magnetic field as the sole energy source of spatial curvature. Then, it is clear that (see also~\cite{TMav})
 \begin{equation}
\mathcal{R}_{ab}=\frac{1}{3}\mathcal{B}^{2}h_{ab}+\frac{1}{2}\pi^{\text{(B)}}_{ab}\hspace{6mm}\text{with}\hspace{6mm}\mathcal{R}_{ab}n^{a}n^{b}=0\,,
\label{eqn:zero-magn-gravity}
\end{equation}
irrespective of the field's strength. In other words, magnetic fields do not self-gravitate or do not `feel' their own gravity, no matter how strong the latter may be.\\

\section{Gravitational collapse of a magnetised fluid}
\label{sec:Grav-collapse}

The gravitational collapse of compact stellar objects, like white dwarfs, neutron stars, black holes, as well as that of protogalactic clouds usually involves (weak or strong) magnetic fields. In the context of general relativity, independent studies have pointed out the unconventional tendency of the $B$-fields  to resist their own gravitational implosion. The same works have also raised the question as to whether the magnetic presence and the resulting Lorentz forces could actually halt the contraction of the surrounding collapsing matter~\cite{T23}-\cite{Th}. In addition, alternative studies of charged collapse, this time employing the repulsive (electrostatic) Coulomb forces, have found that the latter could also prevent the formation of spacetime singularities~\cite{N3}--\cite{R3}. The present section probes the gravitational collapse of a highly conductive charged medium by means of the Raychaudhuri equation and along the lines of~\cite{T23}-\cite{TMav}. Making a step further, we take advantage of a $1+2$ spatial splitting and arrive at a simple criterion which could decide the ultimate fate of homogeneously contracting magnetised media.\\

\subsection{Using the Raychaudhuri equation}\label{ssec:Raychaudhuri}

Traditionally, theoretical studies of gravitational collapse make use of the Raychaudhuri equation which has been made famous as a keystone of singularity theorems. Besides, in general terms, the formula in question covariantly describes the volume evolution of a self--gravitating fluid element. In this first subsection, we revisit the problem of gravitational implosion of a highly conducting (magnetised) fluid with the aid of the Raychaudhuri equation\footnote{Apart from its conventional application to timelike worldlines of real (or hypothetical) observers, the aforementioned equation has been applied to spacelike and null curves as well (e.g.~see~\cite{GT,KS,AV}).}, and in light of our new knowledge regarding the behaviour of the associated magnetic field (more specifically of relation~\eqref{eqn:bexp}), as well as of our new developments in the context of the 1+1+2 covariant formalism. Unlike previous independent works, our study builds upon past research (see~\cite{TMav}-\cite{T33}) and leads to a simple and clear criterion determining the fate of homogeneous and magnetised gravitational collapse.

Before proceeding to the analysis, let us have in mind two crucial points. Firstly, magnetic-line deformations are usually caused by electrically charged particles, however relativistic spacetime curvature (gravity) also potentially behaves as a deforming agent~\cite{T23, TMav}. Secondly, the magnetic tension reflects the elasticity of the field lines and their tendency to react against any agent that distorts them from equilibrium~\cite{TMav, T33, KT}.

Let us start with the Raychaudhuri equation, which we have already written in the form of~\eqref{Raych-eqn1}. To proceed, we need to calculate the 3-divergence of the acceleration vector (i.e. ${\rm D}^{a}\dot{u}_{a}$) which gives rise to magneto-curvature tension terms, of crucial importance for our relativistic study.  In order to facilitate the analytic calculations, we assume that the contracting fluid has nearly homogeneous matter\footnote{Note that the homogeneity of the matter fields is a rather common approximation. In fact, spatial homogeneity is a standard assumption in all typical singularity theorems~\cite{W3, HE}. Besides, the assumption of homogeneous matter distribution does not essentially affect the validity of our argument, since gradients in the fluid and in the magnetic density distribution tend to inhibit gravitational contraction, even within Newtonian physics.} and magnetic energy density distributions (${\rm D}_{a}\rho\simeq 0\simeq {\rm D}_{a}P\simeq {\rm D}_{a}B^{2}$, where an equation of state of the form $P=P(\rho)$ has been considered). However, we allow for $B^{b}{\rm D}_{b}B_{a}\neq 0$, so that we can study effects caused by distortions of the magnetic forcelines (see the following discussion). Subsequently, taking the 3-divergence of~\eqref{Euler's-MHD} in combination with the 3-Ricci identities (eq.~\eqref{3-Ricci}) and Maxwell's equations (eq.~\eqref{magn-div}) we arrive at:
\begin{equation}
    {\rm D}^{a}\dot{u}_{a}=c^{2}_{\mathcal{A}}\mathcal{R}_{ab}n^{a}n^{b}+2(\sigma^{2}_{B}-\omega^{2}_{B})\,,
    \label{3-div-of-dot-u}
\end{equation}
where the scalars $\sigma^{2}_{B}={\rm D}_{\langle b}B_{a\rangle}{\rm D}^{\langle b}B^{a\rangle}/2(\rho+P+B^{2})$ and $\omega^{2}_{B}={\rm D}_{[b}B_{a]}{\rm D}^{[b}B^{a]}/2(\rho+P+B^{2})$ represent the magnetic analogues of the shear and the vorticity respectively. Of special interest is the purely relativistic (magneto-geometric) term $\mathcal{R}_{ab}n^{a}n^{b}$ which describes 3-D distortions of the magnetic forcelines due to the curvature of the host spacetime. Note that all the terms on the right-hand side of ~\eqref{3-div-of-dot-u} are tension stresses triggered by the deformation of the magnetic field lines. Each of these terms acts against the agent that caused the deformation in the first place (e.g. the magneto-vorticity $\omega^{2}_{B}$ is caused by rotational effects, $\omega^{2}$, and it tends to counterbalance them. Observe the opposite signs of the pairs $\omega^{2}$, $\omega^{2}_{B}$ and $\sigma^{2}$, $\sigma^{2}_{B}$ in~\eqref{Raych-eq2}). Substituting expression~\eqref{3-div-of-dot-u} into the Raychaudhuri equation~\eqref{Raych-eqn1}, describing the convergence/divergence of the timelike worldlines tangent to the $u^{a}$ field, our equation reads:
\begin{equation}
    \dot\Theta+\frac{1}{3}\Theta^{2}=-R_{ab}u^{a}u^{b}+c^{2}_{\mathcal{A}}\mathcal{R}_{ab}n^{a}n^{b}-2(\sigma^{2}-\sigma^{2}_{B})+2(\omega^{2}-\omega^{2}_{B})+\dot{u}^{a}\dot{u}_{a}\,,
    \label{Raych-eq2}
\end{equation}
where $R_{ab}u^{a}u^{b}=(\rho+3P+B^{2})/2>0$ represents the total (gravitational) energy density of the system. Note that if $\dot\Theta+\frac{1}{3}\Theta^{2}<0$, the above equation implies that an initially contracting congruence of worldlines will focus at a point ($\Theta\rightarrow -\infty$) within finite proper time. Hence, positive terms on the right-hand side of the Raychaudhuri formula act against gravitational collapse whilst negative ones in the inverse way.\\
Concerning the convergence/divergence of spacelike curves tangent to the $n^{a}$ field (i.e. the magnetic forcelines), it can be described by the following version of the Raychaudhuri equation (e.g. see~\cite{GT} or~\cite{TMav}):
\begin{equation}
    \tilde{\Theta}'=-\frac{1}{2}\tilde{\Theta}^{2}-\mathcal{R}_{ab}n^{a}n^{b}-2(\tilde{\sigma}^{2}-\tilde{\omega}^{2})+\tilde{D}^{a}n'_{a}-n'^{a}n'_{a}\,.
    \label{eqn:Raych2}
\end{equation}
In the above, $\tilde{\sigma}^{2}=\tilde{D}_{\langle b}n_{a\rangle}\tilde{D}^{\langle b}n^{a\rangle}/2$ and $\tilde{\omega}^{2}=\tilde{D}_{[ b}n_{a]}\tilde{D}^{[b}n^{a]}/2$ refer to the magnitudes of the $2$-dimensional counterparts of the shear and the vorticity tensor respectively. The prime denotes spatial differentiation along the direction $n^{a}$. It worth considering the pure magnetic field case (i.e. only magnetic field filling spacetime). Recalling that the magnetic field does not self-gravitate (i.e. $\mathcal{R}_{ab}n^{a}n^{b}=0$), eq.~\eqref{eqn:Raych2} recasts into:
\begin{equation}
    \tilde{\Theta}'=-\frac{1}{2}\tilde{\Theta}^{2}\hspace{4mm}\text{accepting the solution}\hspace{4mm} \tilde{\Theta}(l)=\frac{\tilde{\Theta}_{0}}{2+\tilde{\Theta}_{0}l}\,,
    \label{eqn:Raych2-solution}
\end{equation}
where $l$ denotes the proper length along the magnetic direction. Therefore, we consider the following cases. First, if $\tilde{\Theta}_{0}>0$, the magnetic forcelines keep diverging. Second and opposite, if $\tilde{\Theta}_{0}<0$, the magnetic forcelines converge within finite proper length (i.e. $\tilde{\Theta}_{0}\rightarrow -\infty$ as $l\rightarrow -2/\tilde{\Theta}_{0}$). Last but perhaps most interestingly, if the magnetic fieldlines happen to be at natural length (i.e. $\tilde{\Theta}_{0}=0$), will remain so indefinitely unless an external agent interferes. Note that the last case profoundly reveals, once again, the elasticity of magnetic field lines, which fully counterbalance their gravity via their tension.\\

\subsection{A non--collapse criterion}\label{ssec:1+2-split-collapse}

Having in mind the strong gravity conditions which characterise collapsing compact stellar objects (and the counterbalancing relation of the paired terms in~\eqref{Raych-eq2}), we choose to focus our attention on the purely relativistic-curvature terms\footnote{Note that $\dot{u}^{a}\dot{u}_{a}>0$ always, and therefore it resists contraction in any case.} (i.e. $c^{2}_{\mathcal{A}}\mathcal{R}_{ab}n^{a}n^{b}$ which is positive in all cases of realistic gravitational collapse and thus tends to inhibit the gravitational pull of local matter, as encoded by the expression $R_{ab}u^{a}u^{b}$). Besides, we expect that gravity ultimately dominates over kinematics at the final stages of the collapse.\\
Regarding the magneto-geometric tension stress $c^{2}_{\mathcal{A}}\mathcal{R}_{ab}n^{a}n^{b}$, it grows strong with increasing curvature distortion during the collapse, in analogy with the resisting power of a compressed elastic medium. In particular, if at some time during implosion the magneto-curvature tension overwhelms gravity, namely if the following condition holds,
\begin{equation}
    c^{2}_{\mathcal{A}}\mathcal{R}_{ab}n^{a}n^{b}>R_{ab}u^{a}u^{b}\,,
    \label{Grav-coll-cond}
\end{equation}
we expect that contraction will be halted. In addition, if $\dot{\Theta}+\Theta^{2}/3 > 0$ overall, it will turn into expansion. Making use of the Gauss-Codacci formula (e.g. see expression (1.3.39) in~\cite{TCM3}), the above condition transforms into:
\begin{equation}
    2c^{2}_{\mathcal{A}}(\rho-\frac{1}{3}\Theta^{2})+3\alpha_{a}\alpha^{a}+3c^{2}_{\mathcal{A}}(E_{ab}-\frac{1}{3}\Theta\sigma_{ab}+\sigma_{ca}\sigma^{c}{}_{b}-\omega_{ca}\omega^{c}{}_{b})n^{a}n^{b}>\frac{3}{2}(\rho+3w\rho+\mathcal{B}^{2})\,,
\end{equation}
where the first of the two parentheses in the left-hand side represents the isotropic part of the tension stress whilst the second one, the anisotropic. It turns out that the latter must be nonzero which implies that the gravitational collapse has to be anisotropic, if the tension stress is to outbalance the gravitational pull of matter.\\
Subsequently, we will simplify our non-collapse condition by using a $1+2$ split of the spatial components and calculating the double projection of the various quantities along $n^{a}$: $\mathcal{E}\equiv E_{ab}n^{a}n^{b}$, $\Sigma=\lambda\Theta$, $\Sigma_{a}=\epsilon_{ab}\Omega^{b}+\alpha_{a}$, $\sigma_{ca}\sigma^{c}{}_{b}n^{a}n^{b}=\Sigma^{2}+\Sigma^{a}\Sigma_{a}=\lambda^{2}\Theta^{2}+\Omega^{a}\Omega_{a}+\alpha_{a}\alpha^{a}$ and $\omega_{ca}\omega^{c}{}_{b}n^{a}n^{b}=\Omega^{a}\Omega_{a}$. Overall, spatial distortions along the magnetic direction are encoded by the associated projection of the Gauss-Codacci formula (recall (1.3.39) in~\cite{TCM}):
\begin{equation}
    \mathcal{R}_{ab}n^{a}n^{b}=\frac{2}{3}\rho+\mathcal{E}+\left(\lambda^{2}-\frac{\lambda}{3}-\frac{2}{9}\right)\Theta^{2}+\alpha_{a}\alpha^{a}\,.
    \label{eqn:3D-curv-along-n}
\end{equation}
Recall that the anisotropic matter pressure  (i.e. $\Pi/2\equiv(\pi_{ab}n^{a}n^{b})/2$) has been set equal to zero on assuming a perfect fluid model. It is worth noting that the effects of rotation, as expressed by the vorticity vector, and included in the term $\mathcal{R}_{ab}n^{a}n^{b}$, exactly cancel out. Also, the last term in the right-hand side of the above is always positive (i.e. $\alpha_{a}\alpha^{a}>0$), acting thus against contraction in all cases. Taking then into account that:
\begin{equation}
\frac{1}{2c^{2}_{\mathcal{A}}}\left(\rho+3P+\mathcal{B}^{2}\right)=\frac{1}{2}\left(1+4w+3w^{2}\right)\left(\frac{\rho}{\mathcal{B}}\right)^{2}+(2w+1)\rho+\frac{1}{2}\mathcal{B}^{2}\,,
\label{eqn:auxil-rel-non-collapse}
\end{equation}
and substituting~\eqref{eqn:3D-curv-along-n} into the non-collapse criterion~\eqref{Grav-coll-cond}, the latter becomes:
\begin{equation}
   \mathcal{E}+\left(\lambda^{2}-\frac{\lambda}{3}-\frac{2}{9}\right)\Theta^{2}+\alpha_{a}\alpha^{a} > \frac{1}{2}\left(1+4w+3w^{2}\right)\left(\frac{\rho}{\mathcal{B}}\right)^{2}+\frac{1}{3}\left(1+6w\right)\rho+\frac{1}{2}\mathcal{B}^{2}\,.
   \label{eqn:non-collapse-crit2}
\end{equation}
 Subsequently, we consider, as application, some specific values of $\lambda$ parameter within the interval $(-1/3\,,~2/3)$ (determining the rate of change for $\mathcal{B}$-recall the introduction to Part~II). Given the rate of change of matter density (i.e. $\rho\propto a^{-3(1+w)}$), and allowing sufficient time for the collapse to evolve, we write down the dominant terms composing the non-collapse criterion~\eqref{eqn:non-collapse-crit2}:
 \begin{eqnarray}
 &&\mathcal{E}+\alpha_{a}\alpha^{a} > \frac{1}{2}\mathcal{B}^{2}+\frac{26}{225}\Theta^{2}\hspace{8mm}\text{for}\hspace{10mm} \lambda=-\frac{1}{5}:\hspace{2mm} \mathcal{B}\propto a^{-2.6}\,;\nonumber\\ [10pt]
&&\mathcal{E}+\alpha_{a}\alpha^{a} > \frac{1}{2}\mathcal{B}^{2}+\frac{5}{36}\Theta^{2} \hspace{10mm}\text{for}\hspace{10mm} \lambda=-\frac{1}{6}:\hspace{2mm} \mathcal{B}\propto a^{-2.5}\,;\nonumber\\ [10pt]
&&\mathcal{E}+\alpha_{a}\alpha^{a} > \frac{1}{2}\mathcal{B}^{2}+0.21\Theta^{2}\hspace{8mm}\text{for}\hspace{10mm} \lambda=-\frac{1}{30}:\hspace{2mm} \mathcal{B}\propto a^{-2.1}\,;\nonumber\\ [10pt]
&&\mathcal{E}+\alpha_{a}\alpha^{a} > \frac{1}{2}\mathcal{B}^{2}+\frac{2}{9}\Theta^{2}\hspace{12mm}\text{for}\hspace{10mm} \lambda=0:\hspace{2mm} \mathcal{B}\propto a^{-2}\hspace{2mm}\text{and}\hspace{3mm} w < \frac{1}{3}\,;\nonumber\\ [10pt]
&&\mathcal{E}+\alpha_{a}\alpha^{a} > f(w)\left(\frac{\rho}{\mathcal{B}}\right)^{2}+\frac{2}{9}\Theta^{2}\hspace{10mm}\text{for}\hspace{10mm} \lambda=0:\hspace{2mm} \mathcal{B}\propto a^{-2}\hspace{2mm}\text{and}\hspace{3mm} w > \frac{1}{3}\,;\nonumber\\ [10pt]
&&\mathcal{E}+\alpha_{a}\alpha^{a} > f(w)\left(\frac{\rho}{\mathcal{B}}\right)^{2}+\frac{1}{9}\Theta^{2}\hspace{10mm}\text{for}\hspace{10mm} \lambda=\frac{1}{6}:\hspace{2mm} \mathcal{B}\propto a^{-1.5}\,;\nonumber\\ [10pt]
&&\mathcal{E}+\alpha_{a}\alpha^{a} > f(w)\left(\frac{\rho}{\mathcal{B}}\right)^{2}+\frac{2}{9}\Theta^{2}\hspace{10mm}\text{for}\hspace{10mm} \lambda=\frac{1}{3}:\hspace{2mm} \mathcal{B}\propto a^{-1}\,;\nonumber\\ [10pt]
&&\mathcal{E}+\alpha_{a}\alpha^{a} > f(w)\left(\frac{\rho}{\mathcal{B}}\right)^{2}+\frac{5}{36}\Theta^{2}\hspace{8mm}\text{for}\hspace{9mm} \lambda=\frac{1}{2}:\hspace{2mm} \mathcal{B}\propto a^{-0.5}\nonumber\,,
\label{eqn:non-collapse-conditions}
\end{eqnarray}
where $f(w)\equiv\left(1+4w+3w^{2}\right)/2$. The crucial quantities involved in magnetised contraction are, according to our relativistic criterion: the tidal stress tensor along the magnetic direction, i.e. $\mathcal{E}$, the magnetic energy density, the ratio of (squared) matter to magnetic energy density and the squared volume scalar. At this point let us recall that $\mathcal{E}$ comes from $\mathcal{R}_{ab}n^{a}n^{b}$, and therefore describes tidal distortions of the magnetic forcelines. Similarly, $c^{2}_{\mathcal{A}}\mathcal{E}$, originating from $c^{2}_{\mathcal{A}}\mathcal{R}_{ab}n^{a}n^{b}$, represents magneto-tidal tension stresses, triggered by the increasing tidal deformation of the magnetic fieldlines during contraction, and acting against the aforementioned magnetic bending. On the other hand, the various density terms, appearing in~\eqref{eqn:non-collapse-conditions}, contribute to the total gravitational energy (i.e. $R_{ab}u^{a}u^{b})$, reinforcing thus the implosion process. It is worth noting that the case $\lambda=-1/3$, for which the criterion becomes particularly simple, refers to two dimensional contraction (see appendix~\ref{appB-after} for further details).\\
To illustrate further the meaning of $\mathcal{E}$, let us recall that in terms of Newtonian gravity, $E_{ab}$ is associated with the second-order derivative of the gravitational potential $\Phi$ (precisely the Newtonian tidal tensor) or equivalently with the first-order derivative of the tidal forces $F^{a}$, in accordance with (e.g. see~\cite{EMM3}):\footnote{In the context of Newtonian theory, studying tidal forces presupposes the consideration of at least two distinctive massive bodies. However, from a relativistic point of view, we can envisage tidal forces as a result of the different curvature effects (caused by the fluid's spacetime energy distribution) experienced by distinctive particles of the magnetised fluid.}
\begin{equation}
    E^{\text{(Newt)}}_{ab}=\partial_{a}\partial_{b}\Phi-\frac{1}{3}(\partial^{c}\partial_{c}\Phi) h_{ab}\hspace{15mm} \text{and} \hspace{15mm} \mathcal{E}^{\text{(Newt)}}=\mathcal{F}'-F^{a}n'_{a}\,,
    \label{Weyl-Newtonian}
\end{equation}
where the latter relation comes from the double projection of the former along $n^{a}$, and $\mathcal{F}\equiv F^{a}n_{a}$, $F^{a}n'_{a}$ correspond to tidal stresses acting along and normal to the magnetic forcelines respectively.

Predicting actually the fate of nearly homogeneous gravitational collapse of a highly conducting (magnetised) fluid remains an open question. Our results indicate that the latter question reduces to whether the combination of the tidal tensor along the magnetic field lines, and the squared volume scalar ${\Theta}^{2}$, increases faster than the magnetic or the matter to magnetic energy density ratio. The answer seems to depend on the geometric background in hand, and potentially on the problem's initial conditions. Moreover, our non-collapse criterion assumes the availability of sufficient time for the contraction to evolve. Therefore, the possibility that the magnetic fieldlines are broken meanwhile (before a potential halt of the process), is bypassed. We address in detail the aforementioned crucial issue within the following sections.\\

\section{The law of magnetic elasticity under (volume) gravitational distortions}
\label{sec:Magnetic-elasticity-law}

With the present section we move on to the essential part of our work. In detail, we discuss the law of magnetic elasticity, describe an enlightening analogy to magnetised collapse, calculate the magnetic fracture limit and apply it to the problem of gravitational collapse of compact stellar objects. The present section should be studied along with appendix~\ref{appB-after}.\\\\
In reference to equation~\eqref{Raych-eq2} we particularly observe that the magneto-curvature tension stress:
\begin{equation}
    s^{*}=-c^{2}_{\mathcal{A}}u\,,
    \label{el-law2}
\end{equation}
counteracts the magnetised fluid's gravity, $R_{ab}u^{a}u^{b}=(1/2)(\rho+3P+\mathcal{B}^{2})>0$, namely the cause of magnetic (volume) distortion in the form of $u\equiv-\mathcal{R}_{ab}n^{a}n^{b}$ (so that $u<0$ is associated with closed spatial sections and compression). We plausibly require that $\mathcal{R}_{ab}n^{a}n^{b}>0$ ($u<0$) at all times during gravitational contraction. In complete analogy with~\eqref{Raych-eq2}, the symmetric-trace-free and the antisymmetric counterparts of~\eqref{el-law2} can be obtained via the shear and the vorticity propagation formulae respectively, along with Euler's equation of motion. However, here we examine the magnetic elasticity against gravitationally induced volume distortions (i.e. gravitational contraction).\\

\subsection{Insight into the law of magnetic elasticity}

The meaning of~\eqref{el-law2}\footnote{The expression in question has appeared several times in past works (e.g. see~\cite{T23}-\cite{MT}) but it was not recognised or envisaged as an expression of Hooke's law of elasticity and therefore was not given its full interpretation presented here.} is that the tension stress $s^{*}$, tending to restore the magnetic field into its initial (undeformed) state, is proportional to the (gravitationally induced) volume distortion of the magnetic forcelines (see also the appendices~\ref{appB-after} and~\ref{app:magneto-curvature-tension})\footnote{Written here for an ideal (magnetised) fluid.}. According to eq.~\eqref{eqn:3D-curv-along-n} (on assuming that the magnetic direction remains constant during contraction, i.e. $\alpha_{a}=0$), the latter reads:
\begin{equation}
u\equiv-\mathcal{R}_{ab}n^{a}n^{b}=-\frac{2}{3}\rho-\mathcal{E}+g(\lambda)\Theta^{2}\,,
\end{equation}
where $g(\lambda)\equiv -(\lambda^{2}-\lambda/3-2/9)=-(\lambda-2/3)(\lambda+1/3)$.
 The minus sign in the right-hand side of~\eqref{el-law2} implies that $s^{*}$ acts against the increasing magnetic distortion $u$. Note that in deriving the above, we have taken into account eq. (1.3.41) from~\cite{TCM3}, as well as relations~\eqref{eqn:appB1}, \eqref{eqn:Sigma-vector} from the appendix~\ref{appB-after}. It is straightforward to see that condition $\mathcal{E}=-(2/3)\rho+g(\lambda)\Theta^{2}$ corresponds to the natural (undeformed) `volume' state of the magnetic field, where $s^{*}=0$. The proportionality factor $0<c^{2}_{\mathcal{A}}<1$ (note the difference to~\eqref{Magn-curv-tensor}) is always positive and its definition implies that the greater the magnetic density contribution to the total fluid's density, the more rigid the magnetic fieldlines are (or the more they resist to their deformation). In other words, eq~\eqref{el-law2} is a relativistic expression of Hooke's law of elasticity for a gravitationally distorted magnetic field, frozen into a highly conducting fluid. Nevertheless, in contrast to an elastic spring, the proportionality  factor $c^{2}_{\mathcal{A}}$ is not a constant but a variable quantity (a function of the ratio $\rho/\mathcal{B}^{2}$). Moreover, although Hooke's law is an approximate relation valid for sufficiently small deformations, eq~\eqref{el-law2} seems to be valid for any deformation, given that the Ricci identities~\eqref{3-Ricci} hold. Therefore, from our point of view, magnetic fields appear to keep their elastic behaviour as well as to satisfy Hooke's law of elasticity no matter how big their deformation is.

Even if magnetic forcelines do not present an elastic limit\footnote{The elastic limit refers to that value of distortion beyond which the elastic medium is unable to return to its initial state. Mathematically speaking, on setting the external forces equal to zero, the deformation becomes zero as well. Of course we do not know any such example of material in nature.} under their gravitational bending, one expects that they can support a finite amount of distortion. Thus, we expect that there must be at least a fracture limit of the magnetic fieldlines, predicted by the exact formula~\eqref{el-law2}\footnote{In contrast, Hooke's law for elasticity, being a linear approximation-valid for small values of deformation-, does not and could not contain such information.}. Beyond that limit, the magnetic field should stop being frozen into the fluid, in the sense of magnetohydrodynamics (recall Alfv\'{e}n's theorem). The significance of such a limit becomes clear on considering for instance the astrophysical/cosmological phenomenon of magnetised gravitational collapse. In particular, magnetic fields are known not to self-gravitate as well as to have the potential to impede gravitational implosion from reaching a spacetime singularity. Before proceeding to a definition and theoretical calculation of the magnetic fracture limit under gravitational distortions, we present our approach to magnetised contraction through an analogy.\\
We study the gravitational collapse of a magnetised fluid. In other words, we study the contraction of a medium which behaves elastically along a specific direction $n^{a}\parallel B^{a}$. The above statement points out the crucial difference between `ordinary' and magnetised gravitational collapse. We liken the latter to the contraction of a waterballoon filled with water and surrounded by water. Through this perspective, water stands for the astrophysical/cosmic fluid whilst the elastic medium (balloon containing the water) stands for the magnetic field. Both the balloon and the magnetic field, force, in a sense, the water/fluid to behave elastically. In other words, the water/fluid resists its contraction along the waterballoon/magnetic direction via stresses increasing linearly in proportion to its deformation (Hooke's law for elasticity). Finally, we expect that magnetic forcelines, like the waterballoon, can support a finite amount of contraction, which we estimate in the following subsections.\\

\subsection{Magnetic fracture limit and gravitational contraction}\label{ssec:fr-limit-contraction}

In the first place, we claim that the fracture limit must correspond to a maximum of the magnetic deformation with respect to proper time. However, given that as the deformation of an elastic medium increases, so do the internal tension stresses acting against it; it often happens that the maximum deformation coincides with the maximum resisting tension stress (see the following subsection~\ref{ssec:further-frac-limit}). In reference to our case, because the law of magnetic elasticity, eq.~\eqref{el-law2}, is valid at all times during the collapse (for any deformation), it is necessary that the fracture limit corresponds to a double maximum, of the magnetic deformation and the magneto-curvature tension as well (i.e. $\dot{u}_{\text{fr}}=0=\dot{s}^{*}_{\text{fr}}$)\footnote{The monotonic increase of $u$ and $s^{*}$ during contraction is given by the problem's nature.}. Besides, a maximum of spatial deformation along a direction $n^{a}$, i.e. $u=-\mathcal{R}_{ab}n^{a}n^{b}$, does not make physical sense by itself within the problem of magnetised gravitational collapse. The quantity $u$ should increase monotonically without ever reaching a maximum value. It is only through the magnetic field presence (with its elastic behaviour encoded by~\eqref{el-law2}), appearing in $s^{*}$ but not directly in $u$, that a maximum of the latter acquires physical beingness together with a parallel maximum of the former. Beyond that maximum value, the fieldlines of the magnetised fluid are expected to be broken, namely to lose their coherence and stop connecting the same fluid particles. Within the present subsection, we comment on our gravitational collapse criterion. In the following subsection, we proceed to the determination of the magnetic fracture limit.\\
Now in reference to the problem of magnetised gravitational collapse, we face the following question. \textit{Do magnetic fieldlines affect the fate of gravitational collapse? In particular, will they manage to impede contraction towards a singularity or will they be inevitably broken beforehand?} Let us envisage magnetic elasticity under increasing gravity through the following causal perspective. The magneto-curvature tension $s^{*}=c^{2}_{\mathcal{A}}\mathcal{R}_{ab}n^{a}n^{b}$ is considered an exclusive result of the system's gravity $R_{ab}u^{a}u^{b}$ (recall eq~\eqref{eqn:Raych2})\footnote{Besides, during the principally relativistic phenomenon of collapse, gravity is plausibly pointed out as the dominant term, cause of magnetic deformation.}. Hence, our argument can be stated this way: If $s^{*}$ turns out to be smaller than the fluid's gravity at the fracture limit, then it should have also been smaller earlier during the collapse. In such a case, magnetic fieldlines are not able to impede contraction towards a singularity (refer also to~\cite{MT}). Oppositely, if $s^{*}$ overwhelms the fluid's gravity at the fracture limit, then magnetic forcelines halt the collapse. In other words, our collapse criterion reads:
\begin{equation}
        s^{*}=-(c^{2}_{\mathcal{A}}u)_{\text{fr}} < (R_{ab}u^{a}u^{b})_{\text{fr}}\,.
    \label{eq:grav-coll-cond}
\end{equation}
Note that for (magnetised) astrophysical objects in equilibrium, the total gravitational energy density is generally by many orders of magnitude greater than the magneto-curvature tension. This basically happens because the ratio of the magnetic over the total energy density is for most (if not all) applications very small, so that the Alfv\'{e}n speed is many orders of magnitude smaller than unity.\\
Whether the criterion in question is satisfied or not, seems to depend on the precise value of $\lambda$ parameter. In other words, the question whether the magnetic fieldlines manage to prevent contraction or not, reduces to how fast the magnetic density increases.\\

\subsection{Specification of the magnetic fracture limit--Applications to neutron stars and white dwarfs}\label{ssec:further-frac-limit}

In the present subsection we determine the magnetic fracture limit as a double maximum, of the magnetic deformation $u$ and the magneto-curvature tension stress $s^{*}$ as well. In practice, taking the dot derivative of~\eqref{el-law2} under the condition $\dot{u}_{\text{fr}}=0$, we get:
\begin{equation}
\dot{\left(s^{*}\right)}_{\text{fr}}=\dot{\left(c^{2}_{\mathcal{A}}\right)}_{\text{fr}}u_{\text{fr}}=\frac{1}{\mathcal{B}^{2}(1+\beta)^{2}}\left[\dot{\rho}+\dot{P}-\frac{2}{\mathcal{B}}\dot{\mathcal{B}}(\rho+P)\right]_{\text{fr}}u_{\text{fr}}\,,
    \label{eq:dot-c}
\end{equation}
where $\beta\equiv(\rho+P)/\mathcal{B}^{2}$ (see appendix~\ref{app:auxiliary-calculations} for details). Assuming a polytropic equation of state (i.e. $P=k\rho^{\gamma}$, where $k$ and $\gamma$ are constant parameters), and setting~\eqref{eq:dot-c} equal to zero, we arrive at the condition:
\begin{equation}
    \left(\frac{\gamma}{\rho}\right) P^{2}+\left(2\lambda+\gamma-\frac{1}{3}\right)P+\left(2\lambda-\frac{1}{3}\right)\rho=0\,.
    \label{eq:P-condition}
\end{equation}
Note that in deriving the above, we have taken into account the propagation equation for the magnetic field at the magnetohydrodynamic limit, namely $\dot{\mathcal{B}}=(\lambda-2/3)\Theta\mathcal{B}$~(see the introduction to Part~II). Moreover, it is simply verified that expression~\eqref{eq:dot-c} is positive for $P< (1/3-2\lambda)\rho/\gamma$, and negative in the opposite case. The real solutions of the above quadratic are:\footnote{It is easy to check that $c^{2}_{\mathcal{A}}$ is an increasing function of proper time ($\Theta<0$ is always assumed) for $P<\rho/\gamma$ whilst decreasing for $P>\rho/\gamma$.}
\begin{equation}
    P_{\text{fr}}=k\rho^{\gamma}_{\text{fr}}=\left(\frac{\frac{1}{3}-2\lambda}{\gamma}\right)\rho_{\text{fr}}\Leftrightarrow\rho_{\text{fr}}=\left(\frac{\frac{1}{3}-2\lambda}{k\gamma}\right)^{\frac{1}{\gamma-1}}\hspace{15mm} \text{and}\hspace{15mm} P_{\text{fr}}=-\rho_{\text{fr}}\,,
    \label{eq:quadr-sols}
\end{equation}
of which we accept the former and reject the latter, under the consideration of ordinary collapsing matter. Note that eq~(\ref{eq:quadr-sols}a) provides the values (in geometrised units) of matter density and pressure at the magnetic fracture limit. Overall, we have associated the magnetic fracture limit with the unique (double) maximum of $u$ and $s^{*}$ with respect to proper time. The aforementioned point (a theoretical approach to the magnetic fracture limit) is certainly expected to slightly differ, in practice, from the actual instant in which the magnetic fieldlines are broken (i.e. $s^{*}=0$).\\
We can make use of the information provided by~(\ref{eq:quadr-sols}a), together with an initial setting, in order to predict how much has the volume of a given magnetised (collapsing) star changed until reaching its magnetic fracture limit. In particular, with the aid of~\eqref{density-evol-polytrope} (see appendix~\ref{app:polytropic-eq-of-state}), we deduce that the relation between matter density/scale factor at the fracture limit (i.e. $\rho_{\text{fr}}$ and $a_{\text{fr}}$) and their counterparts at an initial (hydrostatic equilibrium) state of the star (i.e. $\rho_{0}$ and $a_{0}$), is
\begin{equation}
    \rho_{\text{fr}}=\rho_{0}\left(\frac{\frac{C}{k}a^{-3(1-\gamma)}_{0}-1}{\frac{C}{k}a^{-3(1-\gamma)}_{\text{fr}}-1}\right)^{\frac{1}{\gamma-1}}\hspace{10mm} \text{or} \hspace{10mm} a^{-3(1-\gamma)}_{\text{fr}}=\frac{k}{C}\left[\left(\frac{\rho_{0}}{\rho_{\text{fr}}}\right)^{\gamma-1}\left(\frac{C}{k}a^{-3(1-\gamma)}_{0}-1\right)+1\right]\,.
    \label{eq:fr-0}
\end{equation}
The above formula points out the sensible conclusion that the more dense a star is, the earlier it reaches its fracture limit. In the following we consider, as an application, two illustrative examples concerning a neutron star and a white dwarf.\\

\subsubsection{Neutron star of mass $M=1.5M_{\odot}$, radius $R=10$~km and average magnetic field $\mathcal{B}\sim 10^{12}$~G}\label{sssec:neutron-star}

A neutron star with the above mentioned characteristics has matter and magnetic densities $\rho\sim 10^{-14}~\text{cm}^{-2}$ and $\mathcal{B}^{2}\sim 10^{-27}~\text{cm}^{-2}$, in geometrised units (i.e. $c=1=G$). For conversion between cgs and geometrised units see e.g. the appendix of~\cite{JBH}). Assuming that the matter of the star consists in average of non-relativistic neutrons, the parameters of its polytropic equation of state are $\gamma=5/3$ and $k=5.3802\cdot 10^{9}~$cgs$~\sim 10^{6.7}~\text{cm}^{4/3}$ (e.g. refer to~\cite{ST} or~\cite{JSB} for details regarding the parameters' values). Finally, from~\eqref{density-evol-polytrope} written in the initial conditions, we determine the value of the integration constant, $C=(\rho^{2/3}_{0}+k)/a^{2}_{0}\sim 10^{9.3}a^{-2}_{0}$. Taking into account that $\mathcal{B}^{2}\propto a^{6\lambda-4}$ (refer to the introduction to Part~II), and substituting all the aforementioned values in~(\ref{eq:fr-0}b), we find out that\\\\
\begin{tabular}{ |p{1.5cm}||p{2cm}|p{2cm}|p{2cm}||p{2cm}|p{2cm}||p{2cm}| }
 \hline
 \multicolumn{7}{|c|}{} \\
 \hline
 \hspace{4mm}$\lambda$ & \hspace{4mm}$a_{\text{fr}}/a_{0}$ &$\hspace{4mm}\rho_{\text{fr}}/\rho_{0}$&$\rho_{\text{fr}}~(\text{cm}^{-2})$&$\hspace{4mm}\mathcal{B}^{2}_{\text{fr}}/\mathcal{B}^{2}_{0}$&$\mathcal{B}^{2}_{\text{fr}}~(\text{cm}^{-2})$&$\hspace{4mm}(\mathcal{B}^{2}/\rho)_{\text{fr}}$\\
 \hline
 $-1/30$  &  $1.0\cdot 10^{-1.3}$&$2.4\cdot 10^{6.3}$&$2.4\cdot 10^{-7.7}$&$2.9\cdot 10^{5}$&$2.9\cdot 10^{-22}$&$6.1\cdot 10^{-15}$  \\
 $-1/6$  &  $1.8\cdot 10^{-1.3}$&$2.3\cdot 10^{3}$&$2.3\cdot 10^{-11}$&$1.7\cdot 10^{5}$&$1.7\cdot 10^{-22}$&$7.4\cdot 10^{-11}$\\
 \hspace{4mm}$0$  &  $1.2\cdot 10^{-1}$&$7.7\cdot 10^{2}$&$8.0\cdot 10^{-12}$&$4.8\cdot 10^{3}$&$4.8\cdot 10^{-24}$&$6.0\cdot 10^{-13}$\\
 $+1/10$  &  $3.6\cdot 10^{-1.3}$&$2.0\cdot 10^{2}$&$2.0\cdot 10^{-12}$&$3.4\cdot 10^{2}$&$3.4\cdot 10^{-25}$&$1.7\cdot 10^{-13}$\\
 \hline
\end{tabular}\\\\
Therefore, we expect that the magnetic forcelines will be broken when the neutron star's radius becomes a hundred to ten times smaller (that is $1$~km) than its initial (equilibrium) value. In parallel, the magnetic/matter density ratio will have grown by zero to five orders of magnitude (recall that $(\mathcal{B}^{2}/\rho)_{0}\sim 10^{-13}$).

\subsubsection{White dwarf of mass $M=0.6M_{\odot}$, radius $R=1.4\cdot 10^{-2}$~$R_{\odot}$ and average magnetic field $\mathcal{B}\sim 10^{6}$~G}\label{sssec:white-dwarf}

The white dwarf in question has matter and magnetic densities $\rho\sim 10^{-23}~\text{cm}^{-2}$ and $\mathcal{B}^{2}\sim 10^{-39}~\text{cm}^{-2}$, in geometrised units. Assuming that the stellar fluid mainly consists of ultra relativistic electrons, the parameters of its polytropic equation of state read $\gamma=4/3$ and $k\sim 10^{15}$~cgs$\sim 10^{4.3}~\text{cm}^{2/3}$. Under the above mentioned initial conditions, the integration constant of~\eqref{density-evol-polytrope} gives $C\sim 10^{7.6}~a^{-1}_{0}$. Overall, (recalling once again that $\mathcal{B}^{2}\propto a^{6\lambda-4}$) we find out that:
\\\\
\begin{tabular}{ |p{1.5cm}||p{2cm}|p{2cm}|p{2cm}||p{2cm}|p{2cm}||p{2cm}| }
 \hline
 \multicolumn{7}{|c|}{} \\
 \hline
 \hspace{4mm}$\lambda$ & \hspace{4mm}$a_{\text{fr}}/a_{0}$ &$\hspace{4mm}\rho_{\text{fr}}/\rho_{0}$&$\rho_{\text{fr}}~(\text{cm}^{-2})$&$\hspace{4mm}\mathcal{B}^{2}_{\text{fr}}/\mathcal{B}^{2}_{0}$&$\mathcal{B}^{2}_{\text{fr}}~(\text{cm}^{-2})$&$\hspace{4mm}(\mathcal{B}^{2}/\rho)_{\text{fr}}$\\
 \hline
 $-1/30$  &  $1.9\cdot 10^{-3}$&$3.4\cdot 10^{8}$&$3.4\cdot 10^{-15}$&$2.7\cdot 10^{11}$&$2.7\cdot 10^{-28}$&$7.9\cdot 10^{-14}$  \\
 $-1/6$  &  $1.4\cdot 10^{-3}$&$1.6\cdot 10^{9}$&$1.6\cdot 10^{-14}$&$1.9\cdot 10^{14}$&$1.9\cdot 10^{-25}$&$1.2\cdot 10^{-11}$\\
\hspace{4mm} $0$  &  $2.2\cdot 10^{-3}$&$2.0\cdot 10^{8}$&$2.0\cdot 10^{-15}$&$4.2\cdot 10^{10}$&$4.2\cdot 10^{-29}$&$2.1\cdot 10^{-14}$\\
 $+1/10$  &  $4.8\cdot 10^{-3}$&$1.3\cdot 10^{7}$&$1.3\cdot 10^{-16}$&$7.7\cdot 10^{7}$&$7.7\cdot 10^{-32}$&$5.9\cdot 10^{-16}$\\
 \hline
\end{tabular}\\\\
We deduce that the stellar fluid reaches its magnetic fracture limit when its radius shrinks to approximately one hundred thousand times its initial value (that is the fracture radius is some hundreds of meters). As for the magnetic/matter density ratio, it increases by about an order of magnitude, i.e. $(\mathcal{B}^{2}/\rho)_{\text{fr}}\sim 10^{0.8}~(\mathcal{B}^{2}/\rho)_{0}\sim 10^{-15}$.\\

\section{Discussion-Concluding remarks}

On projecting Faraday's equation at the MHD limit along the magnetic direction $n^{a}$, and using a $1+2$ (covariant) spatial decomposition, we have shown that it recasts into a solvable form. Its general solution for the magnetic magnitude, i.e. $\mathcal{B}\propto a^{3\lambda-2}$, depends on a real parameter $\lambda$, coming from the introduced physically senseful relation $\Sigma\equiv\sigma_{ab}n^{a}n^{b}=\lambda\Theta$, between the shear and the volume scalar. For a magnetic field decreasing/increasing during expansion/contraction, the aforementioned real parameter obeys the constraint $\lambda<2/3$. For orthogonal (magnetised) Bianchi cosmologies, the constraint in question strictens to $-2/3\leq\lambda<2/3$.

Applying our magnetic evolution formula into the study of contracting worldlines in nearly homogeneous magnetised gravitational collapse, we suggest a non-collapse criterion. According to that, the contraction's fate at advanced stages is essentially determined by: the tidal scalar $\mathcal{E}$ (the tidal tensor projected twice along the magnetic direction), the magnetic density $\mathcal{B}^{2}$, the ratio $(\rho/\mathcal{B})^{2}$ ($\rho$~: density of matter) and the squared volume scalar $\Theta^{2}$. The actual form of the criterion depends of course on the particular magnetic evolution rate (i.e. on $\lambda$ parameter). Moreover, the precise evolution formulae for $\mathcal{E}$ and $\Theta^{2}$ depend on the geometric background in hand.

After distinguishing and describing the kinematically and gravitationally induced magnetic tension stresses, we focus our attention to a specific magneto-curvature tension stress component, the one triggered by volume distortions due to gravity. We point out the aforementioned tension as the dominant stress counteracting (magnetised) gravitational collapse. The essence of our reasoning lies in that gravitational deformation of magnetic forcelines is governed by a particular form of Hooke's law of elasticity, originating from the magneto-curvature coupling monitored by the Ricci identities. However, there are two basic features distinguishing gravitational (relativistic) distortions of magnetic forcelines from mechanical distortions of elastic materials. Firstly, unlike mechanical distortions of elastic materials, Hooke's law in the form of~\eqref{el-law2} is not an approximate expression only valid for small magnetic deformations (thus magnetic forcelines do not seem to have an elastic limit). In contrast, as long as Ricci identities are an appropriate definition of spatial curvature for large values of the latter (advanced stages of gravitational collapse), the law in question consists of an exact expression, valid for any size of distortion. Secondly, the proportionality factor in the elasticity law~\eqref{el-law2} is a variable instead of a constant quantity.

Based on the aforementioned expression for gravito-magnetic elasticity, we have suggested a calculation of the magnetic fracture limit under gravitational volume distortions. Two explicit cases have been considered as application, of a neutron star and a white dwarf. The results have been summarised in two tables for some representative values of the $\lambda$ parameter. Addressing also the question whether magnetic forcelines manage to impede contraction before being broken, our answer depends on the magnetic density's rate of change (i.e. on the precise value of $\lambda$ parameter). Considering our results as new, we raise the fundamental problem regarding the observational-experimental (and further theoretical) verification of the gravito-magnetic elasticity law. Although magnetic elasticity under great gravitational distortions is practically not a subject offered for study in earthly laboratories, progress towards the experimental path could alternatively and in the first place be achieved by examining magnetic distortions under progressively increasing rotations.

Overall, our general law of variation for the magnetic field of an expanding or contracting highly conducting fluid, is expected to provide magnetohydrodynamics with a valuable theoretical tool. Interestingly, knowing how the magnetic fieldlines explicitly behave during gravitational implosion, could hopefully shed light on the various evolution phases of astrophysical objects. The motivation for the above sentence essentially comes from considering that many stars or protogalactic clouds are associated with (even small) magnetic fields which are rapidly increasing during their collapse. Concerning the potential applications in cosmology, our exact (not approximate) evolution formula for the magnetic field could fortunately refresh the question regarding the contribution of magnetic fields to the kinematics of the early universe.\\

\begin{subappendices}

\section{Propagation equations and constraints for the kinematically induced magnetic tension stresses}\label{appA-prop-eqns-magn-tension-stresses}

In order to arrive at the propagation equation for $\sigma^{(B)}_{ab}$, we make the following steps. First, project eq.~\eqref{4-Ricci} along the timelike 4-velocity $u^{a}$; second, project orthogonal to $u^{a}$ with the aid of $h_{ab}$ and with respect to both indices (removing thus timelike terms); third, take the symmetric and trace-free part of the resulting relation. The equation in question finally reads:\footnote{An index with bar denotes that the associated component has been projected orthogonal to $u^{a}$.}
\begin{eqnarray}
    \dot{\sigma}^{(B)}_{\bar{a}\bar{b}}&=&-\Theta\dot{u}_{\langle a}B_{b\rangle}+2\dot{u}_{\langle a}\omega_{b\rangle}{}^{c}B_{c}+{\rm D}_{\langle a}\dot{B}_{\bar{b}\rangle}-\left(\sigma_{c\langle a}+\omega_{c\langle a}+\frac{1}{3}\Theta h_{c\langle a}\right)\sigma^{c(B)}{}_{b\rangle}\nonumber\\
    &&-\left(\sigma_{c\langle a}+\omega_{c\langle a}\right)\omega^{c(B)}{}_{b\rangle}-\frac{1}{2}B_{\langle a}q_{b\rangle}\,,
    \label{prop-B-shear}
\end{eqnarray}
where $\dot{u}_{a}=u^{b}\nabla_{b}u_{a}$ is the fluid's acceleration and $q_{a}$ its flux vector. On deriving the above we have taken into account eq. 1.3.1 of~\cite{TCM}, as well as that
\begin{equation}
    \nabla_{a}B_{b}={\rm D}_{a}B_{b}-u_{a}\dot{B}_{b}+u_{b}(\nabla_{a}u^{d})B_{d}+u_{a}u_{b}\dot{u}^{c}B_{c} \hspace{5mm} \text{and} \hspace{5mm} R_{a\langle bc\rangle d}u^{a}B^{d}=\frac{1}{2}B_{\langle b}q_{c\rangle}\,,
\end{equation}
where eqs 1.2.6, 1.2.8 and 1.2.11 of~\cite{TCM} have been used on finding the latter of the above. Following a similar procedure but taking the antisymmetric part of~\eqref{4-Ricci} (via contraction with the 3-D Levi-Civita pseudotensor $\epsilon_{abc}$) this time, we arrive at the propagation equation for the magnetic tension induced by twisting effects:
\begin{eqnarray}
    \dot{\omega}^{(B)}_{\bar{a}}&=&-3\epsilon_{abc}\dot{u}^{b}\sigma^{c}{}_{d}B^{d}-\epsilon_{abc}{\rm D}^{b}\dot{B}^{\bar{c}}-\epsilon_{abc}\left(\sigma^{bd}+\omega^{db}+\frac{1}{3}\Theta h^{bd}\right){\rm D}_{d}B^{c}\nonumber\\
    &&-(\dot{u}^{b}B_{b})\omega_{a}+(\dot{u}^{b}\omega_{b})B_{a}+H_{ab}B^{b}-\frac{1}{2}\epsilon_{abc}B^{b}q^{c}\,,
    \label{prop-B-vorticity}
\end{eqnarray}
where $H_{ab}$ is the magnetic Weyl component and we have taken into account that:
\begin{equation}
    \dot{B}_{\bar{a}}=-\frac{2}{3}\Theta B_{a}+(\sigma_{ab}+\epsilon_{abc}\omega^{c})B^{b}\hspace{5mm} \text{and} \hspace{5mm} \epsilon_{abc}R^{ebcd}u_{e}B_{d}=H_{ab}B^{b}-\frac{1}{2}\epsilon_{abc}B^{b}q^{c}\,.
    \label{Farad}
\end{equation}
Note that eq~(\ref{Farad}a) is an expression of Faraday's law at the MHD limit. On the other hand, the spacelike part of~\eqref{4-Ricci} leads to the divergence conditions for the aforementioned quantities. In detail, we start from the 3-D Ricci identities~\eqref{3-Ricci}. Subsequently, we take either its trace or its contraction with $\epsilon_{abc}$. The former case leads to:
\begin{equation}
    {\rm D}^{b}\sigma^{(B)}_{ab}=\text{curl}\omega^{(B)}_{a}+2\omega_{ab}\left(-\frac{2}{3}\Theta B^{b}+\sigma^{b}{}_{c}B^{c}\right)+\mu\omega_{a}-2\omega^{2}B_{a}+\mathcal{R}_{ba}B^{b}
    \label{Div-B-shear}
\end{equation}
whilst the latter to
\begin{equation}
    {\rm D}^{a}\omega^{(B)}_{a}=\frac{1}{6}\Theta\mu-2\sigma_{ab}\omega^{a}B^{b}\,.
    \label{Div-B-vorticity}
\end{equation}
In~\eqref{Div-B-shear} $\mathcal{R}_{ab}$ represents the 3-D Ricci tensor.
On deriving the above we have made use of~(\ref{Farad}a) as well as of 
\begin{equation}
    \omega^{a}B_{a}=\mu/2 \hspace{5mm} \text{and} \hspace{5mm} \epsilon^{abc}\mathcal{R}_{dcba}B^{d}=-\frac{2}{3}\Theta\mu-4\sigma_{ab}\omega^{a}B^{b}\,,
    \label{Gauss}
\end{equation}
where $\mu$ is the charge density and~(\ref{Gauss}a) is an expression of Gauss's law at the MHD limit. It is worth noting that for zero rotational distortions (i.e. $\omega_{ab}=0$) of the magnetic field eqs~\eqref{prop-B-shear} and~\eqref{Div-B-shear} significantly simplify to\footnote{An ideal fluid (i.e. $q_{a}=0$) has been assumed in the first equation.}
\begin{equation}
    \dot{\sigma}^{(B)}_{\bar{a}\bar{b}}=-\Theta\dot{u}_{\langle a}B_{b\rangle}+{\rm D}_{\langle a}\dot{B}_{\bar{b}\rangle}-\left(\sigma_{c\langle a}+\frac{1}{3}\Theta h_{c\langle a}\right)\sigma^{c(B)}{}_{b\rangle}\hspace{5mm} \text{and} \hspace{5mm} {\rm D}^{b}\sigma^{(B)}_{ab}=\mathcal{R}_{ba}B^{b}\,.
\end{equation}
Overall, equations~\eqref{prop-B-shear},~\eqref{prop-B-vorticity},~\eqref{Div-B-shear} and~\eqref{Div-B-vorticity} determine the kinematics of the magnetic tension stresses triggered by shear and vorticity effects. Note that due to Gauss' law at the MHD limit (i.e. ${\rm D}^{a}B_{a}=0$), there are no magnetic tension stresses triggered by volume changes.\\

\section{$1+2$ Covariant approach}\label{sec:1+2-split}

In some cases, a further 1+2 decomposition of the 3-dimensional space (leading to an overall 1+1+2 spacetime splitting--see~\cite{GT},~\cite{CB} and~\cite{CMBD} for some introductory information) in one specific spatial direction and a 2-dimensional surface orthogonal to it, may reveal additional useful information about the problem in hand. This is more likely to happen when the geometry, or the physics select a preferred spatial direction. For instance, one could consider the radial component of a spherically symmetric spacetime, or the rotation axis of a magnetised star, which may also happen to be parallel to the direction of the magnetic forcelines. However, a split of the spatial components may reveal valuable information about the problem in hand even there are not any apparent, favorable geometric or physical conditions (e.g. see the decomposition of Maxwell equations in the present piece of work.).\\

\subsection{Spatial splitting}\label{ssec:1+2-backgr}

In what follows, we show how 3-D mathematical objects (vectors, tensors, equations etc.) decompose into a component parallel to a spatial direction and two components lying on a 2-D surface perpendicular to the aforementioned direction (for a detailed presentation see also~\cite{CB}). Let us introduce a space-like unit vector $n^{a}$ orthogonal to $u^{a}$ ($n^{a}n_{a}=1$, $n^{a}u_{a}=0$), which defines a specific spatial direction. Subsequently, we can define the symmetric tensor $\tilde{h}_{ab}\equiv h_{ab}-n_{a}n_{b}$ which projects vectors onto 2-D surfaces orthogonal to $n^{a}$ ($\tilde{h}_{ab}n^{b}=0$, $\tilde{h}^{a}{}_{a}=2$, $\tilde{h}_{a}{}^{c}\tilde{h}_{bc}=\tilde{h}_{ab}$). In analogy with the 1+3 formalism, 3-vectors and the corresponding second-rank, symmetric and trace-free tensors are split in their irreducible components according to the relations:
\begin{equation}
    v^{a}=Vn^{a}+V^{a},
\end{equation}
where $V\equiv v^{a}n_{a}$ and $V^{a}\equiv \tilde{h}^{a}{}_{b}v^{b}$ while
\begin{equation}
    v_{ab}=V(n_{a}n_{b}-\frac{1}{2}\tilde{h}_{ab})+2V_{(a}n_{b)}+V_{ab},
\end{equation}
where $V\equiv v_{ab}n^{a}n^{b}=-\tilde{h}^{ab}v_{ab}$, $V_{a}\equiv \tilde{h}_{a}{}^{b}n^{c}v_{bc}$ and $V_{ab}\equiv (\tilde{h}_{(a}{}^{c}\tilde{h}_{b)}{}^{d}-(1/2)\tilde{h}_{ab}\tilde{h}^{cd})v_{cd}$.
For instance, let us consider the 1+1+2 decomposition of the energy-momentum tensor $T_{ab}=g_{ac}g_{bd}T^{cd}=(\tilde{h}_{ac}-u_{a}u_{c}+n_{a}n_{c})(\tilde{h}_{bd}-u_{b}u_{d}+n_{b}n_{d})$, which leads to:
\begin{equation}
    T_{ab}=\rho u_{a}u_{b}+\tilde{\rho}n_{a}n_{b}+\tilde{P}\tilde{h}_{ab}+2u_{(a}q_{b)}+2n_{(a}\tilde{q}_{b)}+\Pi_{ab},
\end{equation}
where $\tilde{\rho}\equiv T_{ab}n^{a}n^{b}=P+\Pi$ and $\tilde{P}\equiv (\tilde{h}^{ab}/2)T_{ab}=P-\Pi/2$ (therefore $\Pi=(2/3)(\tilde{\rho}-\tilde{P}$)) are the analogues of relativistic energy density and pressure defined in reference to spacelike curves with tangent vector $n^{a}$. Regarding $\tilde{q}_{a}\equiv \tilde{h}_{a}{}^{b}n^{c}T_{bc}= \Pi_{a}$ and $\Pi_{ab}\equiv (\tilde{h}_{(a}{}^{c}\tilde{h}_{b)}{}^{d}-(1/2)\tilde{h}_{ab}\tilde{h}^{cd})T_{cd}$, they represent the (2-D) surface (normal to $n^{a}$) counterparts of the energy flux vector and the viscosity tensor respectively (refer to equation~\eqref{pi-decomp} for the decomposition of the anisotropic stress tensor).
We gather here for reference all of the decomposition relations of vectors and tensors, which we use throughout this thesis:\footnote{Note that $\dot{n}_{a}n^{a}=0$ in eq.~\eqref{dot-n-dec} and therefore $\alpha_{a}n^{a}=0$.}
\begin{equation}
    \dot{u}^{a}=\mathcal{A}n^{a}+\mathcal{A}^{a}
\end{equation}
\begin{equation}
\dot{n}^{a}=\mathcal{A}u^{a}+\alpha^{a}
\label{dot-n-dec}
\end{equation}
\begin{equation}
    \omega^{a}=\Omega n^{a}+\Omega^{a}
\end{equation}
\begin{equation}
    q^{a}=Qn^{a}+Q^{a}
\end{equation}
\begin{equation}
    E^{a}=\epsilon n^{a}+\epsilon^{a}
\end{equation}
\begin{equation}
    B^{a}=\mathcal{B}n^{a}+\mathcal{B}^{a}
\end{equation}
\begin{equation}
    \mathcal{J}^{a}=jn^{a}+j^{a}
\end{equation}
\begin{equation}
    \sigma_{ab}=\Sigma(n_{a}n_{b}-\frac{1}{2}\tilde{h}_{ab})+2\Sigma_{(a}n_{b)}+\Sigma_{ab}
    \label{sigma-1+2}
\end{equation}
\begin{equation}
    \pi_{ab}=\Pi(n_{a}n_{b}-\frac{1}{2}\tilde{h}_{ab})+2\Pi_{(a}n_{b)}+\Pi_{ab}
    \label{pi-decomp}
\end{equation}
\begin{equation}
    E_{ab}=\mathcal{E}(n_{a}n_{b}-\frac{1}{2}\tilde{h}_{ab})+2\mathcal{E}_{(a}n_{b)}+\mathcal{E}_{ab}\,.
\end{equation}
In the last equation, $E_{ab}$ is the electric component of the Weyl (long-range) curvature tensor. There is also the magnetic tensor component $H_{ab}$. Weyl curvature is associated with tidal forces and gravitational waves (e.g. refer to~\cite{TCM}). The aforementioned decomposition relation will be used only once when discussing the gravitational collapse of a magnetised fluid in section~\ref{sec:Grav-collapse}. Finally, for some details concerning the meaning of the shear's scalar and vector components see the appendix section~\ref{appB-after}. 

Regarding the derivatives of a general tensor field $T_{ab...}{}^{cd...}$, the one along $n^{a}$ and the other projected on the 2-surface normal to $n^{a}$, these are defined respectively as:
\begin{equation}
    T'_{ab...}{}^{cd...}\equiv n^{e}{\rm D}_{e}T_{ab...}{}^{cd...}
\end{equation}
and
\begin{equation}
    \tilde{{\rm D}}_{e}T_{ab...}{}^{cd...}\equiv \tilde{h}_{e}{}^{s}\tilde{h}_{a}{}^{f}\tilde{h}_{b}{}^{p}\tilde{h}_{q}{}^{c}\tilde{h}_{r}{}^{d}...{\rm D}_{s}T_{fp...}{}^{qr...}\,.
\end{equation}
Finally, the 2-D Levi-Civita pseudotensor can be defined via the contraction of its 3-D counterpart along the spatial direction of $n^{a}$, $\epsilon_{ab}\equiv\epsilon_{abc}n^{c}$. It follows that:
\begin{equation}
    \epsilon_{ab}n^{b}=0 \quad \text{and} \quad \epsilon_{ab}\epsilon^{cd}=2\tilde{h}_{[a}{}^{c}\tilde{h}_{b]}{}^{d}
\end{equation}
as well as that $\epsilon_{abc}=n_{a}\epsilon_{bc}-n_{b}\epsilon_{ac}+n_{c}\epsilon_{ab}$.\\

\subsection{2-Dimensional kinematic quantities}\label{ssec:2-D-kinematics}

In analogy with its 3-D counterpart, the motion on the 2-D surface orthogonal to $n^{a}$ is characterised by a set of kinematic quantities which come from the decomposition of the gradient of $n^{a}$. In other words, we have:
\begin{equation}
    {\rm D}_{b}n_{a}=\tilde{\sigma}_{ab}+\tilde{\omega}_{ab}+\frac{1}{2}\tilde{\Theta}\tilde{h}_{ab}+n_{b}n'_{a},
\end{equation}
where $\tilde{\sigma}_{ab}\equiv {\rm D}_{\langle b}n_{a\rangle}$, $\tilde{\omega}_{ab}\equiv {\rm D}_{[b}n_{a]}$ and $\tilde{\Theta}\equiv {\rm D}^{a}n_{a}$ are respectively the shear and the vorticity tensors, the surface expansion-contraction scalar and $n'_{a}\equiv n^{b}{\rm D}_{b}n_{a}$ the spatial derivative of $n^{a}$ along its own direction. The sum $\tilde{{\rm D}}_{b}n_{a}=\tilde{\sigma}_{ab}+\tilde{\omega}_{ab}+\frac{1}{2}\tilde{\Theta}\tilde{h}_{ab}$ describes the relative motion of neighbouring spacelike curves orthogonal to the surface in question.

It is worth comparing the 2-D version of the shear $\tilde{\sigma}_{ab}\equiv {\rm D}_{\langle b}n_{a\rangle}$ with those of the individual 1+2 components of its 3-D version $\sigma_{ab}\equiv {\rm D}_{\langle b}u_{a\rangle}$. Concerning the 2-D vorticity tensor, it has only one independent component (i.e. it consists of a vector along the one of the two independent directions defining the 2-D surface), so that it can be written as $\tilde{\omega}_{ab}=\tilde{\omega}\epsilon_{ab}$, where $\tilde{\omega}^{2}=(1/2)\tilde{\omega}^{ab}\tilde{\omega}_{ab}$. Finally, the condition $n'^{a}=0$ implies that the $n^{a}$ field is tangent to a congruence of spacelike geodesics.\\

\subsection{Deriving the $2$-dimensional form of the Ricci identities}\label{appssec:2-D-Ricci}

We begin with writing the Ricci identity in its $3$-dimensional form for a vector lying on the $2$-surface normal to $n^{a}$, e.g. for $k^{a}$ ($k^{a}n_{a}=0)$:
\begin{equation}
    2{\rm D}_{[a}{\rm D}_{b]}k_{c}=-2\omega_{ab}\dot{k}_{\langle c\rangle}+\mathcal{R}_{dcba}k^{d}\,.
\end{equation}
Projecting the above orthogonal to the $n^{a}$-field, yields:
\begin{equation}
    \tilde{h}_{d}{}^{a}\tilde{h}_{e}{}^{b}\tilde{h}_{f}{}^{c}~{\rm D}_{a}{\rm D}_{b}k_{c}-\tilde{h}_{d}{}^{a}\tilde{h}_{e}{}^{b}\tilde{h}_{f}{}^{c}~{\rm D}_{b}{\rm D}_{a}k_{c}=-2\tilde{h}_{d}{}^{a}\tilde{h}_{e}{}^{b}\tilde{h}_{f}{}^{c}~\omega_{ab}\dot{k}_{\langle c\rangle}+\tilde{h}_{d}{}^{a}\tilde{h}_{e}{}^{b}\tilde{h}_{f}{}^{c}\tilde{h}_{g}{}^{i}~\mathcal{R}_{icba}k^{g}\,,
    \label{eqn:proj-2Ricci}
\end{equation}
where $\tilde{h}_{d}{}^{a}\tilde{h}_{e}{}^{b}\tilde{h}_{f}{}^{c}~{\rm D}_{a}{\rm D}_{b}k_{c}=\tilde{{\rm D}}_{d}{\rm D}_{e}k_{f}$. Subsequently, we take into account that $\tilde{{\rm D}}_{a}k_{c}=\tilde{h}_{a}{}^{b}\tilde{h}_{c}{}^{d}~{\rm D}_{b}k_{d}$ which, after expansion of the projection tensors, leads to:
\begin{equation}
    {\rm D}_{a}k_{c}=\tilde{{\rm D}}_{a}k_{c}+n_{a}k'_{c}-n_{c}({\rm D}_{a}n_{d})k^{d}-n_{a}n_{c}(n^{d}k'_{d})\,.
    \label{eqn:expansion}
\end{equation}
Using the aforementioned expressions, eq~\eqref{eqn:proj-2Ricci} becomes:
\begin{equation}
    \tilde{{\rm D}}_{d}{\rm D}_{e}k_{f}-\tilde{{\rm D}}_{e}\tilde{{\rm D}}_{d}k_{f}-(\tilde{{\rm D}}_{e}n_{d})k'_{f}+(\tilde{{\rm D}}_{e}n_{f})(\tilde{{\rm D}}_{d}n_{g})k^{g}=-2\Omega\epsilon_{de}\dot{k}_{\bar{f}}+\tilde{h}_{g}{}^{i}\tilde{h}_{f}{}^{c}\tilde{h}_{e}{}^{b}\tilde{h}_{d}{}^{a}~\mathcal{R}_{icba}k^{g}\,,
    \label{eqn:2Ricci-3}
\end{equation}
where we have made use of $\omega_{ab}=\epsilon_{abc}\omega^{c}$ and $\epsilon_{abc}=n_{a}\epsilon_{bc}-n_{b}\epsilon_{ac}+n_{c}\epsilon_{ab}$, and considered that the sequence according to which the projection tensors act on $\mathcal{R}_{ifed}k^{g}$, does not matter. Expanding now the first term in the left-hand side of~\eqref{eqn:2Ricci-3} by deploying eq~\eqref{eqn:expansion}, and projecting once again the resulting equation with $\tilde{h}_{a}{}^{d}\tilde{h}_{b}{}^{e}\tilde{h}_{c}{}^{f}$ , we arrive at: 
\begin{equation}
2\tilde{{\rm D}}_{[a}\tilde{{\rm D}}_{b]}k_{c}=2\tilde{{\rm D}}_{[b}n_{a]}k'_{c}-2\Omega\epsilon_{ab}\dot{k}_{\bar{c}}+\left(\tilde{v}_{gb}\tilde{v}_{ca}-\tilde{v}_{ga}\tilde{v}_{cb}\right)k^{g}+\tilde{h}_{g}{}^{i}\tilde{h}_{c}{}^{f}\tilde{h}_{b}{}^{e}\tilde{h}_{a}{}^{d}~\mathcal{R}_{ifed}k^{g}\,,
\end{equation}
where $\tilde{v}_{ab}\equiv \tilde{{\rm D}}_{b}n_{a}=\tilde{\sigma}_{ab}+\tilde{\omega}_{ab}+(\tilde{\Theta}/2)\tilde{h}_{ab}$. Finally, employing our definition for the $2$-D Riemann tensor $\tilde{\mathcal{R}}_{abcd}$ (see eq~\eqref{eqn:2-D Riemann} in the following subsection), the Ricci identity for a vector lying on the 2-surface normal to $n^{a}$, reads:
\begin{equation}
2\tilde{{\rm D}}_{[a}\tilde{{\rm D}}_{b]}k_{c}=2\tilde{\omega}_{ab}k'_{c}-2\Omega\epsilon_{ab}\dot{k}_{\bar{c}}+\tilde{\mathcal{R}}_{dcba}k^{d}\,.
    \label{2-D_Ricci_identity}
\end{equation}
We observe that the $2$-D form of the Ricci identity includes two vorticity terms, describing rotation of the 2-surface normal to $n^{a}$; these are $\tilde{\omega}_{ab}\equiv\tilde{\omega}\epsilon_{ab}\equiv \tilde{D}_{[b}n_{a]}$ and $\Omega\epsilon_{ab}$ with $\Omega\equiv\omega_{a}n^{a}\equiv -n^{a}\text{curl}u_{a}=-(1/2)\epsilon_{ab}\tilde{{\rm D}}^{a}u^{b}$, where $\Omega$ denotes the norm of the vorticity vector along direction $n^{a}$. As far as we know, a $2$-dimensional form of the Ricci identities has not appeared before in the literature.\\

\subsection{Finding the 2-D Ricci tensor and scalar in terms of the fluid dynamic quantities}\label{appsec:2-Ricci-kinematics}

We define the Riemann tensor $\tilde{\mathcal{R}}_{abcd}$ of a 2-surface orthogonal to a vector $n^{a}$, as follows:
\begin{equation}
\tilde{\mathcal{R}}_{abcg}\equiv\tilde{h}_{a}{}^{d}\tilde{h}_{b}{}^{e}\tilde{h}_{c}{}^{f}\tilde{h}_{g}{}^{j}\mathcal{R}_{defj}-\tilde{v}_{ac}\tilde{v}_{bg}+\tilde{v}_{ag}\tilde{v}_{bc},
     \label{eqn:2-D Riemann}
\end{equation}
where $\mathcal{R}_{defj}$ is the 3-D Riemann tensor and $\tilde{v}_{ab}=\tilde{{\rm D}}_{b}n_{a}=\tilde{\sigma}_{ab}+\tilde{\omega}_{ab}+(\tilde{\Theta}/2)\tilde{h}_{ab}$. The above definition is analogous to its counterpart for the $3$-Riemann tensor (e.g. see~(1.3.34) in~\cite{TCM3}). The 2-D Ricci tensor is derived from relation~\eqref{eqn:2-D Riemann} by contracting the indices $a$ and $c$, namely:
\begin{equation}
\tilde{\mathcal{R}}_{bg}=\tilde{h}^{df}\tilde{h}_{b}{}^{e}\tilde{h}_{g}{}^{j}~\mathcal{R}_{defj}-\tilde{\Theta}\tilde{v}_{bg}+\tilde{v}^{c}{}_{g}\tilde{v}_{bc}\,.
\label{eqn:2-Ricci-tensor}
\end{equation}
Projecting the $3$-D Riemann tensor (e.g. see eq.~1.3.35 in~\cite{TCM}) in accordance with~\eqref{eqn:2-Ricci-tensor}, we find out that:
\begin{equation}
\tilde{\mathcal{R}}_{ab}=\frac{1}{2}\tilde{\mathcal{R}}\tilde{h}_{ab}+\Sigma\Sigma_{ab}+\tilde{\sigma}_{c\langle a}\tilde{\sigma}^{c}{}_{b\rangle}+2\tilde{\sigma}_{c[a}\tilde{\omega}^{c}{}_{b]}+\Sigma_{c\langle a}\Sigma^{c}{}_{b\rangle}+2\Omega\Sigma_{c[a}\epsilon^{c}{}_{b]}\,,
\label{eqn:2-D Ricci}
\end{equation}
where
\begin{equation}
    \tilde{\mathcal{R}}=-2\mathcal{E}+\frac{2}{3}\rho-\Pi-\frac{2}{9}\Theta^{2}-\tilde{\Theta}^{2}+\frac{2}{3}\Sigma\Theta-\frac{1}{2}\Sigma^{2}-2\Omega^{2}+\frac{4}{3}\tilde{\sigma}^{2}-2\tilde{\omega}^{2}+\frac{2}{3}\Sigma_{ab}\Sigma^{ab}
    \label{eqn:Ricci-mean-curv}
\end{equation}
is the scalar curvature of a 2-surface.\footnote{Note that $\rho$ and $\Pi$ in~\eqref{eqn:Ricci-mean-curv} include contributions from various forms of matter.} In the above we have used the definitions: $2\tilde{\sigma}^{2}\equiv\tilde{\sigma}_{ab}\tilde{\sigma}^{ab}$ and $\tilde{\omega}_{ab}\equiv\tilde{\omega}\epsilon_{ab}$.\\

\section{The scalar and vector components of the shear tensor}\label{appB-after}

\footnote{The present appendix section is envisaged as a correction of Appendix A in~\cite{MT}. In particular, we point out here that eqs~(102) and~(106) of the aforementioned work are valid, not generally, but within a specific framework.}The present appendix unit provides a technical supplement to calculations spanning the main text of Part~II. Throughout this thesis we encounter several times products of tensors with the spacelike (unit) vector field $n^{a}$ (such that $n_{a}u^{a}=0$ and $n_{a}n^{a}=1$), which is taken parallel to the magnetic forcelines (i.e. $B^{a}=\mathcal{B}n^{a}$). In particular, our calculations often involve the quantities $\Sigma=\sigma_{ab}n^{a}n^{b}$ and $\Sigma_{a}=\tilde{h}_{a}{}^{b}\sigma_{bc}n^{c}$ (with $\tilde{h}_{ab}=g_{ab}+u_{a}u_{b}-n_{a}n_{b}=h_{ab}-n_{a}n_{b}$, satisfying $\tilde{h}_{ab}n^{b}=0$). Concerning the former, it can be written as
\begin{equation}
\Sigma\equiv\sigma_{ab}n^{a}n^{b}\equiv {\rm D}_{\langle b}u_{a\rangle}n^{a}n^{b}={\rm D}_{(b}u_{a)}n^{a}n^{b}-\frac{1}{3}\Theta h_{ab}n^{a}n^{b}=u'_{a}n^{a}-\frac{1}{3}\Theta\,,
\label{eqn:appB1}
\end{equation}
where $u'_{a}=n^{b}{\rm D}_{b}u_{a}$. Note that the term $u'_{a}n^{a}$ does not generally vanish because the prime ($'$) is not an actual spatial derivative operator. This means that $'$ does not generally satisfy the product (Leibniz) associative rule between (the timelike) $u^{a}$ and (the spacelike) $n^{a}$. To illustrate the meaning of the term $u'_{a}n^{a}$, let us consider the double projection of the volume expansion/contraction tensor, $\Theta_{ab}=\sigma_{ab}+(\Theta/3)h_{ab}$, along a spatial direction $n^{a}$, which reads:
\begin{equation}
\Theta_{ab}n^{a}n^{b}=u'_{a}n^{a}\hspace{10mm}\text{with}\hspace{10mm}\Sigma=\Theta_{ab}n^{a}n^{b}-\frac{\Theta}{3}\,.
\label{eqn:theta-tensor}
\end{equation}
Within the main text we work under the consideration $\Sigma=\lambda\Theta$ ($\lambda$: a real number), so that $\Theta_{ab}n^{a}n^{b}=u'_{a}n^{a}$ is a multiple of $\Theta$.\\
On the other hand, in reference to the vector shear component, namely $\Sigma_{a}$, we have
\begin{equation}
\Sigma_{a}\equiv \tilde{h}_{a}{}^{b}\sigma_{bc}n^{c}\equiv \tilde{h}_{a}{}^{b}n^{c}{\rm D}_{\langle c}u_{b\rangle}=\tilde{h}_{a}{}^{b}n^{c}{\rm D}_{(c}u_{b)}=\frac{1}{2}\Sigma_{a}+\frac{1}{2}\epsilon_{ab}\Omega^{b}+\frac{1}{2}\tilde{h}_{a}{}^{b}u'_{b}\,,
\end{equation}
which subsequently leads to
\begin{equation}
\Sigma_{a}=\epsilon_{ab}\Omega^{b}+\tilde{h}_{a}{}^{b}u'_{b}\hspace{8mm}\text{or}\hspace{8mm} \Sigma_{a}=\epsilon_{ab}\Omega^{b}+\alpha_{a}\,.
\label{eqn:Sigma-vector}
\end{equation}
where in deriving~(\ref{eqn:Sigma-vector}b), we have taken into account
the projection of Faraday's law normal to the magnetic direction $n^{a}$ (implying $\tilde{h}_{a}{}^{b}\dot{n}_{b}\equiv\alpha_{a}=\tilde{h}_{a}{}^{b}u'_{b}$).\\
Before closing the present section, let us examine the particular case corresponding to $\Sigma=-\Theta/3$. The condition in question implies that $u'_{a}=0$, which means that the fluid velocity is homogeneous along the magnetic forcelines (for $B^{a}=\mathcal{B}n^{a}$). In other words, the magnetic fieldlines are envisaged as streamlines of the fluid. Moreover, $\Sigma=-\Theta/3$ translates to $\Theta_{ab}n^{a}n^{b}=0$. Under the aforementioned constraint, only accelerated (non-geodesic) motion or motion due to vorticity is allowed along direction $n^{a}$. In reference to the problem of nearly homogeneous magnetised gravitational collapse, recall that there is no acceleration along $n^{a}$ (due to ${\rm D}_{a}\rho=0$). Also, if the net charge density is zero (i.e. $\mu=0$), there is no vortex motion along the magnetic direction, as follows from Gauss's law. Therefore, for the problem in question, condition $\Sigma=-\Theta/3$ or $\Theta_{ab}n^{a}n^{b}=0$ requires the total absence of motion parallel to the magnetic field. In other words, it consists of a special case of $2$-dimensional gravitational contraction.\\

\section{Spatial decomposition of the magnetohydrodynamic constraints and detailed solution of the equation $\dot{\mathcal{B}}=\left(\lambda-\frac{2}{3}\right)\Theta\mathcal{B}$.}\label{app*:Magnetic-solution}

To begin with, let us consider the decomposition (orthogonal and along $n^{a}$) of the constraints~\eqref{MHD-current} and~\eqref{MHD-constraints}. In particular, the former splits into:
\begin{equation}
-\mathcal{B}^{2}\mathcal{A}_{a}-2\mathcal{B}\tilde{\rm{D}}_{a}\mathcal{B}+\mathcal{B}^{2}n'_{a}=\mathcal{B}\epsilon_{ab}j^{b} \hspace{15mm} \text{and} \hspace{15mm} \tilde{\omega}\mathcal{B}=-\frac{j}{2}\,;
\label{eqn:bj}
\end{equation}
As for the scalar equations~\eqref{MHD-constraints}, they are written as
\begin{equation}
\Omega \mathcal{B}=\frac{\mu}{2}\hspace{15mm} \text{and} \hspace{15mm} \mathcal{B}'+\tilde{\Theta}\mathcal{B}=0\,.
\label{b'-Omega-b}
\end{equation}
In case of neutral total charge density (i.e. $\mu=0$), eq~(\ref{b'-Omega-b}a) implies that there is no angular velocity along the magnetic direction. Concerning~(\ref{b'-Omega-b}b), we observe that it is a covariant, linear, partial differential equation of first order. Given that $\tilde{\Theta}\equiv 2l'/l$ (with $l\equiv\alpha\beta$ being\footnote{The total scale factor we write as $a\equiv\alpha\beta\gamma$, where $\alpha$, $\beta$, $\gamma$ represent the anisotropy factors along the three spatial directions.} the scale factor of the surface orthogonal to $n^{a}$), the equation in question accepts the general solution:
\begin{equation}
    \mathcal{B}\propto l^{-2}\,.
    \label{eqn:B-s}
\end{equation}
Subsequently, we provide a detailed solution of Faraday's equation in the form of~\eqref{MHD-B-prop}. In detail, as $\mathcal{B}$ is a scalar quantity, its covariant differentiation is equivalent to its ordinary differentiation, so that
\begin{equation}
    \dot{\mathcal{B}}=u^{a}\nabla_{a}\mathcal{B}=u^{a}\partial_{a}\mathcal{B}=(u^{0}\partial_{0}+u^{1}\partial_{1}+u^{2}\partial_{2}+u^{3}\partial_{3})\mathcal{B}=\left(\lambda-\frac{2}{3}\right)\Theta \mathcal{B}\,.
    \label{eqn:bdot}
\end{equation}
Now by defining new space-time variables $\tilde{x}^{a}$ such that\footnote{Note that here the repeated index $i$ does not imply summation of components.}
\begin{equation}
    \tilde{x}^{i}=\int\frac{dx^{i}}{u^{i}}\,,
\end{equation}
expression~\eqref{eqn:bdot} becomes:
\begin{equation}
(\tilde{\partial}_{0}+\tilde{\partial}_{1}+\tilde{\partial}_{2}+\tilde{\partial}_{3})\mathcal{B}=\left(\lambda-\frac{2}{3}\right)\Theta \mathcal{B}\,,
    \label{tilde-b-dot}
\end{equation}
where $\tilde{\partial}_{i}$ are the new derivative operators with respect to the variables $\tilde{x}^{i}$. Let us try to solve the latter equation by assuming variables separation : $\mathcal{B}=\mathcal{T}(\tilde{x}^{0})U(\tilde{x}^{1})V(\tilde{x}^{2})W(\tilde{x}^{3})$, where $\tilde{x}^{0}$ is the new temporal variable and $\tilde{x}^{1},\tilde{x}^{2},\tilde{x}^{3}$ are the new spatial variables. Relation~\eqref{tilde-b-dot} takes thus the form:
\begin{equation}
    \frac{\tilde{\partial}_{0}\mathcal{T}}{\mathcal{T}}+\frac{\tilde{\partial}_{1}U}{U}+\frac{\tilde{\partial}_{2}V}{V}+\frac{\tilde{\partial}_{3}W}{W}=\left(\lambda-\frac{2}{3}\right)\Theta(\tilde{x}^{0},\tilde{x}^{1},\tilde{x}^{2},\tilde{x}^{3})\,,
    \label{sep-variables2}
\end{equation}
We observe that each of the fractions in the above equation depends only on one of the variables $\tilde{x}^{i}$. Subsequently, equation~\eqref{sep-variables2} holds if and only if $\Theta(\tilde{x}^{0},\tilde{x}^{1},\tilde{x}^{2},\tilde{x}^{3})=\Theta_{0}(\tilde{x}^{0})+\Theta(\tilde{x}^{1})+\Theta_{2}(\tilde{x}^{2})+\Theta_{3}(\tilde{x}^{3})$. Therefore, the original partial differential equation reduces to four ordinary differential equations of the form $(\tilde{\partial}_{1}U/U)=\left(\lambda-\frac{2}{3}\right)\Theta_{1}(\tilde{x}^{1})$, which are integrated directly to give $U=c_{1}e^{\left(\lambda-\frac{2}{3}\right)\int{\Theta_{1}}d\tilde{x}^{1}}$. Hence, it is overall clear to see that the solution for $\mathcal{B}$ can be written as
\begin{equation}
\mathcal{B}=\mathcal{C}e^{\left(\lambda-\frac{2}{3}\right)\left(\int{\Theta_{0}}d\tilde{x}^{0}+\int{\Theta_{1}}d\tilde{x}^{1}+\int{\Theta_{2}}d\tilde{x}^{2}+\int{\Theta_{3}}d\tilde{x}^{3}\right)}=\mathcal{C}e^{\left(\lambda-\frac{2}{3}\right)\left(\int{\frac{\Theta_{0}}{u^{0}}}dx^{0}+\int{\frac{\Theta_{1}}{u^{1}}}dx^{1}+\int{\frac{\Theta_{2}}{u^{2}}}dx^{2}+\int{\frac{\Theta_{3}}{u^{3}}}dx^{3}\right)}\,,
\label{eqn:bexp}
\end{equation}
where $\mathcal{C}$ is an arbitrary constant and we have found out that our variables separation assumption turns out to be true\footnote{Recall that the original equation~\eqref{eqn:bdot} is a partial differential one. However, we have shown that it reduces four ordinary equations (see~\eqref{sep-variables2}). As a consequence, the general solution we have found, eq.~\eqref{eqn:bexp} is actually the only solution of the original equation.}. Equation~\eqref{eqn:bexp}, which is a solution\footnote{As far as we know, it is the first time that the solution in question appears in the literature.} of Faraday's law at the MHD limit, tells us that if $\Theta_{i}(\tilde{x}^{i})$ are continuous functions in a specific closed interval $[\alpha_{1}, \alpha_{2}]$ of their domain and they preserve constant sign (e.g. $\Theta_{i}(\tilde{x}^{i})\leq 0$--implying continuous gravitational contraction) for every value of their variable belonging in the interval, then $\int_{\alpha_{1}}^{\alpha_{2}}{ \Theta_{i}(\tilde{x}^{i})}d\tilde{x}^{i}<0$ and the magnetic field generally obeys an exponential type of increase with respect to the spacetime variables. In fact, the aforementioned exponential type behavior seems to be outward because on defining a scale factor $a(\tilde{x}^{0},\tilde{x}^{1},\tilde{x}^{2},\tilde{x}^{3})$, such that $\Theta=3\dot{a}/a$ (also $\Theta_{0}=3{\rm d}a_{0}/(a_{0}{\rm d}\tilde{x}^{0})$ and $\Theta_{i}=3{\rm d}a_{i}/(a_{i}{\rm d}\tilde{x}^{i}))$, equation~\eqref{eqn:bexp} reduces to:
\begin{equation}
    \mathcal{B}\propto a^{3\lambda-2}=\left(a_{0}(\tilde{x}^{0})a_{1}(\tilde{x}^{1})a_{2}(\tilde{x}^{2})a_{3}(\tilde{x}^{3})\right)^{3\lambda-2}\,.
    \label{Bexp->B-a}
\end{equation}
Finally, we shall keep in mind the following remarks. Firstly, on deriving relations~\eqref{eqn:bexp},~\eqref{Bexp->B-a} we have not adopted a specific coordinate reference frame. Secondly, the evolution of $\mathcal{B}$ in each spacetime direction is independent of its evolution in the other directions with respect to the tilted variables only, where $\mathcal{B}=\mathcal{T}(\tilde{x}^{0})U(\tilde{x}^{1})V(\tilde{x}^{2})W(\tilde{x}^{3})$. The crucial equation~\eqref{eqn:bexp}, or~\eqref{Bexp->B-a}, provides us the keystone for studying magnetic fields in cosmological and astrophysical problems (refer to the following sections).\\
In order to specify the constant $\mathcal{C}$, we observe that the key fluid dynamic quantity related to the magnetic field, is the volume scalar $\Theta$. Therefore, we turn our attention to the relation which describes its evolution, the so-called Raychaudhuri equation (see the chapter's main text or e.g.~\cite{EMM3}),
\begin{equation}
    \dot\Theta=-\frac{1}{3}\Theta^{2}-\frac{1}{2}(\rho+3P+\mathcal{B}^{2})-2(\sigma^{2}-\omega^{2})+D^{a}\dot{u}_{a}+\dot{u}^{a}\dot{u}_{a}\,.
    \label{Raych-eqn1}
\end{equation}
Considering an instant during which the fluid is found in its equilibrium (static) state\footnote{Such an instant could have been either the initial instant-just before the collapse starts-or a transitional instant, during which the collapse stops and the fluid starts expanding.} (setting $\Theta=0=\sigma^{2}$ and $\dot u_{a}=0=\omega^{2}$), we have $\mathcal{B}=\mathcal{C}$, and~\eqref{Raych-eqn1} leads to (the star index refers to equilibrium values in the following)
\begin{equation}
    \mathcal{C}^{2}=-(2\dot\Theta_{*}+\rho_{*}+3P_{*}),
    \label{eqn:C1}
\end{equation}
which means that $\mathcal{C}$ is a real constant if
\begin{equation}
    \dot\Theta_{*}<-\frac{1}{2}(\rho_{*}+3P_{*})<0.
\end{equation}
In other words, the rate of change of the volume scalar in the equilibrium has to be negative and smaller than the gravitational mass of the system due to conventional matter ($\frac{1}{2}(\rho_{*}+3P_{*})>0$).\\

\section{Magneto-curvature tension stresses}\label{app:magneto-curvature-tension}

Following the discussion in subsection~\ref{ssec:magneto-curvature tension}, the magneto-curvature tension stresses associated with shear, rotational and volume curvature distortions are:
\footnote{On deriving eqs~\eqref{s-STF}-\eqref{elast-law1} we make use of the so-called Gauss-Codacci formula (e.g. see eq.~1.3.39 in~\cite{TCM}).}
\begin{eqnarray}
    s_{\langle ac\rangle}&=&\mathcal{B}^{2}\mathcal{R}_{d\langle ac\rangle b}n^{b}n^{d}=\nonumber\\
    &&\mathcal{B}^{2}\left[\epsilon_{\langle a|q|}\epsilon_{c\rangle s}\left(\mathcal{E}^{qs}+\Omega^{q}\Omega^{s}\right)n_{\langle a}n_{c\rangle}+\frac{1}{3}\left(\rho-\frac{1}{3}\Theta^{2}+3\Pi-2\Theta\Sigma-3\Sigma^{2}\right)+\Pi_{\langle a}n_{c\rangle}\right] \nonumber \\
    &&+\mathcal{B}^{2}\left[-\pi_{ac}-2\left(\Sigma+\frac{\Theta}{3}\right)\Sigma_{\langle a}n_{c\rangle}+\frac{\Theta}{3}\sigma_{ac}-\Sigma_{\langle a}\Sigma_{c\rangle}\right]\,,
    \label{s-STF}
\end{eqnarray}
\begin{equation}
s_{[ac]}=\mathcal{B}^{2}\mathcal{R}_{d[ac]b}n^{b}n^{d}=\mathcal{B}^{2}\left[2\left(\frac{\Theta}{3}n_{[c}+\Sigma n_{[c}+\Sigma_{[c}\right)\epsilon_{b]d}\Omega^{d}+\frac{\Theta}{3}\omega_{bc}\right]
    \label{twisting}
\end{equation}
and
\begin{equation}
    s=s^{c}{}_{c}=\mathcal{B}^{2}\mathcal{R}_{bd}n^{b}n^{d}=\mathcal{B}^{2}\left[\frac{2}{3}\rho+\mathcal{E}+\frac{\Pi}{2}+\left(\lambda^{2}-\frac{\lambda}{3}-\frac{2}{9}\right)\Theta^{2}\right]\,,
    \label{appeq:elast-law1}
\end{equation}
where $\pi_{ab}$ and $E_{ab}$ are the anisotropic stress and the tidal (or electric Weyl) tensors respectively. Moreover, we have $\Pi\equiv \pi_{ab}n^{a}n^{b}$, $\Pi_{a}\equiv \tilde{h}_{a}{}^{b}n^{c}\pi_{bc}$, $\mathcal{E}\equiv E_{ab}n^{a}n^{b}$, $\mathcal{E}_{ab}\equiv (\tilde{h}_{(a}{}^{c}\tilde{h}_{b)}{}^{d}-(1/2)\tilde{h}_{ab}\tilde{h}^{cd})E_{cd}$, $\Sigma\equiv \sigma_{ab}n^{a}n^{b}=\lambda\Theta$ and $\Sigma_{a}\equiv \tilde{h}_{a}{}^{b}n^{c}\sigma_{bc}=\epsilon_{ab}\Omega^{b}+\alpha_{a}$ (we use the former of the last two expressions only in writing eq~\eqref{appeq:elast-law1} from the above), with $\epsilon_{ab}\equiv \epsilon_{abc}n^{c}$ being the 2-D counterpart of the Levi-Civita pseudotensor, $\Omega_{a}\equiv\tilde{h}_{a}{}^{b}\omega_{b}$ and $\tilde{h}_{ab}\equiv h_{ab}-n_{a}n_{b}$ an operator projecting orthogonal to the magnetic field direction $n^{a}$. We observe that tidal effects (electric Weyl components) are associated with shape and volume magnetic distortions only. Assuming an ideal fluid model, the anisotropic stress terms in the above vanish. Then, of particular interest is that the deformation due to gravitational compression/expansion in~\eqref{appeq:elast-law1} is determined by the density  of matter and the tidal tensor projected along the magnetic fieldlines.\\

\section{Temporal evolution of the matter density under a polytropic equation of state}\label{app:polytropic-eq-of-state}

Considering an ideal, polytropic (i.e. $P=k\rho^{\gamma}$, with $k$ and $\gamma$ constants) fluid at the MHD limit, the continuity equation, $\dot{\rho}=-\Theta(\rho+P)$ ($\Theta=3\dot{a}/a$, $a$ denoting the scale-factor of the fluid's volume), reads the following explicit Bernoulli form
\begin{equation}
    \frac{d\rho}{da}+\frac{3}{a}\rho+\frac{3k}{a}\rho^{\gamma}=0\,.
    \label{cont-eq-polytrope}
\end{equation}
The equation in question accepts the general solution
\begin{equation}
    \rho=\left[Ca^{-3(1-\gamma)}-k\right]^{\frac{1}{1-\gamma}}\,,
    \label{density-evol-polytrope}
\end{equation}
with $C$ (note that $C>0$ for $k>0$) being the integration constant. In the cases of non-relativistic neutrons ($\gamma=5/3$) and ultra-relativistic electrons ($\gamma=4/3$) the above equation recasts into
\begin{equation}
    \rho=\left(Ca^{2}-k\right)^{-3/2}\hspace{15mm} \text{and}\hspace{15mm} \rho=\left(Ca-k\right)^{-3}
    \label{density-evol-polytrope2}
\end{equation}
respectively. Obviously, the pressure of matter, $P=k\rho^{\gamma}$, increases faster than its density for $\gamma>1$ (i.e. the cases we consider). We employ the above equations in determining the magnetic fracture limit of a neutron star and a white dwarf in the main text (see subsection~\ref{ssec:further-frac-limit}).\\

\section{Some auxiliary calculations}\label{app:auxiliary-calculations}

In reference to eq~\eqref{eq:dot-c} (i.e. temporal derivative of the Alfv$\acute{\text{e}}$n speed) in subsection~\ref{ssec:further-frac-limit}, we employ the continuity equation and the equation of state ($P=k\rho^{\gamma}$) to calculate the dot derivative of matter density and pressure, i.e. $\dot{\rho}=-\Theta(\rho+P)$ and $\dot{P}=\gamma(P/\rho)\dot{\rho}$. Furthermore, we make use of the law of magnetic contraction, $\dot{\mathcal{B}}=(\lambda-2/3)\Theta\mathcal{B}$ (under the ideal MHD approximation of a magnetised fluid, and condition~$\Sigma=\lambda\Theta$). Therefore, eq~\eqref{eq:dot-c} recasts into:
\begin{equation}
\dot{\left(s^{*}\right)}_{\text{fr}}=-\dot{\left(c^{2}_{\mathcal{A}}\right)}_{\text{fr}}u_{\text{fr}}=\frac{1}{\mathcal{B}^{2}(1+\beta)^{2}}\left[\dot{\rho}+\dot{P}-\frac{2}{\mathcal{B}}\dot{\mathcal{B}}(\rho+P)\right]u_{\text{fr}}=\frac{-\Theta(\rho+P)}{\mathcal{B}^{2}(1+\beta)^{2}}\left(\gamma\frac{P}{\rho}+2\lambda-\frac{1}{3}\right)u_{\text{fr}}=0\,,
    \label{eq:c-dot-app1}
\end{equation}
which clearly leads to~\eqref{eq:P-condition}. The fracture limit condition is mentioned within the main text. Alternatively, assuming a barotropic equation of state (i.e. $P=w\rho$ and $\dot{P}=w\dot{\rho}=(P/\rho)\dot{\rho}$ with $w=\text{constant}$), eq~\eqref{eq:dot-c} reduces to:
\begin{equation}
    \dot{\left(c^{2}_{\mathcal{A}}\right)}_{\text{fr}}=-\frac{1}{\mathcal{B}^{2}(1+\beta)^{2}}\left[\dot{\rho}+\dot{P}-\frac{2}{\mathcal{B}}\dot{\mathcal{B}}(\rho+P)\right]=\frac{\Theta(\rho+P)}{\mathcal{B}^{2}(1+\beta)^{2}}\left(\frac{P}{\rho}+2\lambda-\frac{1}{3}\right)=0\,,
    \label{eq:c-dot-app2}
\end{equation}
from which we deduce that:
\begin{equation}
P_{\text{fr}}=\left(\frac{1}{3}-2\lambda\right)\rho_{\text{fr}}\,,\hspace{6mm}\text{with}\hspace{6mm}0\leq \left(w=\frac{1}{3}-2\lambda\right)\leq 1\hspace{2mm}\rightarrow\hspace{2mm} -\frac{1}{3}\leq\lambda\leq\frac{1}{6}\,,
\label{eqn:app-frac-limit-w}
\end{equation}
or that $P_{\text{fr}}=-\rho_{\text{fr}}$ (i.e. $w=-1$). The last solution is directly rejected because ordinary collapsing matter is assumed.

\end{subappendices}

\newpage

\chapter{Magnetised Bianchi I cosmology}\label{chap:cosmic-magnetic-fields}

\section{General introductory remarks on cosmic magnetic fields}

Within the context of the standard cosmological model, large-scale gravitational as well as electromagnetic perturbations are causally produced via the inflationary mechanism. In particular, spacetime distortions initially appear in the form of quantum fluctuations during the so-called Planck epoch. Subsequently, due to the exponential expansion of the inflation era, these quantum fluctuations are forced to pass out of the Hubble horizon, where they freeze out in the form of classical perturbations. After inflation, during reheating and the following radiation era, the electrical conductivity of the initially poorly conducting cosmic medium increases rapidly~\cite{EMM4}. As a consequence, the electric fields gradually vanish and the currents freeze the magnetic fields in with the cosmic fluid. In other words, the post-inflationary universe can be causally described by the ideal magnetohydrodynamical model, within the Hubble scale. Besides, the adoption of the MHD approximation in the standard cosmological framework is in accordance with the fact that only large-scale magnetic (not electric) fields have been observed. In the following, our interest focuses on the evolution of large-scale magnetic fields lying within the Hubble horizon.\\
Let us recall that in the MHD framework, magnetised ideal barotropic fluids obey the continuity equation in the form: $\dot{\rho}=-\Theta(1+w)\rho$, where $0\leq w<1$ is the barotropic index. Therefore, radiation ($w=1/3$) and dust ($w=0$) evolve according to $\rho_{\text{rad}}\propto a^{-4}$ and $\rho_{\text{dust}}\propto a^{-3}$ respectively, where $a$ denotes the average scale factor. Using the magnetic density evolution formula $\rho_{B}\propto a^{6\lambda-4}$ (with $-2/3\leq\lambda<2/3$~-recall eq.~\eqref{eqn:magn-law-of-variation-orthog-Bianchi}), the ratios of the magnetic over the radiation and dust density read:
\begin{equation}
     \frac{\rho_{B}}{\rho_{\text{rad}}}=\left(\frac{\rho_{B}}{\rho_{\text{rad}}}\right)_{p}\left(\frac{a}{a_{p}}\right)^{6\lambda}\hspace{15mm} \text{and} \hspace{15mm} \frac{\rho_{B}}{\rho_{m}}=\left(\frac{\rho_{B}}{\rho_{m}}\right)_{p}\left(\frac{a}{a_{p}}\right)^{6\lambda-1}\,,
     \label{densities-ratio}
 \end{equation}
  where the suffix $p$ indicates the values of the involved quantities at the present, and $a_{p}/a=1+z$, with $z$ being the redshift. When the two forms of energy acquire equal densities, the associated scale factors, $a_{\text{eq}~(B-\text{rad})}$ and $a_{\text{eq}~(B-m)}$, are:
 \begin{equation}
     a_{\text{eq}~(B-\text{rad})}=\left(\frac{\rho_{\text{rad}}}{\rho_{B}}\right)^{\frac{1}{6\lambda}}_{p}a_{p}\sim 10^{-3/\lambda}a_{p}\hspace{10mm} \text{and} \hspace{10mm} a_{\text{eq}~(B-m)}=\left(\frac{\rho_{m}}{\rho_{B}}\right)^{\frac{1}{6\lambda-1}}_{p}a_{p}\sim 10^{-\frac{22}{6\lambda-1}}a_{p}\,.
     \label{aeq}
 \end{equation}
In the above calculation we have taken into account that the present value of intergalactic magnetic fields amounts to the order of $10^{-15}$ Gauss (e.g.~refer to~\cite{Tav4}-\cite{KKT4}). Making use of natural units ($c=\hbar=k_{B}=1$) the intergalactic magnetic energy density today is expressed in terms of GeV's as $\rho_{B}\sim 4\times 10^{-70}~\text{GeV}^{4}$, in accordance with the equivalence: $1~ (\text{Gauss})^{2}/(8\pi)\simeq 2\times 10^{-40}~\text{GeV}^{4}$ (e.g.~see the appendix of~\cite{KolT4}). Moreover, the density of matter today is $\rho_{m}\sim 10^{-30}~\text{gr}/\text{cm}^{3}\sim 4\times 10^{-48}~\text{GeV}^{4}$~($\rho_{m}=\Omega_{m}h^{2}\rho_{\text{crit}}$ with $\rho_{\text{crit}}\sim 10^{-29}~\text{gr}/\text{cm}^{3}$ and $\Omega_{m}h^{2}\simeq 0.14$ today~\cite{Planck}) whilst its radiation counterpart is $\rho_{\text{rad}}=10^{-34}~\text{gr}/\text{cm}^{3}\sim 4\times 10^{-52}~\text{GeV}^{4}$ ($1~\text{GeV}^{4}\simeq 2\times 10^{17}~\text{gr}/\text{cm}^{3}$).\\
Subsequently, it is worth raising an issue related to the constraint that cosmic nucleosynthesis imposes on the magnitude of the magnetic energy density. In particular, magnetic fields are known to increase nuclear reaction/transformation rates\footnote{Besides, magnetic fields contribute to the expansion rate of the universe and thus indirectly affect the rate of nuclear interactions.}, so that a potential domination of the magnetic energy density over radiation, during the early stages of cosmic evolution, may be incompatible with current predictions/observations regarding the nuclei abundance in the universe. We attempt here a first approach to the question by comparing the densities of magnetic fields and radiation during nucleosynthesis. In practice, considering that nuclear binding energies are of the order of some MeV, which correspond (in thermal-statistical equilibrium) to absolute temperatures of the order $T_{\text{NS}}\sim 1~\text{MeV}/(k_{B}=8.61\cdot 10^{-11}~\text{MeV~K}^{-1})\sim 10^{10}$~K ($k_{B}$ is the Boltzmann constant), we can estimate that nucleosynthesis within the standard cosmological model takes place at redshift:\footnote{In the above estimation, we have assumed that the cosmic radiation is found in thermodynamic equilibrium, so that it can be approximated by the black-body radiation model. In particular, the radiation density has to be proportional to the fourth power of the cosmic fluid's absolute temperature $T$, in accordance with the Stefan-Boltzmann law,
\begin{equation}
    \rho_{\text{rad}}=\sigma_{SB}T^{4}\,,
    \label{Stefan-Boltzmann}
\end{equation}
where $\sigma_{SB}=5.670\times 10^{-8}~\text{W}~\text{m}^{-2}~\text{K}^{-4}$ represents the Stefan-Boltzmann constant. Note that the combination of (the familiar) $\rho_{\text{rad}}\propto a^{-4}$ and eq.~\eqref{Stefan-Boltzmann} leads to the familiar relation $T\propto a^{-1}$, for the cosmic temperature evolution.} 
 \begin{equation}
     1+z_{\text{NS}}=\frac{T_{\text{NS}}}{T_{p}}\sim 10^{9}\,,\hspace{15mm} \text{which means that} \hspace{15mm} a_{\text{NS}}\sim 10^{-9}a_{p}\,.
     \label{eqn:nucleos}
 \end{equation}
Therefore, the ratio of magnetic over radiation density at nucleosynthesis reads:
\begin{equation}
     \left(\frac{\rho_{B}}{\rho_{\text{rad}}}\right)_{\text{NS}}=\left(\frac{\rho_{B}}{\rho_{\text{rad}}}\right)_{p}\left(\frac{a_{\text{NS}}}{a_{p}}\right)^{6\lambda}=10^{-18}\cdot \left(10^{-9}\right)^{6\lambda}=10^{-54\lambda-18}\,,
     \label{densities-ratio-nucleosynthesis}
 \end{equation}
in accordance with~(\ref{densities-ratio}a) and~\eqref{eqn:nucleos}. A rough requirement for the magnetic fields not to affect the present abundance of nuclei, is that the magnetic density during nucleosynthesis would be of smaller order of magnitude than the radiation density. In other words, the aforementioned requirement translates, via eq.~\eqref{densities-ratio-nucleosynthesis}, into the condition: $\boldsymbol{\lambda}\boldsymbol{\geq}\boldsymbol{-1/3}$.\\

\section{Introductory remarks on magnetised Bianchi I cosmology}\label{Magn-BianchiI-introduction}

Bianchi cosmologies (see~\cite{Land-Lif}-\cite{Ellis} for a list principal works) have traditionally and thoroughly been studied due to their physically interesting anisotropic features. In particular, the well known dipole anisotropy in the cosmic microwave background along with the observation of large-scale (intergalactic) magnetic fields, have both motivated the study of (homogeneous) anisotropic cosmological models.\\
The simplest of the aforementioned models is the Bianchi I, whose isometry group is produced by three commuting generators, Killing vector fields. Its first exact general solution in vacuum space was found by Kasner~\cite{Kasner} (some variations include~\cite{Weyl}-\cite{Levi-Civita}) while, in the presence of dust, by Heckmann and Schucking~\cite{Heck-Schuck}. The latter behaves like the Kasner solution at the beginning of cosmic evolution while it approaches an isotropic Friedmann regime at an advanced stage of expansion. The Heckmann-Schucking solution has been generalised to other kinds of isotropic perfect fluids~\cite{Jacobs0}-\cite{Jacobs1} and to the case of non vanishing cosmological constant~\cite{Khal-Kam,Kam-Ming}. Finally, solutions in the presence of homogeneous anisotropic fields\footnote{It is worth noting that exact solutions with highly anisotropic geometry exist even in empty space or in spaces filled with isotropic matter.}, in particular magnetic fields, have also been explored, basically during the 60s~\cite{Jacobs},~\cite{Rosen}-\cite{Thorne}, but more recently as well~\cite{LeBl}. Although the magnetised Bianchi I is not envisaged a realistic model of the universe, its study can shed new light on processes taking place in very early cosmic stages. Its essential advantage over the FRW model is that it allows for the natural incorporation of the magnetic field. Besides, it is known that in the FRW model, magnetic fields are approximately incorporated in the form of perturbations.\\
Of the various Bianchi models only I, II, III, $\text{VI}_{-1}$ and $\text{VII}_{0}$ are known to be natural hosts of pure large-scale magnetic fields~\cite{HJ}. As far as we have searched, all past works on magnetised Bianchi I cosmology consider a diagonal spacetime metric. In fact, as we remark within the main text, such a metric restricts by construction the orientation of the magnetic field along one of the three independent spatial directions. Moreover, most of the early relevant works (e.g.~\cite{Jacobs},~\cite{Dorosh},~\cite{Thorne} or the more recent~\cite{TM4}) derive a magnetic evolution formula based on the assumption that the magnetic field is an eigenvector of the shear tensor. The assumption in question translates practically into envisaging the magnetic field as the sole source of spatial anisotropy. A common stronger (more restrictive and rarer) assumption encountered in the modern literature, consists of neglecting the shear contribution to magnetic evolution. The latter leads to the well known inverse square law of magnetic variation (e.g. see~\cite{Barrow-magn} and~\cite{KCo}).  \\
We begin the chapter's main part with determining the magnetic evolution formula and writing down explicitly Einstein equations for the Bianchi I metric spacetime, filled with a magnetic field. We then derive the exact full solution of Einstein equations for the model. Subsequently, we move to the innovative part of this work, which involves the presentation of qualitative solutions, describing in detail the small and large-scale cosmic limit. Finally, we end up the chapter with a section introducing and revealing some evolution features of the non-diagonal magnetised Bianchi I model.\\

\section{Bianchi-I diagonal metric model with spatially homogeneous magnetic field}\label{s:BianchiIB}
Let us consider the simplest anisotropically expanding cosmological model, namely the so-called Bianchi I, which has Euclidean spatial sections and is known to allow for the existence of large-scale magnetic fields~\cite{CKPM}.

\subsection{Metric, energy-momentum tensor and magnetic field}\label{ssec:Metric-T-and-B-BianchiI}

The model's diagonal metric reads:
\begin{equation}
    ds^{2}=dt^{2}-A^{2}(t)dx^{2}-B^{2}(t)dy^{2}-C^{2}(t)dz^{2}\,,
    \label{eqn:BianchiI-metric}
\end{equation}
where the average scale-factor is $a=\sqrt[3]{ABC}$. In covariant terms, the only non-vanishing quantities in Bianchi I cosmologies are the relativistic energy density and pressure, the anisotropic stress tensor, the volume scalar, the shear and the electric Weyl tensor (i.e. $\rho$, $P$, $\pi_{ab}$, $\Theta$, $\sigma_{ab}$ and $E_{ab}$ respectively)~\cite{EMM}. All the remaining terms are zero by construction, namely $\omega_{a}=0=\dot{u}_{a}=q_{a}=H_{ab}=\mathcal{R}_{ab}$ (with $\mathcal{R}_{ab}=0$ implying Euclidean spatial sections). It is worth noting that because of their non-zero anisotropic stress tensor ($\pi_{ab}\neq 0$) Bianchi I models can generally host viscous fluids such as the electromagnetic ones, however under the restriction of zero momentum density ($q_{a}=0$). In case of an electromagnetic fluid the aforementioned limitation translates into a zero Poynting vector, $q^{\text{(em)}}_{a}=\epsilon_{abc}E^{b}B^{c}=0$, which means that on considering large-scale magnetic fields, the associated electric components of the Maxwell field have to vanish. This means that the Bianchi I cosmologies satisfy the MHD approximation by construction.\\
We recall that the Lagrangian and the energy-momentum tensor of the electromagnetic field are (in the Heaviside-Lorentz system):
\begin{equation}
L_{\rm (em)} = -\frac{1}{4}\,F_{ik}\,F^{ik}\hspace{5mm}\text{and}\hspace{5mm}T_{\ k}^{i}
= -F^{il}\,F_{kl}
+\frac{1}{4} \, \delta_k^i\, F_{lm}\, F^{lm}\,,
\label{eqn:EM-Lagrangian-and-energy-tensor}
\end{equation}  
where $F_{ab}$ denotes the Faraday tensor. In particular, we consider that our model's spacetime is filled with a homogeneous magnetic field along the direction $z$. Therefore, the only non-vanishing component of the electromagnetic field tensor is $F_{12}$. The sourceless set of Maxwell equations, $\partial_{i}F_{jk}=0$ (in Riemannian framework), implies then that $\partial_{0}F_{12}=0$, namely $F_{12}$ is constant. For the diagonal metric~\eqref{eqn:BianchiI-metric}, the only non-vanishing component of the fully contravariant electromagnetic field tensor is thus given~by: 
\begin{equation}\label{elec5}
F^{12} = g^{11}\,g^{22}\,F_{12} \sim A^{-2}\,B^{-2}\,.
\end{equation}
The above expression means that all the contributions to the mixed components of the energy-momentum tensor in eq.~(\ref{eqn:EM-Lagrangian-and-energy-tensor}b) are proportional to $A^{-2}\,B^{-2}$.
On choosing a convenient parametrization, the individual energy-momentum components can be written as: 
\begin{equation}\label{elec6}
T_{\ 0}^0 = -T_{\ 1}^1 = -T_{\ 2}^2 = T^3_{\ 3} = \frac{\mathcal{B}_0^2}{A^2 B^2}\,,
\end{equation}
where $\mathcal{B}_0^2$ is a positive constant characterizing the intensity of the magnetic field. It is straightforward to check that the trace $T$ of the energy-momentum tensor~\eqref{elec6} vanishes, as it should. \\

\subsection{Einstein equations and evolution formulae}\label{ssec:Einst-eqs-and-evol-formulae}

We are going to work with Einstein equations in the mixed tensor form: 
\begin{equation}\label{Ein}
G^i_{\ j} \equiv R^i_{\ j} - \frac{1}{2}\, R\,\delta^i_{\ j} = T^i_{\ j}\,.
\end{equation}
Adopting the following parametrization for the scale factors: 
\begin{equation} \begin{split}\label{magn9}
A(t) &= a(t) \, e^{\alpha(t)+\beta(t)}\,, \\ 
B(t) &= a(t) \, e^{\alpha(t)-\beta(t)}\,, \\
C(t) &= a(t) \, e^{-2\alpha(t)}\,,
\end{split} \end{equation}
where $a^{3}=ABC$ and the parameters $\alpha(t)$, $\beta(t)$ determine the model's anisotropy; the Ricci components read then:
\begin{eqnarray}
 R_{\ 0}^0
&\!\!=\!\!&
- 3 \frac{\ddot{a}}{a} -6 \dot{\alpha}^2 - 2 \dot{\beta}^2\,,
 \label{magn10}
 \\
 R_{\ 1}^1
 &\!\!=\!\!&
- \frac{\ddot{a}}{a} - 2 \frac{\dot{a}^2}{a^2} - \ddot{\alpha} - 3 \dot{\alpha} \frac{\dot{a}}{a} - \ddot{\beta} - 3 \dot{\beta} \frac{\dot{a}}{a}\,,
 \label{magn11}
 \\
 R_{\ 2}^2
 &\!\!=\!\!&
 - \frac{\ddot{a}}{a} - 2 \frac{\dot{a}^2}{a^2} - \ddot{\alpha} - 3 \dot{\alpha} \frac{\dot{a}}{a} + \ddot{\beta} + 3 \dot{\beta} \frac{\dot{a}}{a}\,,
 \label{magn12}
 \\
R_{\ 3}^3
&\!\!=\!\!&
- \frac{\ddot{a}}{a} - 2 \frac{\dot{a}^2}{a^2} + 2 \ddot{\alpha} + 6 \dot{\alpha}\frac{\dot{a}}{a}\,,
\label{magn13}
\end{eqnarray}
whilst Einstein equations reduce therefore to:
\begin{equation}\label{magn14}
G_{\ 0}^0 = -G^1_{\ 1} = -G^2_{\ 2} = G_{\ 3}^3
= \frac{B_0^2}{a^4}\, e^{-4\alpha}\,,\hspace{6mm}\text{where}\hspace{6mm}R = - 6 \frac{\ddot{a}}{a} - 6 \frac{\dot{a}^2}{a^2} - 6\dot{\alpha}^2 - 2 \dot{\beta}^2
\end{equation}
is the Ricci scalar. Note that the latter must vanish (implying $\dot{\alpha}^2=\dot{\beta}^2/3$) since $T=0$.\\
Taking the difference of the mixed $11$ and $22${--}components of Einstein's equations~\eqref{magn14}
with the expressions~\eqref{magn11} and~\eqref{magn12} for the Ricci tensor, we obtain:
\begin{equation}
R_{\ 1}^1-R_{\ 2}^2 = - 2\ddot{\beta} - 6\, \dot{\beta}\frac{\dot{a}}{a} = 0\,,\hspace{6mm}\text{accepting the solution}\hspace{6mm}\dot{\beta} = \frac{\beta_0}{a^3}\,,
\label{magn20}
\end{equation}
just like in the case of Kasner and Heckmann-Schucking solutions~\cite{Khal-Kam}. Note that it is due to our simplifying assumption: $B^{a}\parallel z^{a}$, that the magnetic field does not enter the temporal evolution equation for $\beta$ parameter. Likewise, combining eqs~\eqref{magn11}, \eqref{magn12}, and~\eqref{magn13} yields to:
\begin{equation}
R_{\ 1}^1+R_{\ 2}^2-2R_{\ 3}^3
= -6\ddot{\alpha}-18\,\dot{\alpha}\frac{\dot{a}}{a} = -\frac{4B_0^2}{a^4}\, e^{-4\alpha}\hspace{6mm}\text{or}\hspace{6mm}\ddot{\alpha}+3\dot{\alpha}\frac{\dot{a}}{a} = \frac{2\,B_0^2}{3\,a^4}\, e^{-4\alpha}\,.
\label{magn22}
\end{equation}
We observe that the magnetic field presence, introducing a preferred spatial direction, directly affects the evolution of the metric (anisotropic) parameter $\alpha$. Moreover, the combination: 
\begin{equation}
R_{\ 1}^1+R_{\ 2}^2+2 R_{\ 3}^3
= -4 \frac{\ddot{a}}{a} - 8\frac{\dot{a}^2}{a^2} + 6\dot{\alpha} \frac{\dot{a}}{a} +2 \ddot{\alpha} = 0\hspace{6mm}\text{leads to}\hspace{6mm}\ddot{\alpha}+3\dot{\alpha}\frac{\dot{a}}{a} = 2\frac{\ddot{a}}{a}+4\frac{\dot{a}^2}{a^2}\,,
\label{magn24}
\end{equation}
a linear differential equation in terms of the anisotropy factor $\alpha$. By multiplying both sides with the spatial volume, $V=a^3$, the above equation recasts into:
\begin{equation}\label{magn28}
\frac{d}{dt}\left(\dot{\alpha} a^3\right)
=\frac{d}{dt}\left(2 \dot{a}\,a^2\right)\hspace{6mm}\text{which upon integration yields}\hspace{6mm} \dot{\alpha} = \frac{2\dot{a}}{a}+\frac{\alpha_0}{a^3}\,,
\end{equation}
where $\alpha_0$ is a constant. Also, making use of eqs.~\eqref{magn22} and \eqref{magn24}, we arrive at:
\begin{equation}\label{magn29}
\frac{1}{a^3}\,\frac{d^2a^3}{dt^2}
= \frac{\mathcal{B}_0^2}{a^4}\, e^{-4\alpha}\,.
\end{equation}
According to the above, we can determine $\alpha(t)$ for a given scale factor $a(t)$. Note that the second time derivative of the spatial volume, $V=a^3$, must always be positive. Substituting eqs.~\eqref{magn10}, (\ref{magn14}b), \eqref{magn20}, \eqref{magn28} and~\eqref{magn29}
into the $(00)${--}component of Einstein eqs.~\eqref{magn14}, we obtain: 
\begin{equation}\label{magn33}
\frac{1}{a^3}\,\frac{d^2a^3}{dt^2}
+ 9 \frac{\dot{a}^2}{a^2}+12 \frac{\alpha_0}{a^3}\frac{\dot{a}}{a}+3\,\frac{\alpha_0^2}{a^6} + \frac{\beta_0^2}{a^6} = 0\hspace{6mm}\text{or}\hspace{6mm} V\,\ddot{V}+\dot{V}^2+4 \alpha_0 \dot{V} + 3\alpha_0^2 + \beta_0^2 = 0\,,
\end{equation}
which consists of a differential equation in terms of the spatial volume.
Remarkably, this equation is integrable. Before proceeding to its solution however, we will present its detailed qualitative analysis in the first place. To begin with, let us note that not all solutions of eq.~\eqref{magn33} solve
the complete system of Einstein and Maxwell equations.
In fact, eq.~\eqref{magn33} gives:
\begin{equation}\label{magn40}
\ddot{V} = -\frac{1}{V}
\left(\dot{V}^2 + 4 \alpha_0\,\dot{V}+3 \alpha_0^2 + \beta_0^2 \right)\,,
\end{equation}
which has to be positive according to~\eqref{magn29}. 
Since $V$ should always be nonnegative, the positivity of $\ddot{V}$ implies that $\alpha_0 \dot V<0$, with $\alpha_0\neq 0$, and
\begin{equation}\label{magn41}
-2\alpha_0 -\sqrt{\alpha_0^2-\beta_0^2} \, 
\leq \, \dot{V} \leq -2\alpha_0+\sqrt{\alpha_0^2-\beta_0^2}\,,\hspace{4mm}\text{where}\hspace{4mm}\alpha_0^2 \geq \beta_0^2\,.
\end{equation}
According to the above condition, the model's rate of contraction/expansion is constrained between two limiting values.\\

\subsection{Qualitative analysis of the model's small and large-scale limit}\label{ssec:Qualitative-analysis}

In the following we present a detailed, qualitative description of the model's behaviour at its small and large-scale limit. Our analysis is based on the consideration of approximate solutions obeying the above presented evolution constraints.
\subsubsection{Contracting universe:~small scale limit}\label{ssec:}
Let us first consider $\alpha_0>0$, implying a contracting universe
with $\dot V<0$. One can start at a certain moment in time with a positive value of $V$ and a negative value of $\dot{V}$, satisfying the inequality~\eqref{magn41}. Since $\ddot{V}>0$, the time derivative of $V$ grows, remaining negative, and the absolute value of $\dot{V}$ decreases,
always obeying the constraint:
\begin{equation}\label{magn43}
\bigl| \dot{V} \bigr| \, \geq \, 2 \alpha_0 - \sqrt{\alpha_0^2-\beta_0^2}
\equiv W_1 > 0\,.
\end{equation}
The universe will therefore reach the singularity characterised by $V = 0$ in
a finite period of time. Subsequently, let us consider two time instants, $t_1$ and $t_2$, such that 
\begin{equation}\label{time}
V(t_1) = 0 \quad \text{and} \quad 
\dot{V}(t_2) = -W_1\,,
\end{equation}
and let us try to understand which happens first. Suppose that $t_1 < t_2$, so that $V$ vanishes while $\dot{V}$ still satisfies
the inequality~\eqref{magn41}, with $|\dot{V}|$ larger than the critical value $W_1$, in accordance with eq.~\eqref{magn43}. Besides, the time $t_1$ cannot be infinite because the absolute value of the time derivative is larger
than $W_1$ and the function $V=V(t)$ reaches zero in a finite period of time. Therefore, we can approximate the volume function for $t\lesssim t_1$ with expression:
\begin{equation}\label{dif3}
V \simeq \sigma \,(t_1-t)^{\lambda}\,,
\end{equation}  
where $\sigma$ and $\lambda$ are positive constants. For $\lambda > 1$, the velocity becomes: 
\begin{equation}\label{dif4}
\dot{V} \simeq -\lambda\, \sigma \,(t_1-t)^{\lambda -1} \to 0 \quad {\rm for} \quad\ t \to t_1\,,
\end{equation}
which contradicts the condition~\eqref{magn43}. On the other hand, if $\lambda < 1$, the velocity in eq.~\eqref{dif4} diverges for $t\to t_1$, which violates the bound~\eqref{magn41}. The only possible choice left is $\lambda=1$, in which case we need another term in the expansion around $t_1$, namely
\begin{equation} \label{dif5}
V \simeq \sigma \,(t_1-t) + \eta_1 \,(t_1 -t)^{\mu}
\ ,
\end{equation}
where $\eta_1$ is a positive constant and $\mu > 1$. 
Substituting the associated expressions for $V$, $\dot{V}$, and $\ddot{V}$ into eq.~\eqref{magn33}, we have:
\begin{equation}\label{dif8}
\mu \, \eta_1 \,\left(\mu-1\right) \left[\sigma (t_1-t) + \eta_1 (t_1-t)^{\mu}\right]
(t_1-t)^{\mu-2}
\simeq 
-\left[\sigma +\mu\, \eta_1 \, (t_1-t)^{\mu-1} -2\alpha_0 \right]^2 
+\alpha_0^2 - \beta_0^2\,. 
\end{equation}
The leading term in the left-hand side above behaves as $(t_1-t)^{\mu-1}$,
which vanishes for $t\to t_1$. In the right-hand side, the leading term is instead a constant, which should therefore vanish, so that
\begin{equation}\label{dif9}
\sigma^2-4\alpha_0\sigma +3\alpha_0^2 + \beta_0^2 = 0\hspace{4mm}\text{for}\hspace{4mm} t\rightarrow t_{1}\,.
\end{equation}
One of the solutions of this equation is $\boldsymbol{\sigma} \boldsymbol{=} \boldsymbol{W_1}$,  which means that $\dot{V}$ reaches the critical value $W_1$ at the same time when the 
volume $V$ vanishes, so that $t_1=t_2$. Next, we equate terms of order $(t_1-t)^{\mu-1}$ in eq.~\eqref{dif8},
\begin{equation}\label{dif11}
\mu\, \eta_1 (\mu-1)\, \sigma
= -2\mu\, \eta_1 (\sigma -2 \alpha_0)\,,\hspace{6mm}\text{which gives}\hspace{6mm} \mu = 1 + \frac{2\sqrt{\alpha_0^2-\beta_0^2 \,}}{2\alpha_0-\sqrt{\alpha_0^2-\beta_0^2 \,}}\,,
\end{equation}
where $\boldsymbol{1} \boldsymbol{<} \boldsymbol{\mu} \boldsymbol{\leq} \boldsymbol{3}$. As for the constant $\eta_1$, note that it takes different values depending on the initial conditions. It is worth noting that we could also consider the case of $\dot V$ reaching the critical value $-W_1$
at the moment $t_2<t_1 $ while the volume $V(t_2) > 0$. Nevertheless, a simple analysis similar to that presented above, shows that this case is excluded.
Thus, we can say that for any contracting evolution, the universe hits the singularity $V=0$ at some finite time $t_1$ when the velocity $\dot{V}$ reaches the critical value $-W_1$.\\
From the (approximate) evolution law of the volume, we can determine the anisotropy factors
$\alpha(t)$ and $\beta(t)$.
Using eqs.~\eqref{magn20} and~\eqref{magn28}, we directly find out that:
\begin{equation}\label{dif14}
\beta(t) = \beta_0 \int \frac{d t}{V} \simeq
-\frac{\beta_0\,\ln(t_1-t)}{2\alpha_0-\sqrt{\alpha_0^2-\beta_0^2}}\hspace{6mm}\text{and}\hspace{6mm} \alpha(t) \simeq
\left(\frac23 -\frac{\alpha_0}{2\alpha_0-\sqrt{\alpha_0^2-\beta_0^2}}\right)\ln(t_1-t)\,.
\end{equation} 
For definiteness, let us set $\beta_0 \geq 0$.
Using definitions~\eqref{magn9}, we can write the three scale factors 
in the Kasner form:
\begin{equation}\label{dif19}
A(t)\!\!\sim\!\!\left(t_1-t\right)^{p_1}\,,\hspace{6mm} B(t)\!\!\sim\!\!\left(t_1-t\right)^{p_2}\,, \hspace{6mm}
C(t)\!\!\sim\!\!\left(t_1-t\right)^{p_3}\,,
\end{equation}
where
\begin{eqnarray}\label{dif20}
p_1
&\!\!=\!\!& 
\frac{\alpha_0-\beta_0-\sqrt{\alpha_0^2-\beta_0^2}}{2\alpha_0-\sqrt{\alpha_0^2-\beta_0^2}} < 0\,,
\nonumber
\\
p_2
&\!\!=\!\!&
\frac{\alpha_0+\beta_0-\sqrt{\alpha_0^2-\beta_0^2}}{2\alpha_0-\sqrt{\alpha_0^2-\beta_0^2}} > 0\,,
\\
p_3
&\!\!=\!\!&
\frac{\sqrt{\alpha_0^2-\beta_0^2}}{2\alpha_0-\sqrt{\alpha_0^2-\beta_0^2}} >0\,.
\nonumber
\end{eqnarray}
It is straightforward to check that the exponents $p_1$, $p_2$, and $p_3$ indeed satisfy the Kasner relations, i.e.
\begin{equation}
p_1+p_2+p_3 = p_1^2+p_2^2+p_3^2=1\,,
\label{dif21}
\end{equation}
which means that the presence of the magnetic field does not change the character of the singularity. Such a behaviour is plausible. In detail, substituting expressions for $V$ and $\ddot{V}$ into eq.~\eqref{magn29},
one obtains that the magnetic field contributes to Einstein's equations the quantity:
\begin{equation}\label{dif22}
\frac{\mathcal{B}_0^2}{A^2 B^2}
\sim \frac{\mu\,\eta_1}{\sigma} \, (\mu-1)\,(t_1-t)^{\mu-3}
\ ,
\end{equation}
where $\mu-3 > -2$.
Therefore, the term~\eqref{dif22} is weaker than the anisotropy, which contributes a term of order
$(t_1-t)^{-2}$ and dominates near the singularity. Besides, the presence of matter less stiff than stiff matter is well known to leave the Kasner type of singularity unaffected~\cite{Bel-Khal}.
\subsubsection{Expanding universe:~small, large-scale limit and transition}\label{sssec:Expanding-universe-limits}
Let us consider now the expanding universe.
In this case, we have $\alpha_0<0$ and therefore the time derivative $\dot{V}$ will be positive.
The expansion will last for $t\to\infty$ with the time derivative $\dot{V}$ approaching the critical value:
\begin{equation}\label{V2}
W_2 = -2 \alpha_0 + \sqrt{\alpha_0^2-\beta_0^2}
 >0.
\end{equation}
Thus, the behavior of $V=V(t)$ can be approximated in the limit $t\to\infty$ as: 
\begin{equation}\label{dif23}
V \simeq W_2\,t-\eta_2\,t^{\nu}
\ ,
\end{equation} 
where $\eta_2$ is a positive constant and $0 < \nu < 1$\,.
Then, substituting the associated expressions for $V$, $\dot{V}$, and $\ddot{V}$ in eq.~\eqref{magn33}, we find once again that:
\begin{equation}\label{dif26}
\nu = 1 + \frac{2\sqrt{\alpha_0^2-\beta_0^2\, }}{2 \alpha_0 - \sqrt{\alpha_0^2-\beta_0^2}}
\ ,
\end{equation}
with $\dfrac13 \leq \nu < 1$.
The anisotropy factors then read:
\begin{equation}\label{dif29}
\alpha= \left(\frac23+\frac{\alpha_0}{\sqrt{\alpha_0^2-\beta_0^2}-2\alpha_0} \right) \ln{t}\hspace{6mm}\text{and}\hspace{6mm}\beta = \frac{\beta_0\, \ln{t}}{\sqrt{\alpha_0^2-\beta_0^2}-2\alpha_0}\,.
\end{equation}
Overall, the scale factors take once again the familiar Kasner form,
\begin{equation}\label{dif30}
A(t) \sim 
t^{p_1}\,, \quad 
B(t) \sim 
t^{p_2}\,, \quad 
C(t) \sim 
t^{p_3}\,,
\end{equation}
where 
\begin{eqnarray}
p_1
&\!\!=\!\!&
\frac{\sqrt{\alpha_0^2-\beta_0^2}-\alpha_0+\beta_0}{\sqrt{\alpha_0^2-\beta_0^2}-2\alpha_0}\,,
\nonumber
\\
p_2
&\!\!=\!\!&
\frac{\sqrt{\alpha_0^2-\beta_0^2}-\alpha_0-\beta_0}{\sqrt{\alpha_0^2-\beta_0^2}-2\alpha_0}\,,
\label{dif31}
\\
p_3
&\!\!=\!\!&
-\frac{\sqrt{\alpha_0^2-\beta_0^2}}{\sqrt{\alpha_0^2-\beta_0^2}-2\alpha_0}\,,
\nonumber
\end{eqnarray} 
which also satisfy Kasner relations~\eqref{dif21}.
The presence of the magnetic field does not influence the asymptotic structure of the metric at $t \to\infty$,
which therefore does not isotropize, unlike the Heckmann-Schucking solution with dust~\cite{Heck-Schuck,Khal-Kam}. The reason for this phenomenon is clear. 
The energy density of the magnetic field at $t \to\infty$ is given by:
\begin{equation}\label{dif32}
\frac{\mathcal{B}_0^2}{A^2B^2} \simeq \nu\,(\nu-1)\,\frac{\eta_2}{W_2}\,t^{\nu-3},  
\end{equation} 
where $\nu-3 < -2$.
Hence it remains weaker than the anisotropy term.\\
Subsequently, let us examine what happens with the expanding universe in its distant past. One can suppose that it was born from the initial singularity at $t=0$, when its volume was $V(0)=0$, and its time derivative had the smallest critical value, 
\begin{equation}
\dot V(0) = W_3 \equiv -2\alpha_0-\sqrt{\alpha_0^2-\beta_0^2}>0\,.
\label{dif33}
\end{equation}
In this case, the scale factors will be of the form of eq.~\eqref{dif30}, with the Kasner exponents reading
\begin{eqnarray}\label{dif34}
p'_1 
&\!\!=\!\!&
\frac{\alpha_0-\beta_0+\sqrt{\alpha_0^2-\beta_0^2}}{2\alpha_0+\sqrt{\alpha_0^2-\beta_0^2}}\,,
\nonumber
\\
p'_2
&\!\!=\!\!&
\frac{\alpha_0+\beta_0+\sqrt{\alpha_0^2-\beta_0^2}}{2\alpha_0+\sqrt{\alpha_0^2-\beta_0^2}}\,,
\\
p'_3
&\!\!=\!\!&
\frac{\sqrt{\alpha_0^2-\beta_0^2}}{2\alpha_0+\sqrt{\alpha_0^2-\beta_0^2}}\,.
\nonumber
\end{eqnarray} 
They again satisfy the Kasner relations~\eqref{dif21}.\\
Aiming to establish a relation between the set of Kasner indices at the beginning and at the end of the evolution, it is convenient to use the
Lifshitz--Khalatnikov parametrization~\cite{Lif-Khal}. 
If the Kasner indices are ordered as:
\begin{equation}
p_1 \leq p_2 \leq p_3,
\label{dif35}
\end{equation}
they can be represented by means of a real parameter $u\geq 1$ according to:
\begin{eqnarray}
p_1
&\!\!=\!\!&
-\frac{u}{1+u+u^2}\,, \nonumber \\
p_2
&\!\!=\!\!&
\frac{1+u}{1+u+u^2}\,,
\label{dif36} \\
p_3
&\!\!=\!\!&
\frac{u\,(1+u)}{1+u+u^2}\,.
\nonumber
\end{eqnarray}
The ordering~\eqref{dif35} can be obtained, for example, by 
setting the anisotropy parameters like:
\begin{equation}
\alpha_0 < 0\,,
\quad
\beta_0 < 0\,,
\quad
|\beta_0| < \frac35\,|\alpha_0|\,.
\label{dif37}
\end{equation}
In particular, this choice implies that the universe expands in the $y$ and $z$ directions,
but does so more rapidly along the direction $z$ of the magnetic field, while it contracts along the $x$ axis.
Combining eqs.~\eqref{dif34} and~\eqref{dif37}, we obtain:
\begin{equation}
u' = \frac{p_3'}{p_2'} =
\frac{\sqrt{\alpha_0^2-\beta_0^2}}{|\alpha_0|+|\beta_0|-\sqrt{\alpha_0^2-\beta_0^2}}\,.
\label{dif38}
\end{equation}
It is convenient to introduce the parameter $\boldsymbol{\xi} \boldsymbol{\equiv} |\boldsymbol{\beta_0}|/|\boldsymbol{\alpha_0}|$, which, when plugged into eq.~\eqref{dif38}, results in the relation:
\begin{equation}
\xi = \frac{(u'+1)^2-u'^2}{(u'+1)^2+u'^2}\,.
\label{dif39}
\end{equation} 
Note that if the parameter $\xi$ satisfies the conditions~\eqref{dif37}, then 
$u'$ satisfies the conditions $1 < u' < \infty$. 
Let us look now at the Kasner exponents~\eqref{dif31}, describing the final stage of the cosmological evolution. 
In this case, the order of the exponents is: $p_3 \leq p_1 \leq p_2$.  Therefore, their representation in terms of $u$ reads: 
\begin{eqnarray}
p_1
&\!\!=\!\!&
\frac{1+u}{1+u+u^2}\,, \label{dif41} \nonumber \\
p_2
&\!\!=\!\!&
\frac{u\,(1+u)}{1+u+u^2}\,, \\
p_3 
&\!\!=\!\!&
-\frac{u}{1+u+u^2}\,, \nonumber 
\end{eqnarray}
where
\begin{equation}
u = \frac{p_2}{p_1} =
\frac{\sqrt{\alpha_0^2-\beta_0^2}-\alpha_0-\beta_0}{\sqrt{\alpha_0^2-\beta_0^2}-\alpha_0+\beta_0}\,.
\label{dif42}
\end{equation}
Substituting the formulae~\eqref{dif38} and \eqref{dif39} into the equation above, we find:
\begin{equation}
u = \frac{1+u'}{u'} < 2\hspace{6mm}\text{or inversely}\hspace{6mm}u' = \frac{1}{u-1}\,.
\label{dif43}
\end{equation}
Considering the temporal evolution towards the singularity (e.g. like in the oscillating approach towards the cosmological singularity~\cite{Bel-Hen, BKL}), we can see that the universe passes through two
transformations in the transition from the parameter $u$ to the parameter $u'$, according to eq.~(\ref{dif43}b). The first transformation is characterised by the shift $u \to u-1$, in which the roles of the $x$ and $z$ axes, corresponding to the exponents $p_1$ and $p_3$ respectively, are exchanged.
This transformation is called `change of Kasner epoch'~\cite{BKL}.
As a result of this transformation, we arrive at a value of the parameter $u-1<1$, see eq.~\eqref{dif43}. The next transformation is defined by $u-1 \to \dfrac{1}{u-1}$, which exchanges the roles of the axes $y$ and $z$, and is called  `change of Kasner era'.\\
We see that our solution displays a transition between two Kasner regimes
chacterized by the same law which is found in an empty Bianchi-II universe. For a detailed description of the dynamics in the Bianchi-II universe in General Relativity and other gravity models, see Ref.~\cite{Giani-Kam}. In Bianchi-VIII or Bianchi-IX models, the universe passes through an infinite series
of changes of the Kasner epochs and eras. In our case, for the choice of parameters in eq.~\eqref{dif37}, the law of transformation for the Kasner exponents~(\ref{dif43}b) includes one change of Kasner epoch and one change of Kasner era.\\
On the other hand, it is worth considering the opposite relation between the anisotropy parameters, to wit
\begin{equation}
\alpha_0 < 0, \quad \beta_0 < 0,
\quad \frac35\,|\alpha_0| < |\beta_0| < 1\,,
\label{dif45}
\end{equation}
or, in other words: $\dfrac35 < \xi < 1$\,. 
In this case, we have:
\begin{equation}
u' = \frac{p_2'}{p_3'}\,,\hspace{6mm}\text{corresponding to}\hspace{6mm}\xi = \frac{(u'+1)^2-1}{(u'+1)^2+1}\,,
\label{dif49}
\end{equation}
with $u = u'+1$ or inversely $u'=u-1$. The last two relations show that we now have only a change of Kasner epoch.\\
\subsection{Exact evolution}
Let us turn back to eq.~\eqref{magn33}, written now as:
\begin{equation}
\frac{d}{dt}\left(V\,\dot{V}+4\alpha_0\, V\right) = -3\alpha_0^2-\beta_0^2\,,
\label{exact}
\end{equation}
which can be analytically integrated through the variable change $V(t) = X(t)\, t$. Using that, our equation recasts into the solvable form:
\begin{equation} \label{magn46}
\frac{X\,dX}{X^2+4\alpha_0\,X+3\alpha_0^2+\beta_0^2}
= -\frac{dt}{t}\,.
\end{equation}
Upon integration of the above equation, we obtain:
\begin{eqnarray}\label{magn48}
&&\frac{\alpha_0}{\sqrt{\alpha_0^2-\beta_0^2}}
\left[\ln\biggl(1+\frac{X}{2\alpha_0+\sqrt{\alpha_0^2-\beta_0^2}}\biggr)
- \ln\biggl(1+\frac{X}{2\alpha_0-\sqrt{\alpha_0^2-\beta_0^2}}\biggr)
\right] \nonumber\\
&&=-\frac{1}{2}\ln\left(1+\frac{X^2+4\alpha_0X}{3\alpha_0^2+\beta_0^2}\right)
+\ln\left(\frac{t_0}{t}\right)\,,
\end{eqnarray}
where integration constants are chosen so that $t_0>0$ is the time at which the volume $V$ (hence $X$) vanishes. Unfortunately, eq.~\eqref{magn48} cannot be inverted to find $X$ as a function of $t$.
Thus, we cannot make use of it to find precise expressions for the anisotropy factors. However, we can use it to study the approach to the singularity for $t\to t_0$ when $X \to 0$.
Expanding the left-hand side of eq.~\eqref{magn48}, we see that the leading term is proportional to $X^2$, and the equation reduces to 
\begin{equation}\label{magn49}
X^2(t)
\simeq
2\left(3\alpha_0^2 + \beta_0^2 \right)\frac{t_0-t}{t_0}\,,
\end{equation} 
which implies that
\begin{equation}\label{magn50}
V(t) \simeq \sqrt{2\bigl(3\alpha_0^2 + \beta_0^2 \bigr)t_0\,(t_0-t)}\hspace{6mm}\text{and}\hspace{6mm}\dot{V} \simeq -\sqrt{\frac{\bigl(3\alpha_0^2 + \beta_0^2 \bigr)t_0}{2(t_0-t)}} \to -\infty\,.
\end{equation}
Note that~(\ref{magn50}b) breaks the inequality~\eqref{magn41}. This means that solution~\eqref{magn49} corresponds to initial conditions
which are incompatible with the existence of a homogeneous magnetic field.
However, solution~\eqref{magn48} is associated with a different regime, corresponding to a volume singularity given by $t \to 0$, with $X(0)$ finite.
In this case, the function $X(t)$ tends to a positive constant which can be found
from a quadratic equation, and corresponds to the extremal points of the
inequality~\eqref{magn41}. Hence, the regime of singularity approach (vicinity of $t=0$) described above, is compatible with the general solution of eq.~\eqref{exact}.\\

\section{Magnetised Bianchi I cosmology with non-diagonal metric}\label{sec:Non-Diagonal-Bianchi-I}

In the present section we address the following question: \textit{What happens to cosmic evolution if we consider a magnetic field oriented along a general spatial direction, instead of $z$?}. To begin with, let us remark that in general, the anisotropic-magnetic pressure $\pi_{ab}=-B_{\langle a}B_{b\rangle}$, contributes some non-diagonal terms in the energy-momentum tensor. Therefore, it is clear that a magnetic field of arbitrary direction is incompatible with the diagonal Bianchi I metric, which predicts, via Einstein equations, a diagonal energy-momentum tensor. Wishing to consider an arbitrary magnetic field, we study the non-diagonal Bianchi I case.\\

\subsection{Metric, energy-momentum tensor and magnetic field}\label{ssec:Metric-T-and-B-BianchiI-non-diagonal}

In general, a homogeneous-anisotropic (Bianchi type) spacetime model can be described by the synchronous frame metric~\cite{Land-Lif}:
\begin{equation}
    ds^{2}=dt^{2}-\gamma_{\alpha\beta}(t)dx^{\alpha}dx^{\beta}\,,
    \label{eqn:synchr-metric}
\end{equation}
where $\gamma_{\alpha\beta}(t)$ represents the spatial metric. The dot derivative of the latter defines the extrinsic curvature tensor $\kappa_{\alpha\beta}\equiv\dot{\gamma}_{\alpha\beta}$ (with $\kappa\equiv\kappa^{\alpha}_{\alpha}$ we denote its trace). Taking now into account that the contravariant component of the magnetic field is proportional to the inverse square root of the spatial metric's determinant~\cite{Padm}, we determine its density evolution rate\footnote{Note that in this general (non-diagonal) case, we determine evolution rates in terms of the metric determinant, which is now used as a scale factor.}:
\begin{equation}
    B^{a}\propto \frac{\mathcal{B}_{0}}{\sqrt{\gamma}}\hspace{6mm}\text{and}\hspace{6mm} \mathcal{B}^{2}=B^{a}B_{a}=g_{ab}B^{a}B^{b}\propto \frac{\mathcal{B}_{0}}{\gamma}\,.
    \label{eqn:magn-field-non-diagonal-BianchiI}
\end{equation}
Assuming that the spacetime in question is filled with the magnetic field $B^{a}$, Einstein equations for the Bianchi I model read (refer to~\cite{Land-Lif} for details):
\begin{equation}
    R^{0}_{0}-\frac{1}{2}R\delta^{0}_{0}=\frac{\mathcal{B}^{2}}{2}\hspace{2mm}\rightarrow\hspace{2mm} \kappa^{2}-\kappa_{\alpha}^{\beta}\kappa_{\beta}^{\alpha}=4\mathcal{B}^{2}=\frac{4\mathcal{B}^{2}_{0}}{\gamma}
    \label{eqn:Einst-eqs-zero}
\end{equation}
and
\begin{equation}
    R_{\alpha}^{\beta}-\frac{1}{2}R\delta_{\alpha}^{\beta}=T_{\alpha}^{\beta}\hspace{2mm}\rightarrow\hspace{2mm} \dot{\kappa}_{\alpha}^{\beta}-\dot{\kappa}\delta_{\alpha}^{\beta}+\frac{\kappa}{2}\kappa_{\alpha}^{\beta}-\frac{1}{4}\left(\kappa^{2}+\kappa_{\gamma}^{\delta}\kappa_{\delta}^{\gamma}\right)\delta_{\alpha}^{\beta}=\frac{1}{2}B_{\alpha}B^{\beta}+\frac{1}{4}\mathcal{B}^{2}\delta_{\alpha}^{\beta}\,,
    \label{eqn:Einst-eqs-spatial}
\end{equation}
where we have considered the magnetic density $\mathcal{B}^{2}/2$, isotropic and anisotropic pressure, $\mathcal{B}^{2}/6$ and $-B_{a}B_{b}+(1/3)\mathcal{B}^{2}h_{ab}$, respectively. Taking now the trace of~\eqref{eqn:Einst-eqs-spatial}, we find out that
\begin{equation}
    \dot{\kappa}+\frac{1}{8}\kappa^{2}+\frac{3}{8}\kappa_{\alpha}^{\beta}\kappa_{\beta}^{\alpha}=-\frac{3\mathcal{B}^{2}}{8}=-\frac{3\mathcal{B}^{2}_{0}}{8\gamma}\,.
    \label{eqn:Einst-eqs-spatial2}
\end{equation}
Then, multiplying~\eqref{eqn:Einst-eqs-zero} by $3/8$ and adding the result to~\eqref{eqn:Einst-eqs-spatial2}, we arrive at the equation:
\begin{equation}
    \dot{\kappa}+\frac{1}{2}\kappa^{2}=\frac{b^{*}}{\gamma}\hspace{2mm}\rightarrow\hspace{2mm} \gamma \hat{y}+y=\frac{2b^{*}}{\gamma}\,,
    \label{eqn:dot-kappa-y-hat}
\end{equation}
where $b^{*}\equiv(9\mathcal{B}^{2}_{0})/8$ is a constant, $y\equiv\kappa^{2}$ and the hat denotes differentiation with respect to $\gamma$.

\subsection{Evolution formula and limits}\label{ssec:evol-formula-and-limits}

Note that~\eqref{eqn:dot-kappa-y-hat} is a linear differential equation in reference to $\gamma$, the integration of which directly leads to:
\begin{equation}
    y\equiv\kappa^{2}=\frac{1}{\gamma}\ln{\left(c_{1}\gamma\right)}^{2b^{*}}\hspace{5mm}\text{or equivalently to}\hspace{5mm} \dot{\gamma}=\sqrt{2b^{*}\gamma}\sqrt{\ln{\left(c_{1}\gamma\right)}}\,.
    \label{eqn:dot-gamma}
\end{equation}
The last equation takes the integrable form:
\begin{equation}
    \frac{d\gamma}{\sqrt{2b^{*}\gamma}\sqrt{\ln{\left(c_{1}\gamma\right)}}}=dt\hspace{2mm}\rightarrow\hspace{2mm} t=\frac{1}{\sqrt{b^{*}}}\int{\frac{dx}{\sqrt{\ln{\left(\sqrt{c_{1}}|x|\right)}}}}\,,
    \label{eqn:t-x-integral}
\end{equation}
where $x\equiv\sqrt{\gamma}\geq 0$. Subsequently, under the variable change $\omega\equiv\sqrt{\ln{\left(\sqrt{c_{1}|x|}\right)}}$, our integral eventually recasts into the Gaussian form:
\begin{equation}
    t=\sqrt{\frac{\pi}{b^{*}c_{1}}}\int{\frac{2e^{\omega^{2}}}{\sqrt{\pi}}~d\omega}\hspace{2mm}\rightarrow\hspace{2mm} t\approx e^{\omega^{2}}\approx\sqrt{\gamma}\hspace{4mm}\rightarrow\hspace{2mm}\boldsymbol{\gamma}\mathbf{\propto t^{2}}\hspace{4mm} (\text{for}~t\rightarrow 0~\text{or}~\infty)\,.
    \label{eqn:t-gamma-R}
\end{equation}
The above shows that the model's scale factor (i.e. the metric determinant) increases parabolically with cosmic time at the limits. In deriving the last expression, we have made use of the following approximation at the limits:
\begin{equation}
    t=c\int{e^{\omega^{2}}~d\omega}\approx \int{e^{\omega^{2}+\ln{\omega}+\ln{c}}~d\omega}\approx \int{\omega~ e^{\omega^{2}}~d\omega}\approx e^{\omega^{2}}\hspace{4mm}\text{for}~~\omega\rightarrow\pm\infty\,.
    \label{eqn:exp-approx-limits}
\end{equation}
In comparison to the diagonal magnetised model, we observe that once again the behaviour at the limits remains of Kasner type. According to eq.~\eqref{eqn:t-gamma-R}, in the non-diagonal case, cosmic time is an error function of the metric determinant:
\begin{equation}
    t=\sqrt{\frac{\pi}{b^{*}c_{1}}}\text{erfi}\left(\sqrt{\ln{\left(\sqrt{c_{1}\gamma}\right)}}\right)\,,
    \label{eqn:t-gamma-full-solution}
\end{equation}
where $k_{1}$, $k_{2}$ are constants. Subsequently, let us check whether it is possible to get a particular solution for the model's metric. In the first place, taking the dot derivative of~(\ref{eqn:dot-gamma}a), we deduce an expression for $\dot{\kappa}$ and, via eq.~\eqref{eqn:Einst-eqs-spatial2}, an expression for $\kappa_{\alpha}^{\beta}\kappa_{\beta}^{\alpha}$:
\begin{equation}
    \dot{\kappa}=-\frac{1}{2\gamma}\ln{\left(c_{1}\gamma\right)}^{2b^{*}}+\frac{b^{*}}{c_{1}\gamma}\hspace{6mm}\text{and}\hspace{6mm} \kappa_{\alpha}^{\beta}\kappa_{\beta}^{\alpha}=\ln{\left(C\gamma\right)}^{\frac{2b^{*}}{\gamma}}\,,
    \label{eqn:dot-k-and-k_{a}^{b}-square}
\end{equation}
with $c_{1}$ and $C$ constants. From the above formulae we find out that the extrinsic curvature has the following form:
\begin{equation}
\kappa_{\alpha}^{\beta}=\sqrt{\ln{\left(C\gamma\right)}^{\frac{2b^{*}}{\gamma}}}\lambda_{\alpha}^{\beta}\hspace{5mm}\text{with}\hspace{5mm} \lambda_{\alpha}^{\beta}\lambda_{\beta}^{\alpha}=1\hspace{5mm}\text{and}\hspace{5mm} \text{Tr}~\lambda=1\,,
    \label{eqn:k_{a}^{b}}
\end{equation}
where $\lambda_{\alpha}^{\beta}$ is a $3\times 3$ symmetric matrix with real constant elements. Given that any real symmetric matrix is diagonalisable, we can write the diagonal form of $\lambda_{\alpha}^{\beta}$ as $\text{diag}(p_{1}, p_{2}, p_{3})$, where $p_{1}, p_{2}, p_{3}$ (adopting the Kasner notation) are its eigenvalues, satisfying the constraints:
\begin{equation}
    p_{1}+p_{2}+p_{3}=1\hspace{6mm}\text{and}\hspace{6mm} p^{2}_{1}+p^{2}_{2}+p^{2}_{3}=1\,.
    \label{eqn:p_{i}-constraints}
\end{equation}
The aforementioned conditions are a direct consequence of~(\ref{eqn:k_{a}^{b}}c) and~(\ref{eqn:k_{a}^{b}}b). Now lowering indices in~(\ref{eqn:k_{a}^{b}}a), using the $3$-D metric $\gamma_{\alpha\beta}$, and deploying its diagonal form, we get:
\begin{equation}
\kappa_{\alpha\alpha}\equiv\dot{\gamma}_{\alpha\alpha}=p_{\alpha}\sqrt{\ln{\left(C\gamma\right)}^{\frac{2b^{*}}{\gamma}}}\gamma_{\alpha\alpha}=p_{\alpha}\sqrt{\frac{2b^{*}}{\gamma}}\sqrt{\ln{\left(C\gamma\right)}}\gamma_{\alpha\alpha}\,,
\label{eqn:gamma-diagonal}
\end{equation}
where the repeated index $\alpha$ does not imply summation here, it just denotes the diagonal components. Although we can not solve~\eqref{eqn:t-gamma-full-solution} for $\gamma$ in the general case, we can make use of the small and large-scale limit, so that~\eqref{eqn:gamma-diagonal} transforms into
\begin{equation}
    \frac{d\gamma_{\alpha\alpha}}{\gamma_{\alpha\alpha}}=\frac{\mathcal{D}}{t}\sqrt{\ln{\left(Ct^{2}\right)}}~dt\,,\hspace{2mm}\text{which upon integration yields to}\hspace{2mm} \gamma_{\alpha\alpha}=\gamma_{0}e^{\frac{\mathcal{D}}{3}\ln^{3/2}{\left(Ct^{2}\right)}}\,,
    \label{eqn:metric-solution-limits}
\end{equation}
where $\mathcal{D}\equiv p_{\alpha}\sqrt{\frac{2b^{*}}{\gamma_{0}}}$ is a constant. The above approximate solution determines the rate of change of the diagonal metric components at the limits. The question regarding the non-diagonal components remains open.\\

\subsection{Concluding remarks}\label{ssec:concluding-remarks}

Throughout our analysis we have seen that the magnetic field affects cosmic evolution in a double manner; first, directly as an energy density source; second, indirectly as a source of anisotropy (due to its vector nature), contributing to the evolution of the anisotropy metric parameters $\alpha$ and $\beta$. In other words, it manifests its presence in both temporal (as energy density) and spatial Einstein equations (via its anisotropic pressure).\\
In the diagonal case-compatible with magnetic field along a sole spatial direction-, we have derived the full general solution under a clarifying and convenient parametrisation. More importantly, our detailed qualitative analysis of the small and large-scale cosmic limit has revealed the model's Kasner type behaviour at both limits, as well as the transition between those via a `change of Kasner epoch' and a subsequent `change of Kasner era'.\\
In the non-diagonal case-allowing for a generally oriented magnetic field-, the general solution is given by cosmic time as a kind of error function of the average scale factor (recall eq.~\eqref{eqn:t-gamma-full-solution}).

\clearpage
\newpage

\addcontentsline{toc}{chapter}{\protect\numberline{}Last word}

\chapter*{Last word}\label{chap:last-word}

Wherever they appear in our universe, either in astrophysical or in cosmological environments, electromagnetic fields are impelled by gravity to manifest their extraordinary features. The phenomena arising from the gravito-electromagnetic coupling await our exploration.

\end{document}